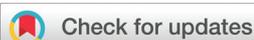

# Hot carriers in graphene – fundamentals and applications

Mathieu Massicotte, [a] Giancarlo Soavi, [b,c] Alessandro Principi [d] and
Klaas-Jan Tielrooij [*e]

Hot charge carriers in graphene exhibit fascinating physical phenomena, whose understanding has improved greatly over the past decade. They have distinctly different physical properties compared to, for example, hot carriers in conventional metals. This is predominantly the result of graphene's linear energy–momentum dispersion, its phonon properties, its all-interface character, and the tunability of its carrier density down to very small values, and from electron- to hole-doping. Since a few years, we have witnessed an increasing interest in technological applications enabled by hot carriers in graphene. Of particular interest are optical and optoelectronic applications, where hot carriers are used to detect (photo-detection), convert (nonlinear photonics), or emit (luminescence) light. Graphene-enabled systems in these application areas could find widespread use and have a disruptive impact, for example in the field of data communication, high-frequency electronics, and industrial quality control. The aim of this review is to provide an overview of the most relevant physics and working principles that are relevant for applications exploiting hot carriers in graphene.

## 1 Introduction

Understanding and controlling the properties of hot carriers – electrons and holes with excess kinetic energy – is a paradigmatic topic in both physics and chemistry. In graphene, hot carriers are particularly consequential because they can be efficiently created, controlled and exploited towards applications, as we will discuss in detail in this review. Hot carriers in graphene have been addressed experimentally[1–4] and explicitly through theory,[5,6] since around 2007, although they have

[a]*Institut Quantique and Département de Physique, Université de Sherbrooke, Sherbrooke, Québec, Canada*
[b]*Institute of Solid State Physics, Friedrich Schiller University Jena, 07743 Jena, Germany*
[c]*Abbe Center of Photonics, Friedrich Schiller University Jena, 07745 Jena, Germany*
[d]*School of Physics and Astronomy, University of Manchester, UK*
[e]*Catalan Institute of Nanoscience and Nanotechnology (ICN2), BIST & CSIC, Campus UAB, 08193 Bellaterra, Barcelona, Spain. E-mail: klaas.tielrooij@icn2.cat*

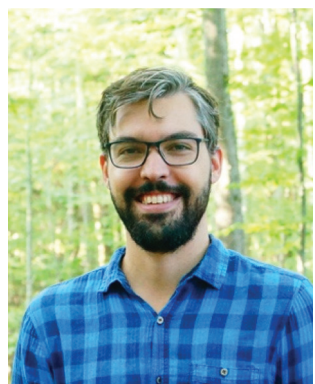

**Mathieu Massicotte**

*Dr Mathieu Massicotte is a postdoctoral research fellow at the Institut Quantique of Université de Sherbrooke in Canada. He completed his PhD in Photonics in 2017 at ICFO – The Institute of Photonic Sciences in Spain after obtaining his master's degree in Physics from McGill University. His research interests lie at the intersection of experimental condensed matter physics and photonics, with a focus on optoelectronic nanodevices based on novel 2D materials.*

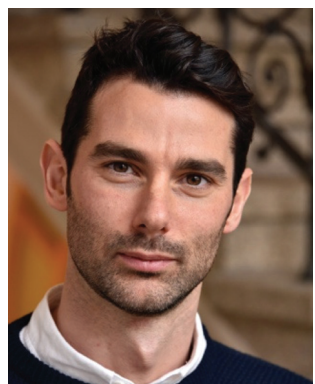

**Giancarlo Soavi**

*Giancarlo Soavi obtained a PhD in Physics from Politecnico di Milano in 2015 and subsequently worked as a Research Associate at the Cambridge Graphene Centre. From 2019 he is a tenure track Professor at the Friedrich Schiller University of Jena. His research interests focus on ultrafast spectroscopy and nonlinear optics in quantum confined systems, including graphene and related layered materials.*

been the subject of intense research in graphene-related materials, such as graphite and carbon nanotubes, since many years *cf.* ref. 7–10. Now, more than a decade after the first studies, the understanding of hot-carrier phenomena in graphene and graphene-based hybrid systems has led to the realization that their properties enable highly promising applications. These insights have already led to a significant number of patents and business creations, besides a large number of scientific publications.

In this review we will focus on the aspects of hot carriers that are most relevant for applications involving light, and on systems where graphene is the active material containing hot carriers. Our aim is not to provide a fully exhaustive review of all relevant literature in this expansive and expanding field. We refer to several reviews with relevant background information: ref. 11 provides an excellent introduction to hot carriers, with a focus on metallic nanostructures. Ref. 12 and 13 discuss photodetection and data communication applications of graphene (and related 2D materials), respectively. In ref. 14 and 15, nonlinear optical effects and applications of 2D materials are reviewed, and ref. 16 and 17 provide reviews on energy dissipation in graphene (and related 2D materials).

We distinguish two types of hot carriers, where in both cases charge carriers have an increased kinetic energy. We will use the term "high-energy charge carriers" for those that have an increased energy, yet do not form part of a thermalized carrier distribution. In this case, it is not possible to define a meaningful temperature, even if the carrier energy can be expressed in units of temperature. We will exclusively use the term "hot carriers" for those high-energy carriers that have thermalized with other charge carriers, thus forming a Fermi–Dirac distribution with an increased temperature – a hot-carrier distribution (see Fig. 1). In this case, the temperature of the electronic system $T_e$ is higher than the (equilibrium) temperature of the lattice $T_L$.

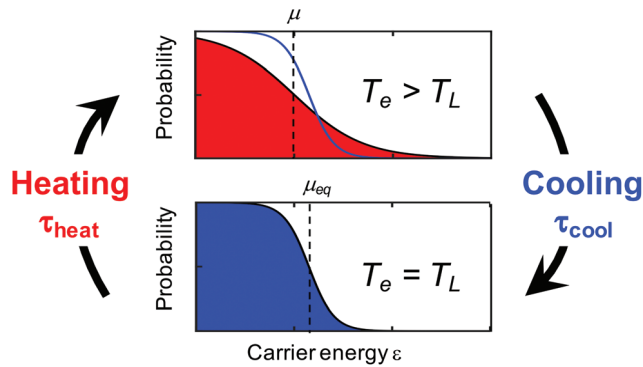

Fig. 1 Illustration of heating–cooling dynamics of graphene carriers, where in the hot state the carrier distribution is characterized by an increased carrier temperature $T_e$ that is larger than the lattice temperature $T_L$ and a "hot" chemical potential $\mu$ that is smaller than the equilibrium chemical potential $\mu_{eq}$.

As a starting point for describing the relevant physics of hot carriers in graphene, we provide the thermoelectric equation describing the time evolution of the electronic temperature $T_e$, assuming that the electron density remains constant over time $t$:

$$C_e \frac{\partial T_e}{\partial t} = \underbrace{\dot{Q}_{ext}}_{\text{heating/cooling}} + \underbrace{\kappa_e \nabla^2 T_e}_{\text{diffusion}} - \underbrace{\nabla \cdot [(V + \Pi)J]}_{\text{thermoelectrics}}, \quad (1)$$

where $C_e$ is the electronic heat capacity. The first term on the right, $\dot{Q}_{ext}$, is the rate at which heat is added by external sources and removed *via* internal or external cooling channels, for example *via* electron–phonon collisions. The other two terms on the right-hand side are related to transport phenomena. The second term corresponds to Fourier's law of heat conduction, and contains the electronic part of the thermal

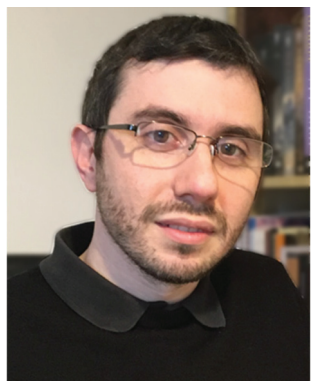

**Alessandro Principi**

*Dr Principi is Lecturer of the Department of Physics and Astronomy of the University of Manchester since September 2017. After his PhD, obtained in 2013 from Scuola Normale Superiore (Pisa, Italy), he worked as postdoc in the United States (University of Missouri, Columbia) and in the Netherlands (Radboud University, Nijmegen). During these years, he has studied the role of many-body interactions* on the transport and dynamical properties of graphene, van der Waals heterostructures and topological materials. He has formed successful collaborations with world-leading experimental research teams studying, among others, hot electron relaxation in graphene.

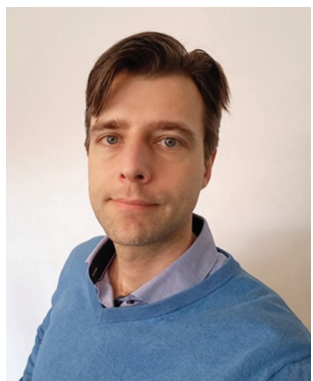

**Klaas-Jan Tielrooij**

*Klaas-Jan Tielrooij obtained his Ph.D. from the University of Amsterdam, the Netherlands, in 2010. He then became a postdoctoral researcher, and later research fellow, at the Institute of Photonic Sciences (ICFO), Spain. In 2018 he became Junior Group Leader of the Ultrafast Dynamics in Nanoscale Systems group at the Catalan Institute of Nanoscience and Nanotechnology (ICN2) in Bellaterra (Barcelona), Spain.* His research interests are in the field of ultrafast dynamics, optoelectronics, nonlinear optics, light–matter interaction, and terahertz technologies, with a current focus on heat and charge transport in nanoscale materials, such as two-dimensional materials.

conductivity $\kappa_e$. The third term describes the energy carried by the electric current density. In this term, $\Pi$ is the local Peltier coefficient of the material, $V$ is the local voltage, and

$$J = \sigma(-\nabla V - S\nabla T_e) \quad (2)$$

is the current density, with $\sigma$ the conductivity. The Peltier coefficient is related to the local Seebeck coefficient of the material *via* the Onsager reciprocity relation $S = \Pi/T_e$. We will discuss the different terms of eqn (1) and (2) in detail in the different sections of this review.

This review is organized as follows: In section 2. *Excitation*, we will discuss different approaches for exciting graphene such that carriers are heated, followed by section 3. *Energy dynamics* on the thermalization and cooling dynamics of the electronic system. These two sections are mainly related to the term $\dot{Q}_{ext}$ of eqn (1). In section 4. *Conductivity*, we will discuss the linear and nonlinear conductivity of hot carriers. We will then discuss transport phenomena in section 5. *Transport*, in particular diffusion and thermoelectric effects, related to the last two terms of eqn (1) and (2). We will then describe several applications enabled by hot carriers in section 6. *Applications*, followed by section 7. *Discussion and outlook*. Finally, we conclude with section 8. *Conclusion*. We also provide two boxes with background information: Box 1. *Boltzmann theory* and Box 2. *Experimental techniques*.

## 2 Excitation

In this section we will describe different ways of exciting graphene, such that hot carriers are created. We then discuss, from a macroscopic thermodynamic viewpoint, how a hot-carrier distribution is established after excitation.

### 2.1 Heating sources

**2.1.1 Thermal excitation.** The first method consists in heating the graphene lattice. This creates hot carriers *via* electron–phonon scattering, and leads to a steady-state condition with equal carrier temperature $T_e$ and lattice temperature $T_L$ in the heated region. The most common way to heat up the graphene lattice locally is by using micro-heaters in the proximity of the graphene sample[18,19] (see Fig. 2a). In this case, heat is created in a metallic wire that heats up due to the Joule effect upon driving a current through the micro-heater. This heating method typically generates non-uniform heating in graphene, with a temperature gradient across the graphene sheet.

**2.1.2 Electrical excitation.** A more direct approach for the generation of hot electrons is *via* Joule heating by driving a current through graphene itself (see Fig. 2b). In this case, the electron system is primarily heated, such that the electron temperature $T_e$ is typically higher than the lattice temperature $T_L$. Electrical heating with power densities up to ~10 kW cm$^{-2}$,[20–22] and even a few hundred kW cm$^{-2}$ (ref. 23–25) have been applied to graphene, and shown to lead to an electronic temperature increase $\Delta T_e$ up to several thousand kelvin. Electrical excitation of graphene can be achieved both with DC power[25,26] or with AC power at kHz,[27] MHz (ref. 28) and GHz (ref. 21 and 22) frequencies.

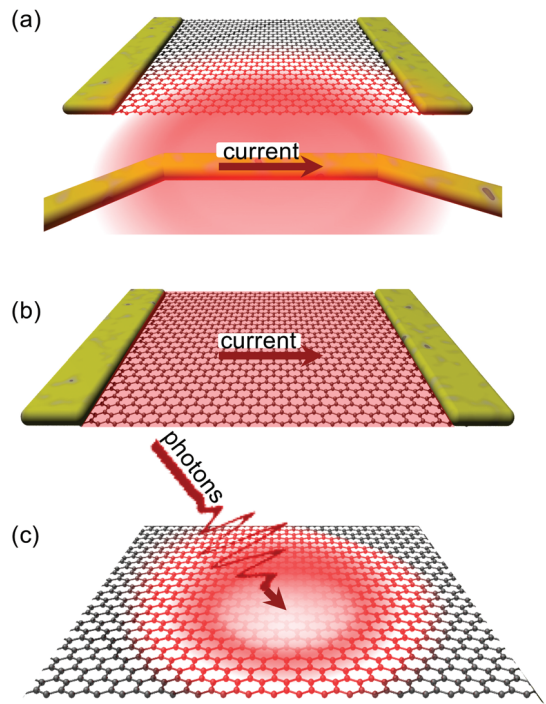

**Fig. 2** Schematic illustrations of (a) thermal excitation, (b) electrical excitation, and (c) optical excitation.

**2.1.3 Optical excitation.** Optical excitation of graphene (Fig. 2c) leads to the generation of high-energy carriers that can turn into hot carriers *via* carrier thermalization (see Fig. 3). The absorption of light by graphene occurs through two main processes, depending on the incident photon energy $\hbar\omega$ with respect to the Fermi energy $E_F$: intraband absorption typically dominates for $\hbar\omega < 2|E_F|$, while interband absorption typically dominates for $\hbar\omega > 2|E_F|$. We will discuss light absorption in section 4. Arguably the most important feature of optical excitation, is that ultrashort pulses with femtosecond duration can be used. This allows for studying the ultrafast dynamics of the electronic system, as well as exploiting these ultrafast dynamics towards novel optoelectronic applications. These dynamics will be the topic of section 3, where we will discuss in more detail how light absorption leads to carrier heating, and how subsequent cooling takes place.

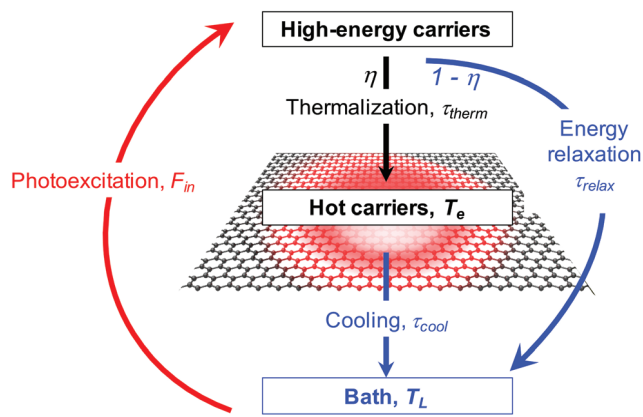

**Fig. 3** Schematic illustration of the formation of hot carriers *via* photoexcitation, followed by thermalization of high-energy carriers. Since high-energy carriers can also directly relax to their environment, the branching ratio between these two processes determines the heating efficiency $\eta$.

### 2.2 Macroscopic picture of carrier heating

**2.2.1. Steady-state heating.** Steady-state heating refers to the situation where graphene is excited either continuously, for example with CW light or a DC current, or by a time-dependent power density input that varies on a timescale that is much slower than the heating–cooling dynamics of graphene charge carriers. Under these excitation conditions, a new equilibrium situation will be established with an increased carrier temperature $T_e > T_L$. In a simple scenario, we can ignore the transport phenomena in eqn (1) and assume one dominant cooling channel. In this case, $\dot{Q}_{ext} = P_{in} - \Gamma_{cool}(T_e - T_L)$, where $P_{in}$ is the absorbed power density of the external source, $\Gamma_{cool} = \frac{C_e}{\tau_{cool}}$ is the heat transfer coefficient in units W m$^{-2}$ K$^{-1}$, and $\tau_{cool}$ is the cooling time. The heat transfer coefficient describes the flow of heat out of the electronic system, for example through electron–phonon coupling. The thermoelectric equation, eqn (1), in this simple scenario reduces to

$$\Delta T_e = T_e - T_L = \frac{P_{in}}{\Gamma_{cool}} = \frac{P_{in}\tau_{cool}}{C_e}. \quad (3)$$

In order to increase the carrier temperature $T_e$, for a given heating power density $P_{in}$, it is beneficial to operate with a small electronic heat capacity and a long cooling time. We will discuss the cooling dynamics in the next section. The electronic heat capacity, or specific heat, in two limiting cases – "doped" and "undoped" – is given by:[29,30]

$$C_{e,doped} = \frac{2\pi E_F}{3(\hbar v_F)^2}k_B^2 T_e = \gamma_{doped} T_e \quad \text{for} \quad k_B T_e \ll E_F, \quad (4)$$

$$C_{e,undoped} = \frac{18\zeta(3)}{\pi(\hbar v_F)^2}k_B^3 T_e^2 = \gamma_{undoped} T_e^2$$
$$\text{for} \quad k_B T_e \gg E_F. \quad (5)$$

Here, $\hbar$ is the reduced Planck constant, $v_F$ the Fermi velocity, $k_B$ Boltzmann's constant, and $\zeta(3) \simeq 1.202$. In the "doped" regime, the heat capacity decreases upon decreasing $E_F$, as there are fewer electrons to share the energy. From the perspective of the specific heat, it is thus advantageous to operate at low Fermi energy, if the objective is to maximize the temperate increase $\Delta T_e$. Upon approaching the Dirac point, the $E_F$-dependence drops out. In this case, $\Delta T_e$ can be further maximized by operating at low temperatures. We note that graphene can reach record-low values of the electronic heat capacity, thus leading to large temperature rises for small excitation. Heat capacities as low as $2 \times 10^{-9}$ J (m$^2$ K)$^{-1}$,[31] and very recently even $1.2 \times 10^{-11}$ J (m$^2$ K)$^{-1}$,[32] have been found using noise thermometry measurements.

**2.2.2 Transient heating.** Transient heating refers to the situation where graphene is excited by a short pulse that varies on a timescale faster than the heating–cooling dynamics. In the simplified case of an infinitely short input pulse of power, absorption is followed by thermalization, where a common temperature of the electronic system is established. We can assign a thermalization time $\tau_{therm}$ to this process. If $\tau_{therm}$ is shorter than the cooling time $\tau_{cool}$, a quasi-equilibrium situation is established after absorption of the pulse and thermalization. In this case, we can use the peak energy density, or input fluence, $F_{in} = \frac{P_{in}}{f_{rep}}$ in order to calculate the peak increase in electron temperature $\Delta T_{e,peak}$. Here, $f_{rep}$ is the repetition rate of the incident excitation, and the input fluence is the absorbed fluence (not the incident fluence). A lower repetition rate gives rise to a larger $\Delta T_{e,peak}$. In order to obtain the peak temperature, we solve the integral $F_{in} = \int_{T_L}^{T_{e,peak}} C_e dT_e$, and obtain the following sub-linear relationships between peak temperature and input fluence:

$$T_{e,peak} = \sqrt{T_L^2 + \frac{2F_{in}}{\gamma_{doped}}} \quad \text{for} \quad k_B T_e \ll E_F, \quad (6)$$

$$T_{e,peak} = \sqrt[3]{T_L^3 + \frac{3F_{in}}{\gamma_{undoped}}} \quad \text{for} \quad k_B T_e \gg E_F. \quad (7)$$

**2.2.3 Heating efficiency.** If the thermalization process – where high-energy carriers interact with other carriers to form a hot-carrier distribution – competes with other energy relaxation processes (see Fig. 3), it can be useful to define a heating efficiency $\eta$. One then replaces $F_{in}$ by $\eta F_{in}$, as done for example in ref. 33. Studies performed with relatively strong excitation typically found lower heating efficiencies than studies done with relatively weak excitation. For example, using an absorbed fluence in the μJ cm$^{-2}$ range, a photoluminescence study found a significantly lower $T_e$ than expected, suggesting a rather low heating efficiency.[30] On the other hand, optical pump – terahertz probe studies with an absorbed fluence in the nJ cm$^{-2}$ range found heating efficiencies well above 50%,[34,35] demonstrating that thermalization within the electronic system dominates over alternative energy relaxation channels. Below, we will discuss the microscopic processes that govern the heating efficiency.

## 3 Energy dynamics

In this section, we describe the microscopic processes governing the energy dynamics of hot carriers. We start with a thermodynamic picture of carrier heating, followed by a description of the carrier–carrier interactions that lead to thermalization of the electronic system, after having absorbed a heating pulse, for example by photoexcitation. During these carrier–carrier interactions the total energy is conserved, and the electronic system typically evolves from a non-thermal to a thermal distribution with elevated carrier temperature. We then discuss the different interactions between carriers and other (quasi-)particles that lead to cooling of the hot-carrier system by damping of heat to a bath. Theoretical details on calculating these scattering events can be found in Box 1. We point out that the process we refer to as "thermalization" of the electronic system is sometimes referred to in literature as "internal thermalization". This is in contrast to "external thermalization", which refers to thermalization of the electronic system with the phonon system.

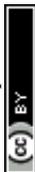

> **Box 1 Boltzmann theory**
>
> The statistical modelling of thermodynamic out-of-equilibrium systems is usually performed by means of a "microscopic" Boltzmann equation, from which eqn (1) can be derived. The latter is commonly used to describe the kinetics of electrons in crystals in the so-called semi-classical regime,[64,65] in which electronic states are described in terms of wave packets.[66,67] When the length scales over which external electric ($E$) and magnetic ($B$) fields vary are much larger than the spread of wave packets themselves,[66] the latter can be treated as point-like particles with well-defined position $r$ (the "centre" of the wave packet) and wavevector $k$. Then, as usual in statistical mechanics, the properties of a large collection of particles can be determined by how they arrange in the phase space. The central object becomes the distribution function, $f(r, k, t)$, which determines at every instant of time $t$ the number of particles in an infinitesimal phase-space volume element centered around the point $(r, k)$. In general, the number of particles therein changes over time. Firstly, when external forces are applied, after an infinitesimal time $dt$, the particles' position and wavevector become $r + \dot{r}dt$ and $k + Fdt/\hbar$, respectively.[66,67] Here, $F = -e(E + \dot{r} \times B)$, while $\dot{r}$ is the particle velocity which, barring Berry-curvature corrections,[67] is equal to $\hbar^{-1}\nabla_k \varepsilon_k$, where $\varepsilon_k$ is their energy dispersion.
>
> The number of particles in a phase-space volume element changes over time also because of collisions, described via the collision integral[64] $I[f(r, k, t)]$. The collision integral accounts for all many-body scattering events a particle undergoes (against, e.g., impurities, phonons or other particles) that lead to a change of its wavevector $k$. After any such scattering event, a given particle transitions to a different phase-space volume element. Therefore, $I[f(r, k, t)]$ is responsible for the time evolution of the occupation function due to collisions.
>
> The time evolution of the distribution function, accounting for both external forces and collisions, is described by the following equation:[64,65]
>
> $$\partial_t f(r,k,t) + \dot{r} \cdot \nabla_r f(r,k,t) + \hbar^{-1} F \cdot \nabla_k f(r,k,t) = I[f(r,k,t)].$$
>
> Eqn (1) can be derived by solving the equation above via the Chapman–Enskog method.[68] Introducing the electron heat capacity and other transport coefficients, the resulting equation can be recast in the form of 1.
>
> Finally, we note that the equation above can be used to describe also the thermalization of non-equilibrium electrons (see section 3.1), not just their subsequent macroscopic cooling dynamics (eqn (1) – see also section 1). To describe the relaxation of non-thermal distribution, the equation above must however be solved numerically and yields a thermal distribution on time scales controlled by strength of the electron–electron collision integral. Once such distribution is reached, electron–electron collisions become ineffective (their collision integral vanishes for thermal distributions) and cooling proceeds via emission of phonons and can be effectively described via eqn (1).

### 3.1 Thermalization

#### 3.1.1 Microscopic thermodynamic picture

*Fermi–Dirac distribution.* Microscopically, the electron temperature $T_e$ and chemical potential $\mu$ define the shape of the Fermi–Dirac distribution of a carrier system (see Fig. 1). This distribution describes the probability of finding an electron (e) or hole (h) with a certain energy $\varepsilon$, and is given by:

$$f_{e,h}(\varepsilon) = \left[1 + e^{(\varepsilon \mp \mu_{e,h})/k_B T_{e,h}}\right]^{-1}. \tag{8}$$

In equilibrium, the temperatures of the electrons in the conduction band and the holes in the valence band are equal, $T_e = T_h$, and so are their chemical potentials, $\mu_e = \mu_h$. The amount of kinetic energy density in units J m$^{-2}$ in the electronic system is given by the sum of the energy density of electrons in the conduction band and holes in the valence band: $\mathcal{E} = \mathcal{E}_e + \mathcal{E}_h$. These energies are given by:

$$\mathcal{E}_{e,h} = \int_0^\infty \varepsilon D(\varepsilon) f_{e,h}(\varepsilon) d\varepsilon, \tag{9}$$

where $D(\varepsilon) = 2\varepsilon/(\pi \hbar^2 \nu_F^2)$ is the density of states of graphene.[36] When a system at equilibrium, with $\mathcal{E}_{eq}$, is excited by an absorbed energy density $F_{in}$, energy conservation dictates that

$$\mathcal{E}_{eq} + F_{in} = \mathcal{E}_{hot}, \tag{10}$$

where the "hot" state has an energy density $\mathcal{E}_{hot}$ that is characterized by "hot" Fermi–Dirac distributions with an increased carrier temperature. The electrons and holes thus have Fermi–Dirac distributions that are broadened. This broadening has been observed using time-resolved ARPES measurements, after pulsed optical heating[37–39] (see Fig. 4a). Typically, it was found that within 150 fs the optically excited non-thermal distribution thermalizes.

A broadening of the Fermi–Dirac distribution is accompanied by a decrease in the chemical potential, as dictated by conservation of the carrier density: $n_e - n_h$ = const.

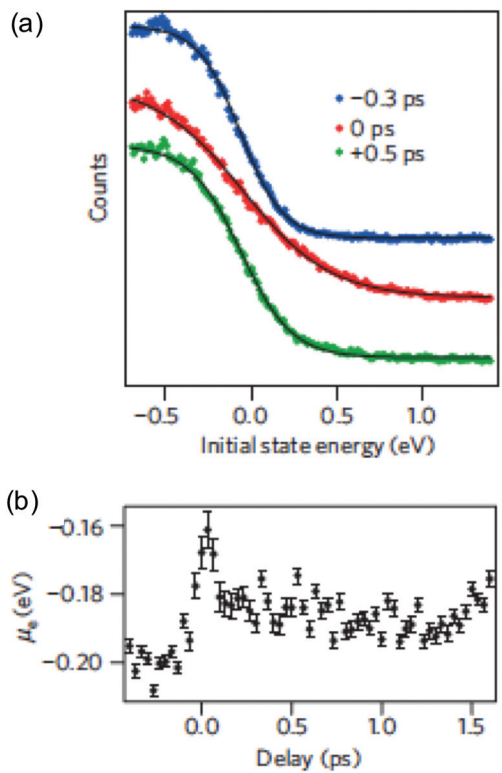

Fig. 4 (a) Carrier distribution measured using time-resolved ARPES, at three different delay times: before (blue), during (red), and after (green) photoexcitation, clearly indicating a broadened (hot) distribution due to photoexcitation. The photon energy was $\hbar\omega$ = 300 meV. (b) Extracted chemical potential from time-resolved ARPES measurements, as function of time delay after photoexcitation. Due to carrier heating, the chemical potential shifts towards the Dirac point. Panels (a) and (b) are adapted with permission from ref. 37 (Copyright 2013 Springer Nature).

The electron and hole carrier densities, $n_e$ and $n_h$, respectively, are given by:

$$n_{e,h} = \int_0^\infty D(\varepsilon) f_{e,h}(\varepsilon) d\varepsilon. \qquad (11)$$

This equation relates the variation in carrier density to the corresponding variation in temperature and, importantly, in chemical potential. Carrier density conservation dictates the following decrease of the chemical potential with carrier temperature, which in both doping regimes gives a decrease with increasing $T_e$:

$$\mu = E_F \left[1 - \frac{\pi^2 k_B T_e^2}{6 E_F^2}\right] \quad \text{for} \quad k_B T_e \ll E_F, \qquad (12)$$

$$\mu = \frac{E_F^2}{4 \ln(2) k_B T_e} \quad \text{for} \quad k_B T_e \gg E_F. \qquad (13)$$

We note that the Fermi energy, which is defined at absolute zero temperature, is related to the carrier density via[36]

$$E_F = \pm \hbar v_F \sqrt{\pi n_{e,h}}. \qquad (14)$$

Time-resolved ARPES studies (see Box 2) experimentally observed that the chemical potential indeed decreases upon carrier heating[37] (see Fig. 4b). The decrease of the chemical potential upon carrier heating has several implications for transport properties (see section 4 on Conductivity and section 5 on Transport) and experimentally observable quantities using the techniques described in Box 2.

---

**Box 2  Experimental techniques**

**Steady-state electrical and optical measurements.** A measurement of the steady-state carrier temperature increase $\Delta T_e$ can be obtained by radio-frequency Johnson noise measurements,[21,22,104] taking advantage of the linear relation between the current noise spectrum and $T_e$,[105] or by detection and fitting with grey body radiation of the graphene hot-electron thermal emission.[25,26] Another steady-state approach for the study of hot electrons is the excitation photomixing scheme.[106]

**Pump-probe.** In ultrafast pump–probe experiments a first light pulse, called "pump", generates high-energy charge carriers. The temporal evolution of these photoexcited carriers is subsequently detected by measuring the absorption, transmission or reflection of a second pulse, called "probe".[107–110] The relative temporal delay between the two pulses is controlled, such that the temporal resolution is determined by the duration of the ultrashort pulses. Tuning of the probe photon energy gives access to the energy distribution of the excited state. The photon energies used for the pump and probe can vary from the terahertz ($\hbar\omega \sim$ meV) to the UV ($\hbar\omega \sim$ eV).

**Time-resolved ARPES.** In an angle-resolved photoemission spectroscopy (ARPES) experiment, high energy photons ($\hbar\omega$ typically above 5 eV) impinge on a crystal in vacuum, ejecting electrons in free space, where their energies and exit angles are measured. These are related to the energy and crystal momentum of the electrons inside the sample. In the case of time-resolved ARPES, the sample is first excited by an ultrashort light pulse which generates high-energy charge carriers and subsequently the ARPES spectrum is measured with a second delayed high photon energy ultrashort pulse.[111]

**Time-resolved photocurrent.** In time-resolved photocurrent (trPC) experiments the sample is excited by two ultrashort laser pulses and the electrical response of the sample (photocurrent or photovoltage) is measured as a function of their delay. If the electrical response scales nonlinearly with the incident laser intensity, such as any hot-electron dominated photocurrent/photovoltage in graphene, the signal at zero/small delays will differ from the signal at large delays, and the recovery dynamics of the signal will reflect the time response of the sample.[33,53,112–114]

*Hot-carrier density.* It will be useful to consider the hot-carrier density $n_{hot}$ – the density of carriers with an energy larger than the chemical potential, given by†

$$n_{hot;e,h} = \int_{\mu_{e,h}}^{\infty} D(\varepsilon) f_{e,h} d\varepsilon$$
$$= \frac{2}{\pi(\hbar v_F)^2} \left( \frac{\pi^2}{12} (k_B T_e)^2 + \ln(2) \mu_{e,h} k_B T_e \right). \quad (15)$$

For intrinsic graphene with $\mu_e = \mu_h = 0$, this reduces to the thermal carrier density $n_{therm} = (\pi/6)(k_{BT_e/\hbar v_F})^2$, often used for graphene under electrical heating.[40,41] Whereas the carrier density $n_e - n_h$ is conserved during thermalization, the hot-carrier density $n_{hot}$ increases. Several studies[16,34,35,42–44] have discussed this effect quantitatively. In the next subsection, we will discuss the microscopic processes that are responsible for this.

*Intraband vs. interband thermalization.* We now discuss the creation of an increased hot-carrier density in the two doping regimes. In the "doped" regime, with $k_B T_e \ll E_F$, only the carrier density of one band needs to be considered for energy and particle conservation. Thus, in the case of electron doping, it is enough to consider conservation of $n_e$, because $n_h \approx 0$. The situation is different closer to the Dirac point, where $k_B T_e \gg E_F$. This is because the conserved quantity $n_e - n_h$ allows for additional electron–hole pair creation across the bands, at least from a thermodynamic viewpoint. We illustrate this following ref. 35, in Fig. 5, where we show the initial Fermi–Dirac distribution directly after absorption of a heating pulse of absorbed power density $P_{in}$, yet before thermalization, and 300 fs later, when thermalization has occurred. These results are based on microscopic charge interactions (see Box 1 and next subsection), yet they provide a clear thermodynamic picture, with broadening of the distributions due to thermalization, *i.e.* an increased $T_e$. They also confirm the decrease in chemical potential associated with thermalization. In the case where $E_F$ is initially far away from the Dirac point ($E_F = 0.4$ eV), this broadening only affects the electrons in the conduction band. The density of holes in the valence band was basically zero before thermalization, and is still zero after thermalization. The density of electrons in the conduction band stays constant, whereas the density of hot electrons in the conduction band $n_{hot,e}$ has clearly increased. Therefore, thermalization in the case of large $E_F$ is referred to as intraband thermalization. In the case where $E_F$ is initially close to the Dirac point ($E_F = 0.05$ eV), the broadening affects both electrons in the conduction band and holes in the valence band. Indeed, before thermalization there was a small density of holes in the valence band, whereas this density is increased after thermalization, as a result of charge and energy redistribution between the two bands. Also the density of electrons in the conduction band has increased. This means that – besides $n_{hot}$ – the density of interband electron–hole pairs, *i.e.* both $n_e$ and

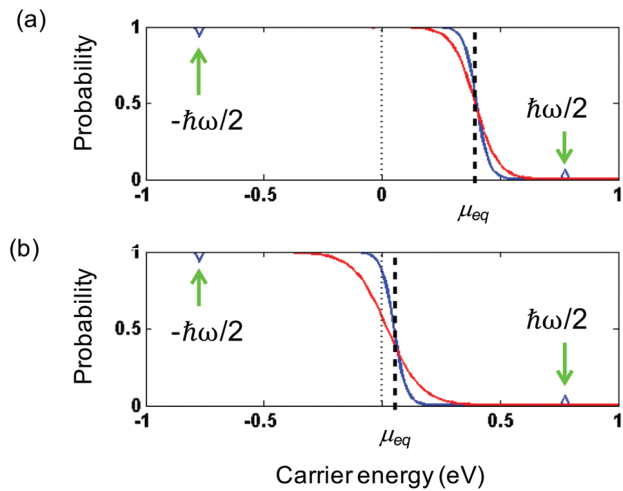

**Fig. 5** (a) Intraband thermalization after photoexcitation with $\hbar\omega = 1.5$ eV, for an initial chemical potential of 0.4 eV. The density of electrons in the conduction band $n_e$ is conserved, while $T_e$ increases. (b) Interband thermalization after photoexcitation, for an initial chemical potential of 0.05 eV. The density of electrons in the conduction band $n_e$, and the density of holes in the valence band $n_h$, increase, while $T_e = T_h$ increases. In both cases, $n_e - n_h$ is conserved, and $n_{hot}$ increases. These figures are based on microscopic simulations performed in ref. 35.

$n_h$, has increased. Therefore, thermalization in the case of small $E_F$ is referred to as interband thermalization.

The thermodynamic picture above suggests that the valence and conduction bands are thermalized with each other, such that the holes in the valence band and electrons in the conduction band have a common chemical potential and carrier temperature. The simulations of ref. 35 show that this is an excellent approximation when the system is examined 300 fs after excitation. At shorter timescales, this picture is not always very accurate. Indeed, several experimental studies with sub-50 fs time resolution have observed short-lived non-thermal distributions.[39,45] A time-resolved ARPES study[37] furthermore showed a situation where the carrier distribution consists of two separate Fermi–Dirac distributions – one for electrons and one for holes. This so-called "inverted" state was found to have a lifetime of ~130 fs. Interestingly, the short-lived situation of two separate distributions was only observed in the case of interband photoexcitation (with photon energy $\hbar\omega > 2E_F$), whereas intraband photoexcitation (with photon energy $\hbar\omega < 2E_F$) directly led to a single, broadened Fermi–Dirac distribution for the valence and conduction bands. These results suggest the occurrence of a rich interplay of dynamical processes during carrier thermalization. We will discuss these dynamics in the following sections.

**3.1.2. Carrier–carrier scattering.** Thermalization occurs primarily *via* Coulomb carrier–carrier scattering events, where carriers exchange energy and momentum by keeping their total energy and momentum constant. We first classify carrier–carrier scattering processes into scattering events within a single band, and scattering events across the bands (see Fig. 6a). In the former case, intraband scattering occurs, which

---

†A. Tomadin, *private communication.*

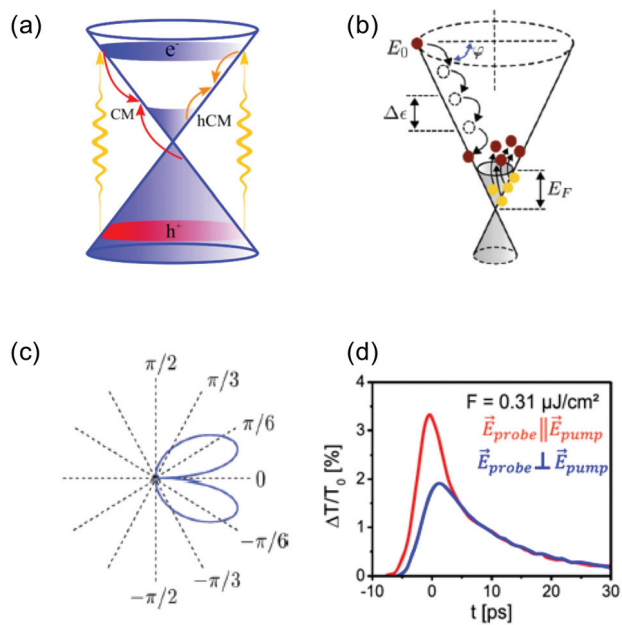

Fig. 6 (a) Schematic illustration of thermalization via Coulomb-induced carrier–carrier scattering processes after photoexcitation. Here, "CM" indicates interband scattering processes (impact ionization and Auger recombination) that can lead to the creation of multiple electron–hole pairs per absorbed photon – carrier multiplication. Similarly, "hCM" indicates intraband scattering processes that can lead to the creation of multiple hot electrons per absorbed photon – hot-carrier multiplication. (b) Schematic illustration of the details of thermalization via intraband carrier–carrier scattering (impact excitation), where one initial high-energy electron leads the creation of multiple hot carriers, i.e. hot-carrier multiplication. In each relaxation step, the initial electron transfer an average mount of energy $\Delta\varepsilon \simeq E_F$ to an electron in the Fermi sea. The angle $\varphi$ indicates the change in momentum of the initial electron. (c) Calculation of the angular distribution of intraband impact excitation scattering for the initial high-energy electron, indicating near-collinear scattering. (d) Experimental observation of anisotropic carrier distribution through pump–probe spectroscopy, with pump and probe polarizations parallel (red) or perpendicular (blue). The photon energy was $\hbar\omega$ = 88 meV. Panel (a) is adapted with permission from ref. 46; panels (b) and (c) are adapted with permission from ref. 42 (Copyright 2013 American Physical Society); panel (d) is adapted with permission from ref. 47 (Copyright 2016 American Physical Society).

gives rise to intraband thermalization. In the latter case, interband scattering occurs, which gives rise to interband thermalization. For interband processes occurring in semiconductor materials, two important types of carrier–carrier interactions are impact ionization and Auger recombination. When impact ionization occurs, a high-energy electron loses energy, while another electron with lower energy gains energy and is promoted from the valence band to the conduction band. The reverse process, where an electron is degraded from the conduction to the valence band, is known as Auger recombination. These processes lead to an increase (impact ionization) or decrease (Auger recombination) in carrier density. In graphene, the same interband impact ionization and Auger recombination processes as in semiconductors can occur, particularly when the Fermi energy is close to the Dirac point.

There are also scattering events that are analogous to impact ionization and Auger recombination, yet occurring within a single band. In this case, the total carrier density stays constant, whereas the hot-carrier density (density of carriers with energy above the chemical potential) increases or decreases, respectively. Therefore, intraband impact ionization is sometimes referred to as impact excitation.[16,42] We note that dynamical screening in the random-phase approximation[48] can lead to a suppression of Auger processes in graphene,[49] thus favoring impact excitation and impact ionization, i.e. leading to efficient carrier heating.

*Hot-carrier multiplication.* In order to examine thermalization via carrier–carrier scattering more closely, we first discuss the situation where high-energy carriers have been created, typically by photoexcitation of graphene with a photon energy $\hbar\omega > 2E_F$ (see Fig. 6b). This creates carriers with an energy around $\varepsilon = \hbar\omega/2$ above the Dirac point. In the "doped" case, where $k_B T_e \ll E_F$, these initial high-energy carriers lose kinetic energy through intraband, inelastic scattering with carriers in the Fermi sea, which gain kinetic energy. Thus, secondary hot carriers are created. This is the process of impact excitation. The amount of energy $\Delta\varepsilon$ that is exchanged between an initial high-energy carrier and a carrier in the Fermi sea follows a distribution function that peaks around $E_F$.[42] Therefore, a single (primary) high-energy carrier can create multiple secondary hot carriers during its cascade, transferring an average amount of $\Delta\varepsilon \approx E_F$ to each secondary hot carrier. This corresponds to the creation of multiple hot carriers per absorbed photon, gradually increasing $n_{hot}$. The occurrence of hot-carrier multiplication was predicted and addressed quantitatively using optical-pump terahertz probe measurements with varying pump photon energy,[34,42] and using time-resolved ARPES measurements.[43] It was furthermore experimentally confirmed when Wu et al. succeeded in electrically collecting multiple hot-carriers per absorbed photon in a photodetector system.[44] The occurrence of hot-carrier multiplication shows that carrier heating in graphene is an exceptionally efficient process that can be exploited for various applications (see section 6).

*Carrier multiplication.* While hot-carrier multiplication can occur in the "doped" regime, where intraband thermalization dominates, "real" carrier multiplication can occur close to the Dirac point, where $k_B T_e \gg E_F$. In this regime interband thermalization occurs, and multiple free carriers can be created per absorbed photon. Winzer et al. showed that there is a competition between the interband processes of impact ionization and Auger recombination, and predicted the occurrence of carrier multiplication for relatively weak photoexcitation.[50] Carrier multiplication was observed experimentally in ref. 51, although no unambiguous results have been presented (yet) showing the electrical collection of multiple electron–hole pairs per absorbed photon. We note that the thermodynamic regimes of interband thermalization and intraband thermalization correspond to the regimes where carrier multiplication and hot-carrier multiplication, respectively, can occur under certain circumstances[35,46] (see Fig. 6a). In semiconductors, the creation of multiple carriers per absorbed photon is very attractive, as it enables efficient photodection and energy harvesting beyond the Shockley–Queisser limit.[52] In graphene, hot-carrier multiplication is arguably more crucial, as there are several ways to exploit the additionally created hot carriers, as we will discuss in section 6.

*Effect of photon energy and power.* For impact excitation, a higher photon energy leads to a higher energy of the primarily excited high-energy carriers. Because the average energy exchange per scattering event $\Delta\varepsilon$ is $\sim E_F$, a higher photon energy leads to a larger number of scattering events during the relaxation cascade, and therefore to the creation of a larger number of secondary hot carriers per absorbed photon, as shown experimentally.[34] Thermodynamically, this corresponds to a larger increase in $T_e$ per absorbed photon. Interestingly, for a given absorbed fluence $F_{in}$, the temperature increase is roughly independent of photon energy. This is because $F_{in} = \hbar\omega n_{\hbar\omega}$, where $n_{\hbar\omega}$ is the absorbed photon density. This means that for a fixed $F_{in}$, an increase in photon energy corresponds to a decrease in photon density, such that the overall heating, in principle, does not depend on photon energy for a given absorbed fluence. Indeed, a constant thermal photocurrent was measured over a broad wavelength range.[53] We note that a larger photon energy can lead to a longer duration of the thermalization cascade,[34] which means that competing energy relaxation processes become more important, leading to a lower heating efficiency $\eta$ for higher photon energies.

So far, we have considered the case of interband photoexcitation, with $\hbar\omega > 2E_F$. For intraband excitation, with $\hbar\omega < 2E_F$, the thermalization dynamics are in many ways very similar. Intraband excitation leads to acceleration of free carriers that gain kinetic energy. These carriers then distribute their energy with the other carriers in the Fermi sea. Indeed, the creation of a thermalized hot state after intraband photoexcitation was observed, even within 30 fs (ref. 37) (see also Fig. 4a). In agreement with this observation of ultrafast thermalization, there are several experimental indications, for example using mid-infrared or terahertz light,[54–56] that intraband photoexcitation leads to efficient carrier heating, at least within the same band.

When keeping the photon energy fixed, while increasing the incident power, or fluence, the heating efficiency $\eta$ typically decreases. Indeed, both the carrier multiplication efficiency[50] and hot-carrier multiplication efficiency[57] were found to decrease with increasing $P_{in}$, and time-resolved optical experiments that require relatively large absorbed fluences (typically above 10–100 μJ cm$^{-2}$), such as time-resolved photoluminescence[30] and time-resolved ARPES[37,38] typically found relatively low heating efficiencies. Microscopically, this is related to the relatively large density of initial high-energy carriers, and relatively high carrier temperature that is being established during thermalization. This slows down the thermalization process, thus making $\tau_{heat}$ longer. This means that the heating efficiency $\eta$ becomes smaller, due to competing energy relaxation processes for high-energy carriers, with relaxation time $\tau_{relax}$ (see Fig. 3). Furthermore, once thermalization has occurred, at sufficiently high carrier temperatures, efficient cooling can occur, for example *via* optical phonon emission (see section 3.2.1).

*Momentum exchange.* Carrier–carrier scattering events do not only lead to exchange of energy; also carrier momentum is modified (see angle $\varphi$ in Fig. 6b). As discussed in several theoretical and experimental works,[42,45,47,49,50,58] carrier–carrier scattering occurs preferentially in the near-collinear direction, due to the kinematic constraints of carrier–carrier scattering processes (see Fig. 6c). Purely collinear scattering, however, is suppressed, because due to their relativistic band dispersion, electrons and holes all share the same group velocity, while backscattering is strongly suppressed because of their chirality. In low dimensionality, the result is a diverging duration of collisions for collinear particles.[59] The preferred near-collinear scattering means that photoexcitation with linear polarization leads to an anisotropic distribution in momentum space, which lasts longer than the interaction time with the photons (see Fig. 6d). In the case of excitation with a photon energy above the optical phonon energy, the distribution becomes isotropic in ~150 fs.[47,58] This is attributed to phonon-mediated non-collinear carrier scattering. However, in the case of excitation with a photon energy below the optical phonon energy, scattering occurs purely through carrier–carrier interactions, and the anisotropy was found to survive for picoseconds at 20 K. This effect can survive up to room temperature.[47,60]

*Controlling carrier–carrier scattering.* We already saw that changing the Fermi energy and the incident photon energy changes the thermalization cascade. However, it is also possible to control the intrinsic carrier–carrier scattering events. For example, the Coulomb interaction strength decreases with temperature, leading to a longer carrier–carrier scattering time at lower temperature, as observed experimentally.[61] Furthermore, one can control carrier–carrier scattering by proximity screening, as demonstrated recently using a high-quality graphene sample with a nearby metal.[62] Finally, the application of a magnetic field will lead to less efficient carrier–carrier scattering.[63]

### 3.2. Carrier cooling

**3.2.1 Optical phonons.** There are many competing cooling pathways in graphene, as summarized in Fig. 7. We first consider the cooling pathway, where a high-energy carrier loses energy by emitting an optical phonon. Graphene has optical phonons at the K-point with an energy $E_{OP,K} = 0.16$ eV, and at the Γ-point with energy $E_{OP,\Gamma} = 0.2$ eV.[69] An important parameter governing emission of optical phonons is the electron–phonon coupling (EPC) constant, whose value has been determined using first principles calculations and Raman measurements.[70–72] According to calculations in ref. 73 and 74, carriers with an energy that is high enough to emit optical phonons do so very efficiently (see Fig. 8a), with scattering times $\tau_{OP}$ well below a picosecond. Scattering of high-energy carriers with optical phonons can therefore compete with Coulomb carrier–carrier interactions that lead to thermalization, and is therefore the main process determining the

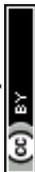

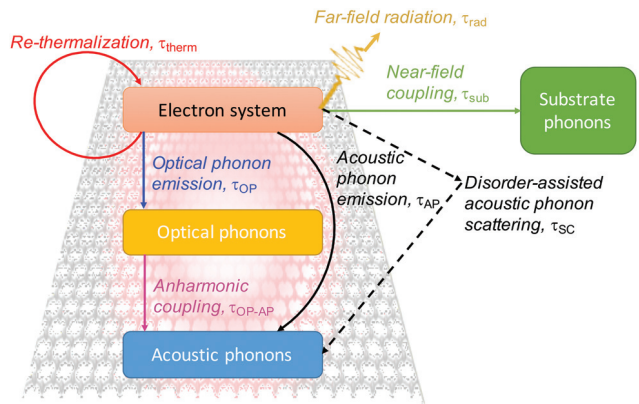

Fig. 7 Schematic overview of hot-carrier cooling mechanisms.

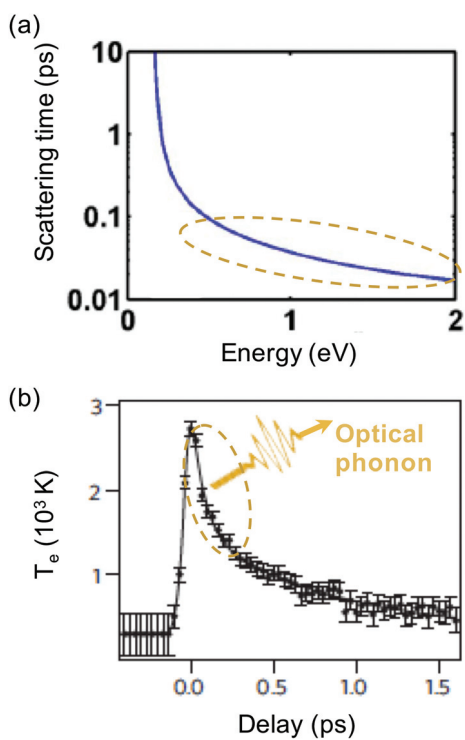

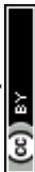

Fig. 8 (a) Calculation of scattering time between a charge carrier with a certain energy and optical phonons, $\tau_{OP}$. (b) Experimental measurement of cooling dynamics using time-resolved ARPES, showing an initial fast decay (∼100 fs) due to optical phonon emission. The slower tail is either ascribed to optical-to-acoustic phonon conversion (hot-phonon bottleneck) or to disorder-assisted acoustic phonon cooling. Panel (a) is adapted with permission from ref. 74 (Copyright 2010 American Institute of Physics); panel (b) is adapted with permission from ref. 37 (Copyright 2013 Springer Nature).

heating efficiency $\eta$ that governs which fraction of the absorbed power density $P_{in}$ ends up in the electronic system (see Fig. 3). The competition between carrier–carrier and carrier-optical-phonon scattering during the thermalization phase was addressed quantitatively in ref. 34. Here, optical phonon emission is associated with the energy relaxation process with timescale $\tau_{relax}$ in Fig. 3.

Time-resolved studies using relatively strong excitation, leading to many high-energy carriers, typically observed ultrafast (few hundred fs) emission of so-called strongly-coupled optical phonons (SCOPs), cf. ref. 74–77, in agreement with the calculations of ref. 74 shown in Fig. 8a. As a result, fast thermalization occurs not only within the electronic system, yet also between the electronic system and the optical-phonon system. This leads to charge carriers having a temperature similar to the temperature of optical phonons, while the acoustic phonon temperature is much lower. This was also observed in electronic transport studies at high electric field.[25,78] Typically, in these studies, bi-exponential cooling dynamics were observed (see Fig. 8b): sub-picosecond decay due to coupling to optical phonons ($\tau_{OP}$), followed by few-picosecond decay that was attributed to cooling of the hot optical phonons to acoustic phonons ($\tau_{OP-AP}$), as the hot phonons are not able to cool fast enough, and re-heat the electronic system. This effect of the hot optical phonons is sometimes referred to as the hot-phonon bottleneck for carrier cooling. The optical phonon lifetime $\tau_{OP-AP}$ has been measured independently, for example through time-resolved Raman measurements, yielding values ranging from 1.2 to 2.55 ps.[30,74,79–81]

For carriers with lower energy, $(\varepsilon + \mu) < E_{OP}$, the timescale of optical phonon emission increases exponentially (see Fig. 8a). Therefore, in the case of relatively weak excitation, or after strong excitation, followed by fast cooling to optical phonons, once $T_e$ is smaller than ∼1000 K, cooling via optical phonons was thought to be rather inefficient.[33] In this case, there would not be enough high-energy carriers to emit optical phonons. Indeed, several experimental studies showed a temporary thermal decoupling between the electronic system and the phonon system, allowing the electronic system to heat up efficiently, before it cools down via phonons.[82] Recently however,[83,84] it was shown that optical phonons can still play an important role in hot-carrier cooling even for carrier distributions with a temperature close to room temperature. This is based on the idea that (at least at room temperature and above) there is a significant fraction of carriers in the tail of the Fermi–Dirac distribution with a kinetic energy that is large enough to couple to optical phonons. Once these high-energy carriers have relaxed by emitting optical phonons, re-thermalization of the electronic system occurs. This leads to "newly excited" carriers with high enough energy to relax by emitting optical phonons. These processes of optical phonon emission and re-thermalization occur simultaneously. Thus, cooling occurs by the combination of optical phonon emission and continuous re-thermalization of the carrier system. A numerical simulation based on this intrinsic cooling mechanism was shown to be consistent with temperature-dependent cooling dynamics obtained with optical-pump THz-probe measurements.[83] More recently, it was shown that this cooling mechanisms leads to bi-exponential decay, with a sub-picosecond initial decay related to direct coupling to optical phonons, and a few-picosecond decay due to the hot-phonon bottleneck.[84] This study showed that a larger peak temperature and smaller $E_F$ typically give rise to slower cooling dynamics for this cooling channel. For graphene where alternative cooling channels (see below) are suppressed, this cooling channel ultimately determines the intrinsic limit of the hot-carrier lifetime of high-quality graphene (with a mobility >10 000 cm$^2$ V$^{-1}$ s$^{-1}$) at room temperature.[84]

**3.2.2 Acoustic phonons.** Charge carriers with a kinetic energy that is not high enough to couple to optical phonons can couple to acoustic phonons. An important parameter governing acoustic phonon emission is the deformation potential, whose value is obtained from first principle calculations, *cf.* ref. 86. Cooling of hot carriers in graphene *via* momentum-conserving scattering with acoustic phonons was first described in ref. 87, where it was found that this cooling channel typically leads to lifetimes $\tau_{AP}$ on the order of nanoseconds. Indeed, it was found to be significantly less efficient than cooling *via* optical phonons for most peak carrier and lattice temperatures. In fact, momentum conservation and the large velocity mismatch between electrons and acoustic phonons ($v_F \gg v_s$, where $v_s$ is the sound velocity) lead to scattering with phonons close to the centre of their Brillouin zone. There, acoustic phonons have nearly zero energy, and therefore cooling requires many successive interactions. This limitation, however, can be overcome through disorder-assisted scattering, as shown in ref. 16 and 85 (see also Fig. 9a). The resultant disorder-assisted cooling mechanism, also known as "super-collision cooling", can give rise to picosecond cooling times $\tau_{SC}$ at room temperature, especially in graphene with a sufficiently large disorder density, and associated low charge mobility. The occurrence of this cooling mechanism was shown experimentally using several experimental techniques shortly after its prediction,[33,88–90] as shown in Fig. 9b. These studies showed that a lower peak temperature and a larger $E_F$ typically give rise to slower cooling dynamics for disorder-assisted supercollision cooling to acoustic phonons.

**3.2.3. Substrate phonons.** Since graphene is typically placed on a substrate, and is atomically thin, carriers can relax by coupling to nearby substrate phonons, with cooling time $\tau_{sub}$. Low *et al.* theoretically addressed this cooling pathway in ref. 93. They found that cooling is an order of magnitude faster in the case of polar substrates, such as SiO$_2$, than in non-polar substrates, such as diamond. This is because of scattering with the surface polar phonon modes. Later, it was shown that if a substrate hosts hyperbolic phonon modes, the coupling between graphene hot carriers and these substrate modes can be even stronger, leading to picosecond cooling times at room temperature[22,91,92] (see Fig. 10a). Hyperbolic phonon polaritons exist in regions where the in-plane and out-of-plane permittivities have opposite signs, and lead to a large photonic density of states.[94] As a result, hot carriers in graphene can cool *via* near-field coupling to this large photonic density of states. This coupling can be seen as super-Planckian radiation of the hot carriers into these optical modes of the substrate. Many layered materials are naturally hyperbolic, meaning that they have some spectral region where hyperbolic modes occur, as studied theoretically in ref. 95. Hexagonal BN (hBN) is arguably the preferred substrate material for graphene

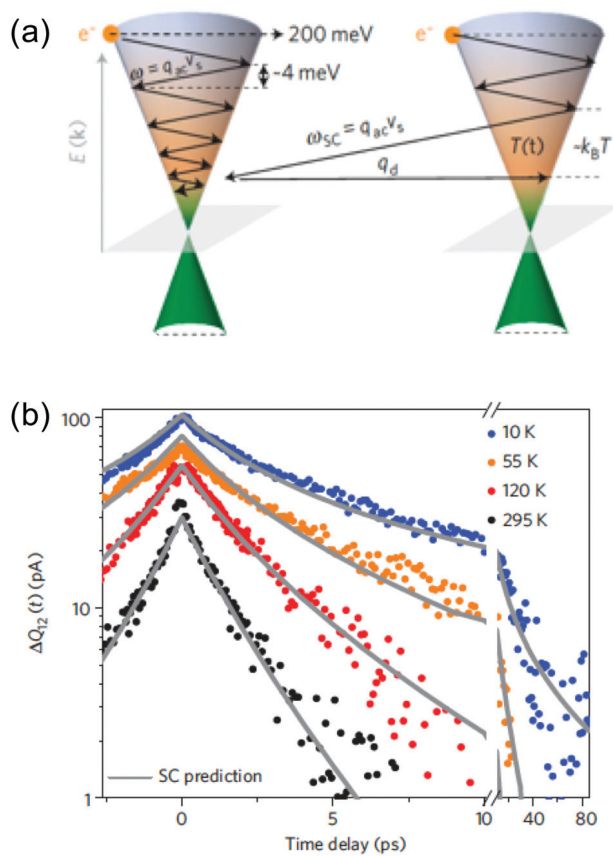

**Fig. 9** (a) Schematic illustration of cooling *via* acoustic phonons without (left) and with (right) disorder, where the presence of disorder speeds up the emission of acoustic phonons, also known as "supercollision cooling".[85] (b) Experimental measurements of cooling dynamics using time-resolved photocurrent microscopy, together with calculations using disorder-assisted acoustic phonon cooling. Panels (a) and (b) are adapted with permission from ref. 33 (Copyright 2013 Springer Nature).

at the moment, because encapsulation of graphene in between hBN leads to very high charge mobility: >50 000 cm$^2$ V$^{-1}$ s$^{-1}$ at room temperature.[96,97] It is also a naturally hyperbolic material, with hyperbolic phonon polaritons occurring in its two Reststrahlen bands, which are the spectral intervals between the longitudinal (LO) and transverse (TO) optical phonon frequencies.[98] Owing to these two spectral regions with hyperbolic phonon polaritons, two optoelectronic experiments have shown that cooling of hot carriers in hBN-encapsulated graphene occurs *via* out-of-plane coupling to hyperbolic phonon polaritons in hBN.[22,91,92] These studies showed that a higher peak temperature and a smaller $E_F$ typically give rise to slower cooling dynamics for super-Planckian cooling to hyperbolic substrate phonon modes (see Fig. 10b).

**3.2.4. Discussion on cooling channels.** The cooling pathways *via* optical phonons, acoustic phonons and substrate phonons described above, and summarized in Fig. 7, can all give cooling times in the ps range at room temperature. However, they do have distinct dependencies on peak carrier temperature and $E_F$, which makes it possible to assess their

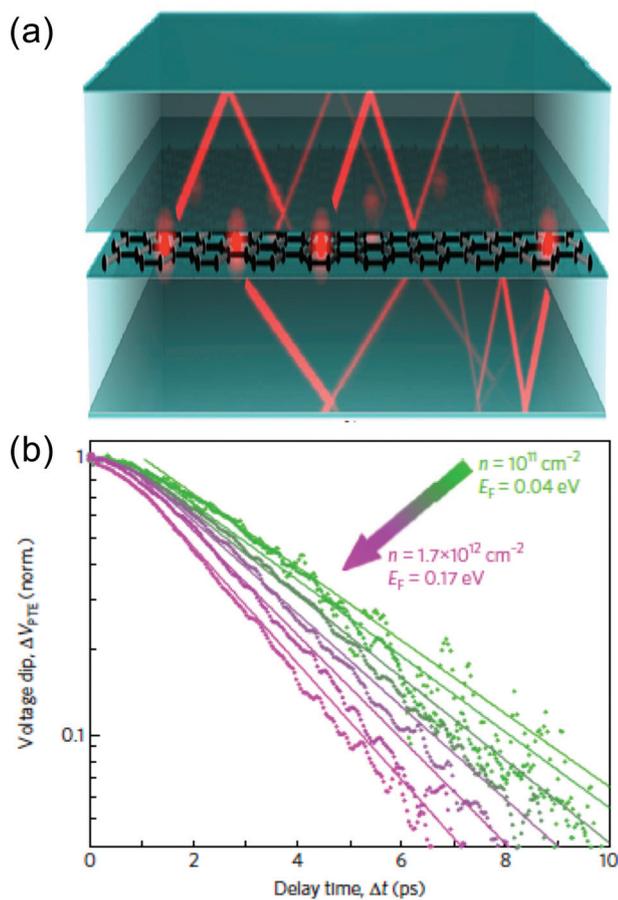

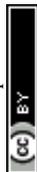

Fig. 10 (a) Schematic illustration of cooling of hot carriers in graphene (red dots) via hyperbolic phonon polaritons in hBN (red lines) via near-field radiation. (b) Experimental measurements of cooling dynamics using time-resolved photocurrent microscopy, showing faster cooling for lower $E_F$, the opposite trend compared to disorder-assisted cooling. Panel (a) is adapted with permission from ref. 91 (Copyright 2017 American Physical Society); panel (b) is adapted with permission from ref. 92 (Copyright 2018 Springer Nature).

respective contributions. What these cooling channels have in common is that they become less efficient upon lowering the lattice temperature, as the phonon occupancy decreases. At sufficiently low lattice temperatures – and depending on external parameters, such as the spatial size of the region with hot carriers – cooling will be dominated by the lateral diffusion of hot carriers, the so-called Wiedemann–Franz cooling (see section 1). This cooling channel was observed using noise thermometry measurements, and – for their specific experiment – dominated for a lattice temperature below 80 K.[99] It's important to note that this is a cooling process that is not captured by the term $\Gamma_{cool}$ in eqn (1), as it is cooling that occurs as a result of transport.

We briefly mention alternative cooling channels. Coupling to graphene plasmons could play a role in the early stages of the energy cascade,[100] and was suggested to explain the short-lived inverted state observed in ref. 37. Experimental signatures of this cooling mechanism were observed recently.[101]

Furthermore, hot-carrier photoluminescence – Planck radiation from the hot-carrier system – has been shown to occur.[30] However, this process is inefficient: it gives rise to a $\Gamma_{cool}$ of $\sim$10 W m$^{-2}$ K$^{-1}$,[92] corresponding to a $\tau_{cool}$ of $\sim$10 ns. A theoretical proposal suggested that exploiting plasmonic near-field effects between neighboring graphene nano-islands could speed radiative heat transfer up to the femtosecond regime.[102] A final cooling mechanism that was identified corresponds to resonant dissipation from individual atomic defects in graphene, as observed in ref. 103. Finally, we note that under certain external conditions – beyond those that create defects or modify the Fermi energy or peak/lattice temperature – the carrier dynamics can be altered. Most notably, the application of a magnetic was shown to lead to significantly slower dynamics.[63]

## 4 Conductivity

The conductivity $\sigma$ governs both the electrical transport and the optical properties of graphene.[115] In the following we will discuss the linear and nonlinear frequency- and temperature-dependent conductivities.

### 4.1 Linear conductivity

There is nowadays a large amount of theoretical[115–123] and experimental[124,125,126,127–130] works dealing with the linear frequency-dependent optical conductivity of graphene $\sigma(\omega)$. Here we will focus on how the interband and intraband transitions of graphene depend on carrier temperature. Typically, for frequencies up to a few THz the intraband conductivity dominates, while for higher frequencies both contributions can play a role. Note that for graphene on a substrate the absorbance (in SI units and normal incidence) is related to $\sigma(\omega)$ by $A(\omega) = 1 - T = \frac{2Z_0}{1+n_s}\mathrm{Re}[\sigma(\omega)]$, where $Z_0 = \frac{1}{c\varepsilon_0} \sim 377\,\Omega$ is the vacuum impedance, $n_s$ the substrate's refractive index and $T$ is the transmission.[130,131] This means that $\sigma(\omega)$ determines the number of photogenerated hot electrons, as it determines the optical absorption, and it's directly related to Joule heating as it determines the resistance of a graphene sheet (see section 2).

#### 4.1.1 Interband conductivity. Graphene's interband optical conductivity can be written as:[125,126,132]

$$\sigma(\omega)_{\text{inter}} = \frac{e^2}{8\hbar}\left[\tanh\left(\frac{\hbar\omega + 2\mu}{4k_B T_e}\right) + \tanh\left(\frac{\hbar\omega - 2\mu}{4k_B T_e}\right)\right]. \quad (16)$$

For large enough photon energies, when $\hbar\omega \gg 2|\mu|$ and $\hbar\omega \gg 4k_B T_e$, interband transitions dominate graphene's absorbance and eqn (16) reduces to $\sigma(\omega)_{\text{inter}} = \sigma_0 = e^2/(4\hbar)$, which thus becomes independent of frequency.[124,125,133] For this value of the sheet conductivity the absorbance for isolated (suspended) graphene ($n_s = 1$) depends only on elementary constants $A = \pi\frac{e^2}{2c\varepsilon_0\hbar} = \pi\alpha \sim 2.3\%$. Note that $\alpha = \frac{e^2}{2c\varepsilon_0\hbar}$ is the fine structure constant. The same "universal conductivity" can be obtained from the Fermi golden rule for two-dimensional Dirac fermions.[124]

For large energies, graphene's band structure deviates from linear dispersion, and excitonic effects at the saddle point singularity at ∼5 eV lead to increased absorption.[121,127] Although theory predicts that the approximation of a linear dispersion should hold only for relatively small values of the incident photon energy ($\hbar\omega$ < 1 eV),[124] experiments observed very small deviations from the universal value of the absorbance $A \sim 2.3\%$ for photon energies up to $\hbar\omega$ = 1.5–2 eV.[124,127] For higher photon energies, the conductivity deviates significantly from the universal value.

Another key aspect of $\sigma(\omega)_{\text{inter}}$ is that the value $2\mu$, which defines the crossing from intraband to interband transitions, can be tuned by doping and external gate voltages,[134] leading to the possibility of designing gate-tuneable optical modulators.[135] According to eqn (16), due to heat-induced broadening of the Fermi–Dirac distribution and the accompanying shift in the chemical potential $\mu$, interband absorption can either increase (for $\hbar\omega < 2E_F$) or decrease (for $\hbar\omega > 2E_F$), as shown in Fig. 11.

**4.1.2 Intraband conductivity.** The intraband (or electrical) conductivity for electrons or holes is given by:[123]

$$\sigma_{\text{intra;e,h}} = (\omega, T_e) = \frac{e^2 v_F^2}{2} \int_0^\infty \partial\varepsilon D(\varepsilon)\left(\frac{\partial f_{\text{e,h}}}{\partial \varepsilon}\right)\frac{\tau_{\text{mr}}}{1 - i\omega\tau_{\text{mr}}}. \quad (17)$$

There are three main approaches that have been used to understand how this intraband conductivity changes upon carrier heating, which is less straightforward than in the case of interband transitions. Experimentally, this issue has been addressed in particular using optical pump – terahertz probe studies, cf. ref. 34, 35, 57 and 136–139. We will now describe the three approaches, where we note that they are not mutually exclusive.

*Drude weight.* In the first approach, the temperature dependence is fully captured by the so-called Drude weight:[131,137]

$$\sigma_{\text{intra;e,h}} = \frac{2e^2\tau_{\text{mr}}}{\pi\hbar^2(1 - i\omega\tau_{\text{mr}})}k_B T_e \ln\left[2\cosh\left(\frac{\mu_{\text{e,h}}}{2k_B T_e}\right)\right]. \quad (18)$$

This result comes from considering spectral weight conservation,[137] which is related to the Thomas–Reiche–Kuhn sum rule for light–matter interaction.[140] For the graphene conductivity, this sum rule indicates that the broadening of the interband conductivity (see previous subsection, and Fig. 18) implies that, in the "doped" regime, the spectrally integrated weight of interband transitions increases. In the "undoped" regime, the integrated interband weight decreases. Spectral weight conservation then dictates that the spectrally integrated weight of intraband transitions should decrease (increase) for the "doped" ("undoped") case. Note that the Drude weight for graphene is different compared to conventional semiconductors or metals.[127,128] This approach predicts that if $k_B T_e \ll E_F$, namely for weak heating and large doping, the hot-carrier conductivity is smaller than the equilibrium conductivity (negative photoconductivity), which is in agreement with several experiments[34,35,57,136–139] (see inset Fig. 11). Furthermore, it explains the experimental observation of an increase in conductivity, *i.e.* positive photoconductivity, close to the Dirac point (where the ratio $T_e/T_F$ is large), as observed in ref. 35, 57, 137 and 139. This approach, however, does not take into account that momentum scattering can depend on Fermi energy and/or carrier temperature. Furthermore, the Drude weight predicts positive photoconductivity at high fluence, which is in contradiction with experiments. In ref. 137, the dependence of $u_{\text{mr}}$ on $T_e$ was inserted phenomenologically, assuming $T_e = T_L$ and momentum relaxation dominated by phonon scattering.

*Sommerfeld expansion.* In ref. 34, the microscopic intraband conductivity of eqn (17) was calculated in the Sommerfeld expansion regime, where the ratio $T_e/T_F$ is small. Here, the energy-dependence of the momentum relaxation time was taken into account, under the realistic assumption that it is dominated by long-range Coulomb scattering with impurities. This corresponds to a momentum relaxation time that typically increases linearly with carrier energy,[123] and is typically the channel that limits charge mobility at room temperature in substrate-supported graphene. This approach correctly gives the negative photoconductivity for $E_F$ away from the Dirac point, as observed experimentally.[34,35,57,136–139] However, it can not be used close to the Dirac point, as the conditions for the Sommerfeld expansion are not met.

*Full microscopic approach including dynamical screening.* Finally, in ref. 35, the microscopic intraband conductivity eqn (17) was calculated explicitly, including the energy-dependent momentum relaxation time, governed by long-range Coulomb scattering with impurities. Dynamical screening in the random-phase approximation was taken into account in the calculation of $\tau_{mr}$. This approach gives positive photoconductivity close to the Dirac point and negative photoconductivity away from the Dirac point, in agreement with experiments.[35,57,137,139] According to ref. 35, the positive photoconductivity close to the Dirac point is the result of the increase in the charge carrier density (interband thermalization with possible carrier multiplication). Away from the Dirac point (where the charge carrier density does not increase, as intraband thermalization occurs), the conductivity is negative, because the heated carrier distribution leads to a decrease in screening of the Coulomb interaction with impurities.[35]

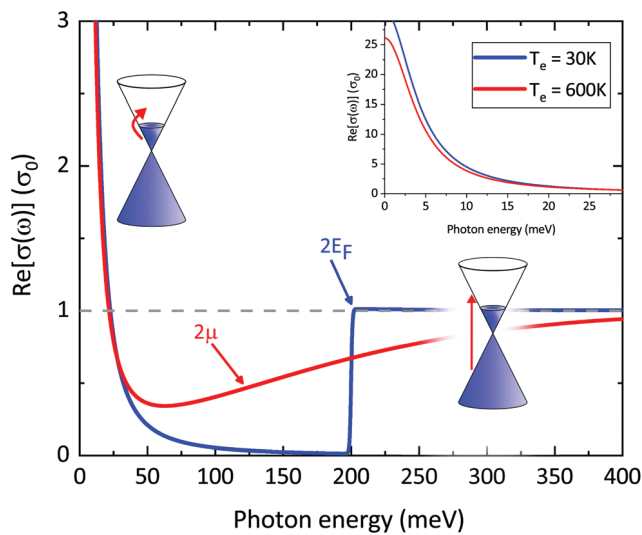

**Fig. 11** Real part of graphene's frequency-dependent linear optical conductivity in units of $\sigma_0$ based on eqn (16) for interband transitions and (18) for intraband transitions (Drude model) with fixed momentum relaxation time $\tau$ = 1 ps and different values of $T_e$ and $\mu$: $T_e$ = 3 K and $\mu$ = 0.1 eV (blue curve), $T_e$ = 600 K and $\mu$ = 0.057 eV (red curve). Note that for interband transitions $\sigma$ at $T_e$ = 600 K can either decrease or increase compared to the value at $T_e$ = 30 K close to $\hbar\omega$ = 0.2 eV. For intraband transitions (inset) $\sigma$ decreases (negative photoconductivity) as expected from the Drude model in the limit of large doping ($T_e < T_F$).

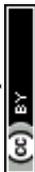

**4.1.3 Nonlinear conductivity.** The nonlinear optical properties of graphene have been investigated extensively, both theoretically[120,122,141–144] and experimentally. Of particular interest are the observation of high-harmonics generation (HHG),[56,145] third-harmonic generation (THG),[146–150] four-wave mixing (FWM)[151–153] and different kinds of intensity-induced changes of the refractive index, such as saturable absorption (SA)[154,155] and the Kerr effect.[156] Being a centrosymmetric crystal, graphene does not display intrinsic even-order nonlinearities in the dipole approximation. For this reason, second-harmonic generation has been observed only due to extrinsic factors, such as breaking of symmetry at an interface[157] and electric currents/fields,[158,159] or at large incidence angles due to the quadrupole nonlinear response.[160] An intuitive picture of the strong nonlinear optical response of graphene can be obtained from a semi-classical model considering the electric field induced sheet current density within the Dirac cone.[161–163] An overview of the nonlinear optical properties of graphene and related layered materials can be found in ref. 15, while here we focus on the effect of hot carriers on the nonlinear optical sheet conductivity.

**4.1.4 Saturable absorption.** Saturable absorption is a non-parametric third-order nonlinear optical process. The nonlinear response induced by SA corresponds to an electric field oscillating at the same frequency of the incoming light, and for this reason SA belongs to the intensity-dependent refractive index nonlinear processes. As discussed in section 6, SA in graphene is widely used for passive mode-locking of ultrafast lasers thanks to its unique combination of broadband absorption and ultrafast dynamics. A detailed theoretical analysis of SA in graphene can be found in ref. 143 and 164, including the formal expression of the field-dependent nonlinear optical conductivity $\sigma_{x,x}(\omega, E_0)$ where $x$ is any in-plane Cartesian direction of the graphene sheet and $E_0$ the incoming electric field responsible for SA.[164] Note that SA is possible for both intraband and interband transitions[143] and in both cases the dynamics of the effect is dominated by the heating and cooling dynamics of hot electrons. A sketch of the SA mechanism in the case of interband transitions is depicted in Fig. 12.[154] The photo-excited hot electrons can inhibit optical transitions by Pauli blocking in a range of $k_B T_e$ around $E_F$, thus reducing the absorption of photons at energy $\hbar\omega \sim k_B T_e$. In the low-excitation regime when $k_B T_e$ is smaller than $E_F$ or the excitation photon energy, the (linear) absorption is independent of the excitation intensity. However, for increasing excitation intensity and higher $k_B T_e$ hot electrons will induce Pauli blocking of interband transitions, leading to a nonlinear intensity-dependent saturable absorption.[143,154] In the case of intraband transitions, SA is due to the photo-induced negative photoconductivity discussed above: an increase of the excitation intensity leads to a reduction of $\sigma_{intra}$ (inset Fig. 11) and thus to a reduction of the absorption. In most materials, SA is well described by the phenomenological absorption law:

$$a(I) = \frac{a_S}{1 + I/I_S} + a_{NS}, \quad (19)$$

where $I_S$ is the saturation intensity, while $a_S$ and $a_{NS}$ are the saturable and non-saturable absorption components.[143,154] This form is fundamentally an exact description for a two-level system. However, the linear band dispersion of graphene produces a qualitatively different intensity dependence of the absorption, leading to the modified expression for SA:[143]

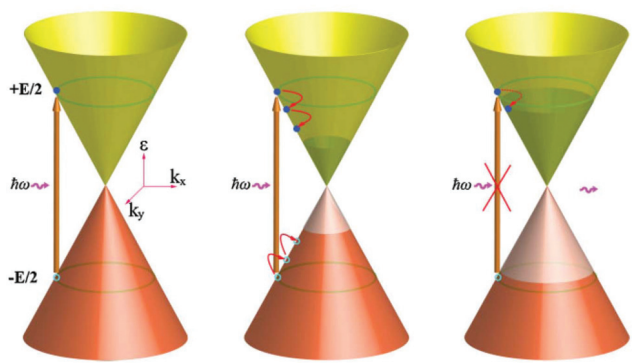

**Fig. 12** Sketch of the SA process in the case of interband transitions driven by hot-electron broadening of the Fermi–Dirac distribution and Pauli blocking. Reprinted with permission from ref. 154 (Copyright 2009 Wiley).

$$a(I) = \frac{a_S}{\sqrt{1 + 3I/I_S}} + a_{NS}. \quad (20)$$

**4.1.5 Near-IR third harmonic generation.** The efficiency of nonlinear processes such as third-harmonic generation (THG) is strongly enhanced due to resonant transitions. This is true also for graphene, where the THG efficiency can be efficiently tuned by more than one order of magnitude when in resonance with vertical multi-photon transitions within the Dirac cone[148,149] (Fig. 13a). In addition, compared to other materials, graphene offers the unique possibility of tuning such transitions over an extremely broadband photon energy range by applying external gate voltages and thus by tuning $E_F$.[148] Since the THG intensity $I_{3\omega}$ scales as $I_{3\omega} \propto |\sigma^{(3)}(\omega, E_F, T_e)|^2$, THG in graphene is fully captured by the nonlinear optical sheet conductivity $|\sigma^{(3)}(\omega, E_F, T_e)|^2$, which can be microscopically calculated at $T_e = 0$ K.[141,164,165] Following symmetry

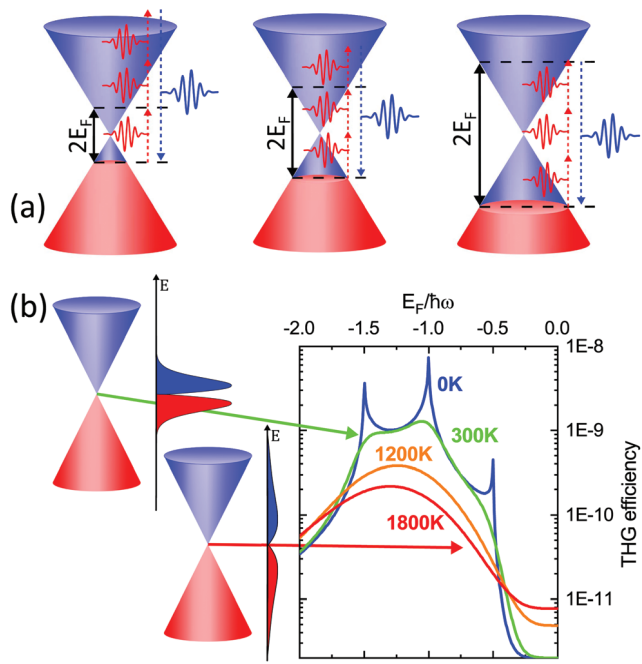

**Fig. 13** (a) THG in graphene is enhanced at specific values of $E_F$ corresponding to one-, two- and three-photon transitions within the Dirac cone. (b) The increase of $T_e$ affects the nonlinear conversion efficiency by broadening the electron distribution. The graph shows the THG efficiency, defined as $I_{3\omega}/I_\omega$, for $\hbar\omega$ = 500 meV and a peak power density of ~230 MW cm$^{-2}$.

considerations, the third-order optical sheet conductivity at $T_e$ = 0 K, which is in general a rank 4 tensor, can be reduced to a single in-plane element:[141,148]

$$\sigma^{(3)}(\omega, E_F) = i\frac{\sigma_0^{(3)}}{24(\hbar\omega)^4}[17G(2|E_F|, \hbar\omega) \\ - 64G(2|E_F|, 2\hbar\omega) + 45G(2|E_F|, 3\hbar\omega)], \quad (21)$$

where $\sigma_0^{(3)} = 4e^4\hbar v_F^2/(32\pi)$ and $G(x,y) = \ln[(x+y)/(x-y)]$. As discussed, $\sigma^{(3)}$ has three sharp resonances at $\hbar\omega m = 3E_F$ with $m$ = 1, 2 and 3 corresponding to one-, two- and three-photon vertical transitions (Fig. 13a).[148,149] Interestingly, for $|E_F|/\hbar\omega \ll 0.5$ the THG efficiency has its minimum value although in principle all multi-photon transitions are resonant at the same time. This is because the signs of the different contributions in eqn (21) sum up to ~0 in this limit, in analogy to the effect of quantum interference for the Raman G mode.[134] Fig. 13b shows $\sigma^{(3)}(\omega, E_F, T_e)$ for different values of $T_e$ between 0 K and 1800 K: for increasing $T_e$ the multi-photon resonances, which are clearly visible at $T_e$ = 0 K, broaden until they merge and the modulation depth of the THG efficiency decreases dramatically. This can be qualitatively understood considering the $T_e$-induced broadening of the Fermi–Dirac distribution. Since THG is a parametric process and the emission occurs only during interaction of the electric field with the nonlinear material, all nonlinear optical experiments on graphene that use ultrashort pulses (<100 fs) should always carefully consider the effect of a high $T_e$. Notably, this is true also when $\hbar\omega < 2|E_F|$ due to broadening of the interband transitions at high $T_e$ and possible intraband transitions mediated by defects and phonons. The increase of $T_e$ is also responsible for large deviations from the typical cubic power dependence of THG experiments, as shown in ref. 150. Similar effects due to an increase of $T_e$ were observed also in FWM experiments[149,153,166] and recently ultrafast all-optical modulation of THG with modulation depth above 90% was obtained in single-layer graphene.[163]

**4.1.6 THz high-harmonic generation.** The case of THz high-harmonic generation (HHG) clearly illustrates the importance of hot electrons in the nonlinear optical response of graphene. Several theoretical works predicted the existence of an intense nonlinear response of graphene at THz frequencies, based on a nonlinearity mechanism relying on coherent electron motion.[122,161,162,167–169] However, experiments either did

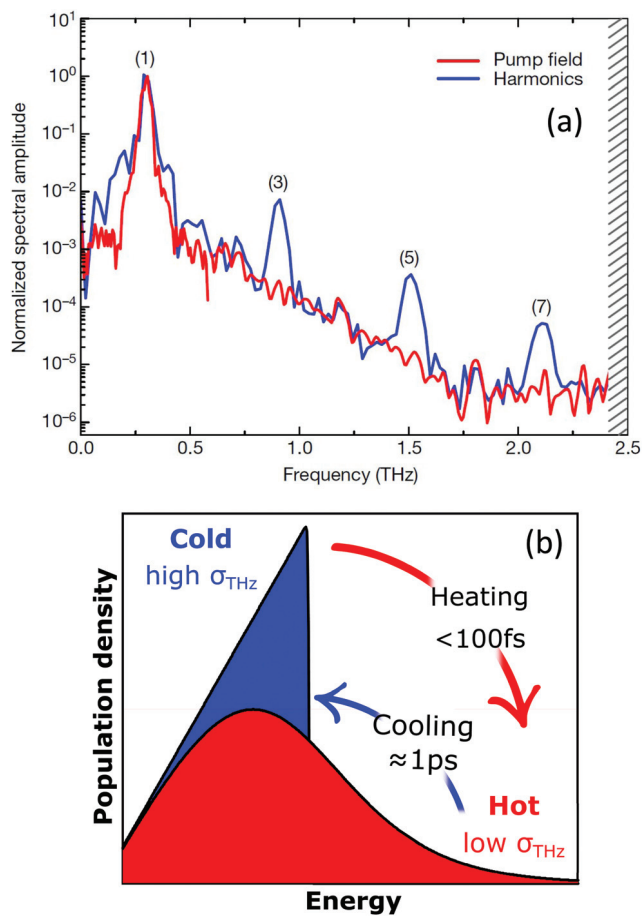

**Fig. 14** (a) THz high-harmonic generation from single layer graphene. (b) Excitation in the THz corresponds to an oscillating period of the electric field of ~1 ps, thus much longer than carrier–carrier scattering and comparable to the cooling time. For this reason, the nonlinear optical response of graphene in this frequency range is driven by incoherent thermodynamic effects arising from the strong temperature dependence of the sheet optical conductivity $\sigma(\omega)$ (inset). Panel (a) is adapted with permission from ref. 56 (Copyright 2018 Springer Nature).

not detect any THz HHG from multi-layer graphene,[170] or only observed weak signatures of THz harmonics at 50 K.[171] These results were explained as the consequence of ultrafast e–e scattering (<100 fs) which suppresses the coherence in the THz-induced electron velocity and current density in materials with linear dispersion.[172]

Recently, HHG (Fig. 14a) with extremely high field-conversion efficiencies of $10^{-3}$, $10^{-4}$ and $10^{-5}$ for third, fifth and seventh harmonics, respectively, was demonstrated at room temperature using doped single-layer graphene.[56] The result was successfully interpreted based on a nonlinearity mechanism that does not rely on coherent electron motion. Rather, the THz nonlinearity was understood from a purely thermodynamic picture of the THz response of doped graphene. In this thermodynamic picture,[56,173] hot carriers play a central role, because the intense nonlinear response of graphene is due to the combination of the THz-induced carrier heating that reduces the THz conductivity,[54] together with the ultrafast heating and cooling dynamics of hot electrons (Fig. 14b). We have discussed the effect of carrier heat on the THz intraband conductivity in section 4.1 and the details of the heating–cooling dynamics in section 3. The nonlinearity arises simply because THz absorption leads to carrier heating, which leads to negative THz photoconductivity, i.e. reduced THz absorption. This effect was discussed in detail in ref. 54. Since heating is highly efficient, this mechanism leads to very high nonlinearity coefficients. The heating–cooling dynamics play a crucial role in observing the THz harmonics created by a multicycle incident THz pulse, as detailed in ref. 56 and 173.

In the future, THz nonlinear optics with graphene could offer a rich playground for on-chip signal processing[174] and for the study of the hydrodynamic regime (see 5.1.4) with Dirac fermions.[175]

## 5 Transport

In this section, we address transport properties related to hot carriers. We first discuss intrinsic properties such as hot-carrier diffusion and thermoelectric effects. Subsequently, we discuss interlayer phenomena, involving transport of graphene hot carriers to a neighboring material.

### 5.1 Diffusion

**5.1.1 Diffusivity.** The diffusion of heat is governed by the first term on the right-hand side of the thermoelectric equation, eqn (1). Ignoring the other terms, we obtain

$$C_e \frac{\partial T_e}{\partial t} = \kappa_e \nabla^2 T_e, \qquad (22)$$

where $\kappa_e$ is the electronic part of the thermal conductivity. This equation takes the form of a standard diffusion equation with a thermal diffusivity $\mathcal{D} = \frac{\kappa_e}{C_e}$, typically given in units of cm$^2$ s$^{-1}$. Since hot charge carriers are transporting the electronic heat, the thermal conductivity $\kappa_e$ and charge conductivity $\sigma$ are directly related in the diffusive regime, namely through the Wiedemann–Franz relation:

$$\kappa = \frac{\pi^2}{3e^2} k_B^2 T_e \sigma. \qquad (23)$$

It can be shown (see ref. 176) that this gives a simple relation between diffusivity and charge mobility $\mu_{\text{charge}}$, which is the same result that can be obtained using the Einstein relation:

$$\mathcal{D} = \frac{\mu_{\text{charge}} E_F}{2e}. \qquad (24)$$

This predicts that the electronic heat diffusivity in graphene at room temperature has a value of 100–5000 cm$^2$ s$^{-1}$, for typical $E_F$ up to 0.2 eV and mobilities varying from a few thousand cm$^2$ V$^{-1}$ s$^{-1}$ for CVD graphene up to 50 000 cm$^2$ V$^{-1}$ s$^{-1}$ for hBN-encapsulated graphene. Microscopic simulations predicted a diffusivity of 360 cm$^2$ s$^{-1}$.[177] The diffusivity of graphene was first studied experimentally using all-optical spatio-temporal scanning,[178] which found a short-lived initial $\mathcal{D}$ of 5500–11 000 cm$^2$ s$^{-1}$, followed by a value on the order of 250 cm$^2$ s$^{-1}$. Recently, a spatio-temporal thermoelectric scanning microscopy experiment on hBN-encapsulated graphene found a value of ∼2000 cm$^2$ s$^{-1}$.[176] This value is in agreement with the measured electrical charge mobility $\mu_{\text{charge}}$ of ∼40 000 cm$^2$ V$^{-1}$ s$^{-1}$ for their sample, thus being in accordance with the Wiedemann–Franz and Einstein relations.

**5.1.2 Cooling length.** An important parameter for several hot-carrier-based applications is the cooling length $\xi_{\text{cool}}$, which describes how far hot charges can travel before they cool down. This is given by:[179]

$$\xi_{\text{cool}} = \sqrt{\frac{\kappa_e \tau_{\text{cool}}}{C_e}}. \qquad (25)$$

For $E_F \gg k_B T_e$, and owing to the Wiedemann–Franz relation, this can be written as:

$$\xi_{\text{cool}} = v_F \sqrt{\frac{\tau_{\text{cool}} \tau_{\text{mr}}}{2}}, \qquad (26)$$

where $\tau_{\text{mr}}$ is the momentum relaxation time that defines the electrical mobility. With typical picosecond cooling times and sub-picosecond momentum relaxation times, the cooling length is typically below 1 μm at room temperature, as observed experimentally for example in ref. 180. Ma et al.[181] measured the cooling length using photocurrent microscopy, and found a non-monotonous trend with lattice temperature, which they ascribed to different regimes where distinctive cooling mechanisms dominated. For device configurations where optical excitation leads to a hot-carrier distribution with a spatial size larger than 1 μm, diffusion typically plays a small role.

**5.1.3 Heat transfer via diffusion.** Spatial diffusion of heat leads to a reduction of the average $T_e$ in the heated region. This can be referred to as "diffusive cooling" or "Wiedemann–Franz cooling", even though it is a transport phenomenon, rather than a cooling phenomenon (associated with $\Gamma_{cool}$). This cooling effect of heat transport was observed in graphene samples where carriers were heated by Joule heating, and cooling occurred by diffusion into the metal leads, *cf.* ref. 88 and 99. Both works obtained the thermal conductivity by measuring the temperature of their system *via* Johnson noise thermometry measurements, and found results that were in agreement with the Wiedemann–Franz law away from the Dirac point. These studies required cryogenic temperatures, such that other cooling and momentum relaxation channels were suppressed.

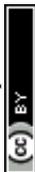

**5.1.4 Non-diffusive transport.** The above derivations – with electronic heat diffusion following Fourier's law of diffusion – hold when the scattering mechanisms affect charge and heat transport in the same way, which is usually the case. In this situation, electronic charge and heat transport are connected *via* the Wiedemann–Franz law. In graphene, however, it was shown that hydrodynamic transport can occur, where heat and charge flow have great similarities with the flow of a liquid in a vessel, being described by the laws of hydrodynamics. Several peculiar charge transport phenomena have been experimentally demonstrated and explained when the system is tuned to the hydrodynamic charge transport regime, such as negative local resistance.[182] This occurs when the carrier–carrier scattering length is shorter than the momentum relaxation length and shorter than the device length. Hydrodynamic flow was also found to occur for phonons in graphite, in the form of the experimental observation of wave-like transport of heat, called second sound.[183]

A special hydrodynamic charge transport regime exists very close to the Dirac point: the Dirac fluid. Here, the Wiedemann–Franz relation breaks down, and electronic heat can travel unimpeded, as electrons and holes travel together along a thermal gradient.[99] This typically occurs very close to the Dirac point, requiring cryogenic temperatures and ultra-clean samples. However in the case of elevated $T_e$ and in the absence of electron–phonon thermalization, it can also be observed further away from the Dirac point, as demonstrated experimentally in ref. 176. In this study, increased diffusivities of hot carriers were found in the first few hundred femtoseconds after photoexcitation with $\mathcal{D}$ up to 70 000 cm$^2$ s$^{-1}$ and thermal conductivities above 10 000 W m$^{-1}$ K$^{-1}$. This transport occurs in the hydrodynamic time window before momentum relaxation occurs (∼350 fs in their case), whereas carrier have already thermalized through carrier–carrier scattering.

### 5.2 Thermoelectric effects

**5.2.1 Seebeck effect.** Hot carriers in graphene can give rise to important thermoelectric and thermomagnetic phenomena.[184–188] Arguably, the most prominent among them is the Seebeck effect, which refers to the conversion of a temperature difference $\Delta T_e$ into an electrical voltage $\Delta V$. According to eqn (2), when $J = 0$, this conversion is proportional to the Seebeck coefficient, or thermopower, $S = -\nabla V/\nabla T_e$. In a diffusive conductor, this effect originates from the net charge imbalance created by the thermal diffusion of carriers that have an energy-dependent DC conductivity $\sigma(\varepsilon) = en(\varepsilon)\mu_{charge}(\varepsilon)$. More specifically, the relation between the electrical conductivity and the Seebeck coefficient is given by the semiclassical Mott formula:[189]

$$S_{Mott} = -\frac{\pi^2 k_B{}^2 T_e}{3e} \frac{1}{\sigma(E_F)} \frac{\partial \sigma(\varepsilon)}{\partial \varepsilon}\bigg|_{\varepsilon=E_F} \quad (27)$$

We note that this formula, which is based on Boltzmann theory (see Box 1), is derived using the Sommerfeld expansion and is thus only valid in the degenerate limit $k_B T_e \ll E_F$, which is the case for most metals and doped graphene. In practice, $S_{Mott}$ can be calculated from the measurement of electrical conductivity of graphene at varying $E_F$, which is conveniently achieved using a gate voltage.

Since 2009, the Mott relation has been experimentally verified several times in graphene.[18,19,190–193] In these experiments, the Seebeck coefficient is typically measured using microfabricated heaters to create a temperature gradient while measuring the thermally induced voltage across the device (see inset Fig. 15). The first measurements of the Seebeck effect in graphene revealed a sizable thermopower value, reaching ∼80 µV K$^{-1}$ close to the Dirac point at room temperature. The sign of the thermopower changes across the Dirac point, as the majority carrier density switches between holes and electrons. At high carrier density, the thermopower is proportional to $1/\sqrt{n}$ and increases linearly with $T_e$, as predicted by the Mott formula. Since the latter depends on the energy derivative of the conductivity, the thermopower strongly depends on the scattering mechanism governing the electrical conductivity $\sigma$. Of particular importance is how the momen-tum relaxation time $\tau_{mr}$ depends on carrier energy $\varepsilon$. As Hwang et al.[194] pointed out, long-range scattering by screened charged impurities appears to be the dominant scattering mechanism in most graphene samples at room temperature. Deviations from the Mott formula were observed close to the Dirac point and at high temperature, where the Sommerfeld expansion is no longer valid. These discrepancies, as well as the nonlinear temperature dependence of $S$ observed in clean graphene devices,[195,196] can be explained theoretically by employing an effective medium theory and by considering the energy-dependence of $\tau_{mr}$ for different momentum scattering mechanisms.[194]

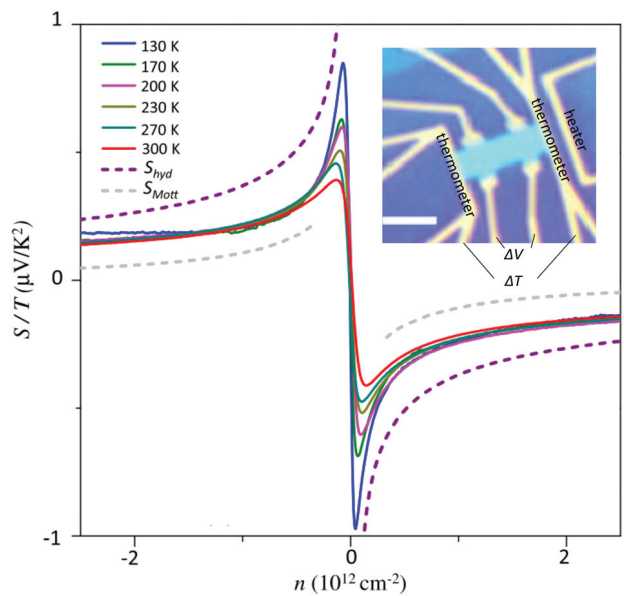

**Fig. 15** Measured $S/T_e$ as a function of charge carrier density $n$ at different temperatures. The grey and purple curves represent $S_{Mott}/T_e$ and $S_{hyd}/T_e$, respectively. The inset shows a typical device used to measure $S$, where the scale bar corresponds to 2 µm. The figure is adapted with permission from ref. 190 (Copyright 2016 American Physical Society).

The situation is quite different in clean graphene samples with ultra-high mobility. As discussed in section 5.1.4, inelastic electron–electron scattering can become the dominant scattering process in clean graphene, and the resulting hydrodynamic behavior of charge carriers can lead to a violation of the Mott formula. This effect was first investigated theoretically in the context of hydrodynamic transport in Dirac fluids.[197,198] It was predicted that in a purely hydrodynamic regime, the Seebeck coefficient is the entropy transported per charge of the carrier, which in the degenerate regime gives

$$S_{hyd} \approx -\frac{2\pi^2 k_B^2 T_e}{3eE_F} \qquad (28)$$

Ghahari et al. provided evidence of hydrodynamic thermoelectric transport in hBN-encapsulated graphene samples with extremely low disorder.[190] At high temperature, the measured thermopower surpasses the calculated $S_{Mott}$, approaching the predicted $S_{hyd}$ (see Fig. 15). This result was explained by considering the inelastic scattering between carriers, as well as the one between carriers and optical phonons. According to a recent theoretical prediction,[199] this enhancement of the Seebeck coefficient in the hydrodynamic regime, combined with the low $\kappa_{hyd}$ in the Fermi-liquid regime should give rise to significant improvement of the heat-to-work conversion efficiency of thermoelectric devices.

5.2.2 **Photo-thermoelectric effect.** The Seebeck effect has been extensively studied using optical excitation as a local heat source. The resulting photo-thermoelectric (PTE) effect has been shown to play an important role in the optoelectronic response of graphene devices. In this case, light absorbed in graphene generates a local distribution of hot carriers characterized by a temperature gradient $\nabla T_e$ (see Fig. 16a). When this gradient overlaps with regions of different Seebeck coefficients ($S_1$ and $S_2$), a voltage $V_{PTE}$ is created, given by:

$$V_{PTE} = -\int S(x)\nabla_x T_e dx = \Delta S \Delta T_e \qquad (29)$$

where $\Delta S = S_2 - S_1$ and $\Delta T_e$ is the temperature increase at the junction between the $S_1$ and $S_2$ regions. In the short-circuit configuration, the photovoltage acts as an electromotive force and generates a photocurrent that can be approximated as $I_{PTE} = \Delta S \Delta T_e / R$, where $R$ is the total resistance of the device. This equation is valid for a rectangular device with a uniformly heated junction. A more accurate estimation of $I_{PTE}$ can be

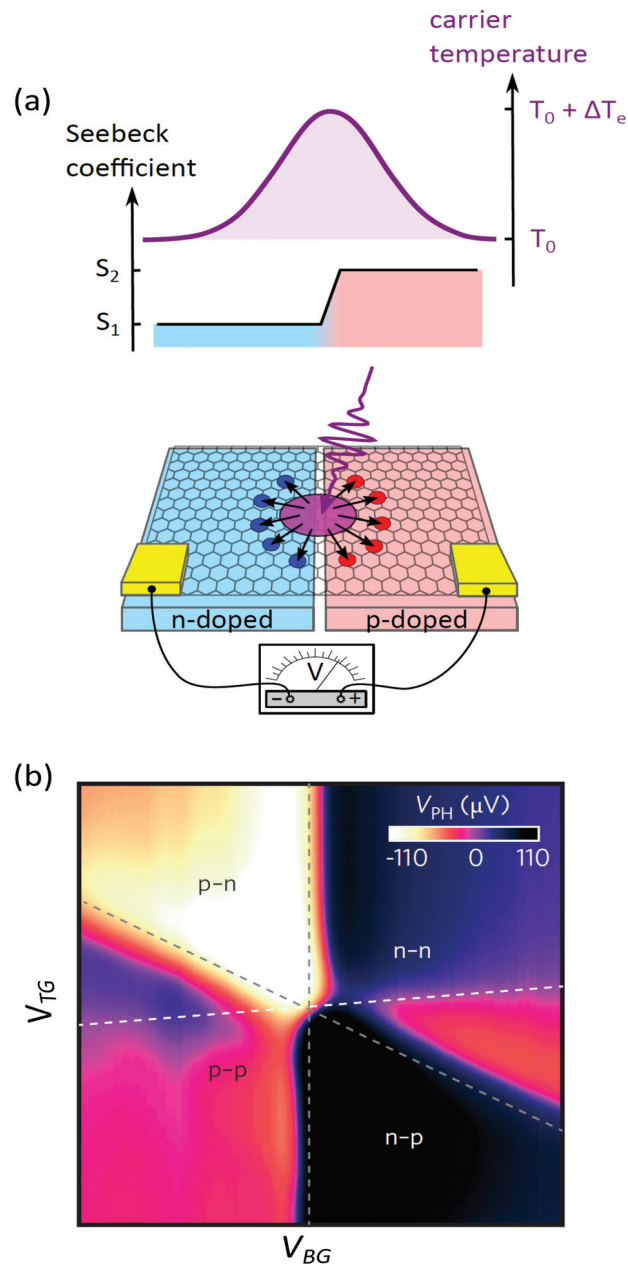

Fig. 16 (a) Schematic representing the generation of a PTE voltage in a graphene p–n junction. Light absorbed at the junction between regions with different Seebeck coefficients ($S_1$ and $S_2$) increases the electronic temperature by $\Delta T_e$ and generates a photovoltage $V_{PTE}$. (b) Photovoltage as a function of back ($V_{BG}$) and top ($V_{TG}$) gate forming a p–n junction. The six-fold change of polarity of $V_{PTE}$ is characteristic of the PTE effect.[82,179] Grey dashed lines represent lines of high resistance taken from transport measurements, while the white dashed line indicates where the carrier density in the two regions is equal. Panel (a) is adapted with permission from ref. 200 and panel (b) is reproduced from ref. 82 with permission from AAAS.

obtained using the Shockley–Ramo theorem[201] or a general Onsager reciprocity approach,[202] which account for the geometrical factors (device dimensions, laser spot size, inhomogeneity) that govern the photoresponse that is measured between electrodes.

The PTE effect was first observed at interfaces of single layer and bilayer graphene[203] and graphene p–n junctions.[82] By varying $E_F$ on both side of the junction, Gabor et al.[82] showed that photovoltage exhibits a six-fold pattern (see Fig. 16b), which is a hallmark of the PTE effect. Indeed, Song et al.[179] predicted that the non-monotonic behaviour of $S(E_F)$ shown in Fig. 15 should lead to multiple sign changes of $V_{PTE}$ (when $S_1 = S_2$ or $E_F \sim 0$ in both regions), in stark contrast with the single sign reversal expected for the photovoltaic effect. The PTE effect has also been shown to dominate the photoresponse of unbiased graphene-metal junctions[114,204,205] and supported-suspended graphene interfaces.[206] We note however that the photovoltaic and photo-bolometric effect can become predominant in biased graphene devices.[206] In section 6.1, we discuss in more details the performance of photodetectors based on the PTE and photo-bolometric effects.

#### 5.2.3 Other thermoelectric effects.
In addition to the (photo-) thermoelectric effect, several other thermoelectric and thermomagnetic processes can take place in graphene, including the Peltier, Thomson and Nernst effects. Both the Peltier and Thomson effect are included in the last term of eqn (1). The Peltier effect, which is the reciprocal effect of the Seebeck effect, refers to the heating or cooling that occurs at junction between regions or materials with different Peltier coefficient $\prod = ST_e$ when an electric current is flowing through it. The Peltier heat rate generated or absorbed at the junction is given by $\dot{Q} = (\prod_2 - \prod_1)I$. This effect has been observed at graphene-metal junctions[20,207] and around geometrical constrictions,[208,209] typically using scanning thermal microscopy. These results, in combination with the record high thermoelectric power factor $\sigma S^2$ measured in clean graphene devices,[196] demonstrate the advantage of using graphene as an active thermoelectric cooler. We also note that according to recent theoretical studies, the Thomson effect becomes stronger in the hydrodynamic regime[199] and can lead to cooling of the Fermi liquid.[210]

Finally, in the presence of an out-of-plane magnetic field $B_z$ (with respect to the 2D graphene plane), hot carriers diffusing under $\Delta T$ are deflected by a Lorentz force, producing a transverse voltage $V_y$. This thermomagnetic effect analogous to the Hall effect is called the Nernst effect and is quantified by the Nernst coefficient

$$N = -\frac{\nabla V_y}{B_z \nabla_x T_e} = \frac{S_{xy}}{B_z} \quad (30)$$

where $S_{xy}$ is the transverse component of the thermopower. The Nernst effect has been measured in graphene using microfabricated heaters and was found to agree with the generalized Mott relation, except near the charge neutrality point.[18,19,191] Laser-induced electron heating was also used to probe the "photo-Nernst" effect. This effect gives rise to a photocurrent located along the free edges of the device and which exhibits a peak at the charge neutrality point.[211] A large photo-Nernst current has been observed in graphene/hBN superlattices due the enhanced Nernst coefficient of the Moiré minibands, demonstrating the collection of multiple hot carriers.[44]

### 5.3 Interlayer transport

Recent advances in the layer-by-layer assembly of 2D materials has led to the birth of a very active field of research on the so-called van der Waals heterostructures (vdWH).[212–214] Interfacing graphene with other bulk or 2D materials has emerged as a powerful way to engineer material properties of hybrid systems. This is perhaps best exemplified by the recent discovery of anomalous superconductivity in twisted bilayer graphene.[215] These heterostructures also provide the ability to control and exploit the properties of hot carriers in graphene. For instance, as discussed in section 3.2.3, encapsulating graphene using hBN gives rise to a new cooling pathway mediated by hyperbolic phonon polaritons in hBN.[92] In addition to influencing the heat transfer, these heterostructures can enable the transport of carriers in the out-of-plane direction (see Fig. 17a). This interlayer charge transport generally takes

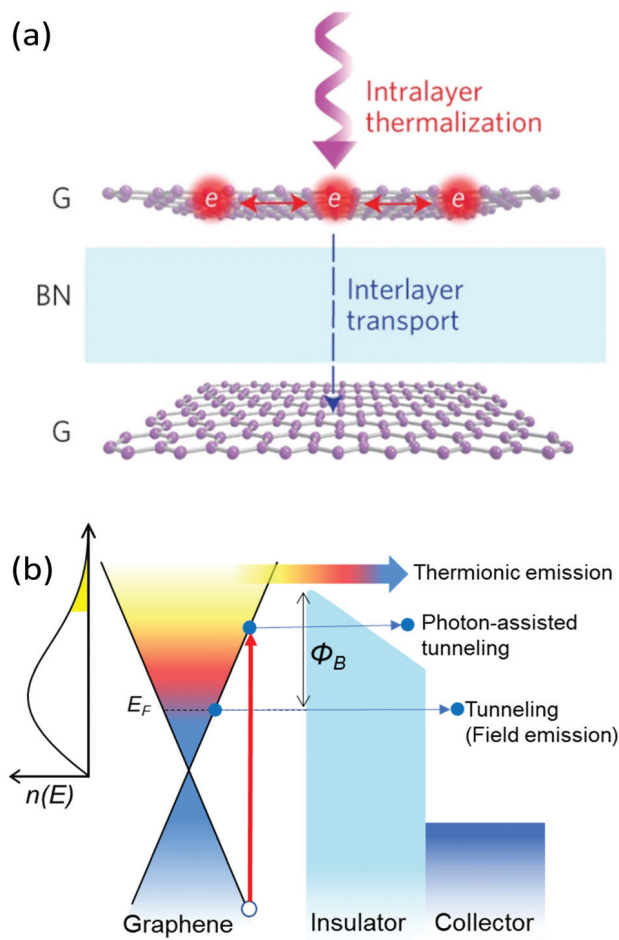

**Fig. 17** (a) Schematic of interlayer transport and intralayer thermalization of photoexcited carriers in a g/hBN/g heterostructure. (b) Band diagram of a graphene heterostructure illustrating various interlayer transport mechanisms. The yellow shaded area in the hot-carrier distribution $n(E)$ represents the carriers with an energy larger than the barrier $\Phi$ that can generate thermionic emission. Panel (a) is reproduced with permission from ref. 216 (Copyright 2016 Springer Nature).

place across g/X heterojunctions, where g denotes graphene and X represents a semiconducting or insulating material, and involves transport mechanisms that are drastically different from intralayer transport. Indeed, inside individual layers, transport is generally diffusive and carried out by the delocalized carriers of the band structure. In the out-of-plane direction, carriers are localized on each layer and their interlayer transport is controlled by the potential energy barrier $\Phi$ formed at the heterojunction. Thus, charge transport between layers typically entails processes such as tunneling and thermionic emission.

**5.3.1 Interlayer transport mechanism.** The theories of thermionic and field emission – also called Richardson's[217,218] and Fowler–Nordheim's[219] laws, respectively – were formulated almost a century ago to describe the emission of electrons from bulk materials into vacuum. As illustrated in Fig. 17b, field emission is produced by the field-induced tunneling of carriers with energy $\varepsilon < \Phi$ across the barrier, which leads to a current that scales with $\exp(-\beta\Phi^{3/2}/F)$, where $\beta$ is a constant on the material and geometry of the barrier and $E$ the electric field across the barrier. Thermionic emission describes the process where carriers in the high-energy tail of the Fermi–Dirac distribution with $\varepsilon < \Phi$ are injected over the barrier, generating a current that scales with $\exp[-\Phi/(k_B T_e)]$. These early theories were later combined into a generalized model[220] and further developed to describe transport in solid-state systems such as on thin films[221] or III–V heterostructures.[222] However, these models usually assume parabolic energy-dispersions relations for carriers in each layer.[223,224]

Recently, several vertical transport models have been proposed that take into account the linear dispersion relation of graphene.[223,225–228] While these models deviate from the classic Richardson's and Fowler–Nordheim laws, the aforementioned exponential behaviours hold true. These transport mechanisms have been observed in g/Si Schottky junctions[229,230] and in several graphene heterostructures where layers of hBN or TMDs are employed as potential barriers.[231–234] In these devices, the crystal lattices of the component layers were not aligned, so the in-plane momentum of the carriers was not conserved during transport. Indeed, the presence of disorder and phonons can lead to relaxation of the momentum conservation condition. Remarkably, by aligning the crystallographic orientation of the layers, momentum and energy conservation leads to resonant tunneling. It was experimentally shown that this can result in a strong negative differential resistance that persists up to room temperature.[235,236] This coherent interlayer transport can be modelled using more advanced techniques such as *ab initio* simulations,[224] and nonequilibrium Green's function[237] and transfer Hamiltonian[238] formalisms.

**5.3.2 Interlayer transport of photoexcited carriers.** These interlayer transport processes also control the interlayer transport of out-of-equilibrium carriers in graphene heterostructures. Rodriguez-Nieva *et al.* predicted that hot carriers created by photon absorption could significantly enhance thermionic emission in graphene heterostructures[239] and Schottky junctions.[240] They showed that in some cases the heat transported by thermionic emission could even dominate other electronic cooling channels, resulting in PTI photodetectors with large responsivities. This photo-thermionic (PTI) effect was observed soon after in g/hBN/g[216] and g/WSe$_2$/g heterostructures.[241] In both studies, the PTI effect is evidenced by the superlinear dependence of the photocurrent on the laser power with photon energies below the barrier bandgap (see Fig. 18a). This is a direct consequence of the exponential thermal activation of carriers over the potential energy barrier $\Phi$. These studies also showed that the PTI signal has a decay time of ~1 ps, which is consistent with the cooling time $\tau_{cool}$ of hot carriers in graphene. PTI emission was also observed in g/Si Schottky junctions and was shown to lead to a significant enhancement of the responsivity as compared to the metal/Si interface.[242]

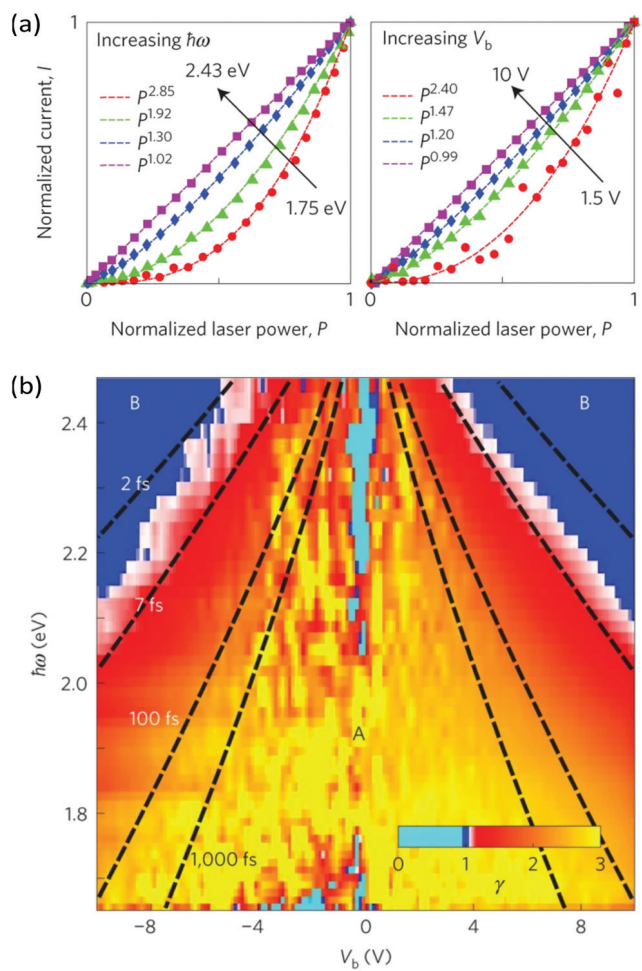

Fig. 18 Competing interlayer transport processes of photoexcited carriers in a g/hBN/g heterostructure. (a) Photocurrent $I$ as a function of excitation laser power $P$, at constant interlayer bias $V_b$ but increasing photon energies $\hbar\omega$ (left panel), and at constant photon energy $\hbar\omega$ but increasing bias (right panel). The data are fitted with a power law $I \sim P^\gamma$. (b) Colour map of $\gamma$ as a function of $V_b$ and $\hbar\omega$. The black dashed lines correspond to different tunnelling times. Regions A ($\gamma > 1$) and B ($\gamma \sim 1$) indicates where the interlayer current is dominated by photo-thermionic emission and photon-assisted tunneling, respectively. This figure is reproduced with permission from ref. 216 (Copyright 2016 Springer Nature).

By increasing the excitation photon energy or the interlayer bias voltage on their g/hBN/g devices, Ma et al.[216] detected a second interlayer transport mechanism which they identified as photon-assisted tunneling, a process related to internal photoemission.[243] In this transport regime, photoexcited carriers in graphene tunnel through the hBN barrier before they thermalize with other carriers. This competition between PTI emission and photon-assisted tunneling provides a way to manipulate electron thermalization in graphene and to estimate the thermalization time of carriers ($\tau_{heat} \sim 10$ fs, see Fig. 18b). Both PTI emission and photon-assisted tunneling were also shown to contribute to the photocurrent generation in g/SiC Schottky junctions. By modelling the dependence of the photocurrent on the laser pulse duration, a tunneling time as short as ∼0.3 fs was extracted, which offers exciting perspectives for applications in ultrafast electronics.[244]

The dynamics of interlayer charge transport in g/TMD junctions have recently been the focus of several optical pump–probe studies.[245–247] While all studies have observed the generation of photocarriers in the TMD layer after sub-bandgap photon energies, its origin is still debated. Yuan et al.[246] suggested that in addition to hot carrier injection, charges in graphene can be photoexcited directly to the adjacent TMD layer via charge transfer states resulting from strong interfacial electronic coupling. In contrast, Chen et al.[247] performed similar measurements and proposed that photoexcited charges in graphene are transferred to the TMD layer after carrier intraband scattering but before electron–hole interband thermalization. In a recent study combining ultrafast terahertz and visible probes, Fu et al.[245] observed a superlinear dependence on the pump fluence for excitation below the TMD bandgap, consistent with PTI emission observed in photocurrent experiments.

**5.3.3 Interlayer thermoelectric transport.** Under constant thermal excitation, the interplay between heat flow and carrier transport across the g/X interface determines the out-of-plane thermoelectric properties of the junction. In analogy to the in-plane thermoelectric effect (see section 5.2.1), it is possible to define an interfacial Seebeck coefficient relating the voltage induced by a temperature difference across the junction. In this case, this thermoelectric voltage is due to thermionic emission of hot carriers over a thin barrier rather than bulk diffusive transport. Using electrical heating, Seebeck coefficients of ∼−220 and −70 μV K$^{-1}$ have been measured for g/hBN interfaces[248,249] and Au/g/WSe$_2$/g/Au heterostructures,[250] respectively. Inspired by the concept of multilayer thermionic devices,[251,252] several theoretical studies have proposed vdW heterostructures for thermionic energy conversion,[228,253–256] showing that they could provide better or comparable power generation and refrigeration efficiency than traditional bulk thermoelectric devices. Thermionic energy converters based on g/vacuum interfaces have also been investigated theoretically and electronic conversion efficiency of 9.8% have been measured,[257] demonstrating the potential of graphene as a hot-carrier emitter.

## 6 Applications

The exceptional properties of graphene, such as its broadband absorption and ultrafast photoresponse, have led to several promising applications in photonics and optoelectronics. Indeed, graphene-based devices display many desirable features, such as high-speed operation, wide spectral response, gate tunability and compatibility with silicon photonics and CMOS platforms. Here, we focus on photonic and optoelectronic devices where hot-carrier effects play a central role, in particular photodetectors, nonlinear optical devices and light emitters.

### 6.1 Photodetectors

Over the past decade, significant efforts have been dedicated to the development of graphene-based photodetectors, as illustrated by the many review articles published on this topic.[12,13,258–262] Graphene photodetectors can convert absorbed photons into an electrical signal through various mechanisms, including mainly the photo-thermoelectric (PTE), photo-bolometric (PB), and photovoltaic (PV) effects. While the latter is based on photoexcited carriers, PTE and PB mechanisms are driven by the excess carrier temperature $\Delta T_e$. As discussed in section 5.2.2, PTE photodetectors produce a photovoltage (or photocurrent) when hot carriers are generated at the junction between regions with different Seebeck coefficients. A hot-carrier bolometer detects incident radiation by measuring the light-induced temperature increase of the electronic bath.[263] This is typically achieved by measuring changes in the temperature-dependent resistance ($dR/dT$) of the device or by measuring thermal noise.

Both PTE and PB detectors take advantage of properties of hot carriers in graphene. As explained in section 3.1, light-induced carrier heating in graphene is efficient because the electron–phonon coupling is weak compared to the strong electron–electron interactions, possibly even leading to the generation of multiple hot carriers per incident photon. Due to the record small heat capacity of graphene, this photo-generated heat gives rise to a substantial increase in $T_e$, which in turns generates a large PB and PTE response. Finally, carrier cooling occurs on a picosecond timescale, enabling high-speed photodetection. These unique thermal properties, combined with the broadband photon absorption of graphene, make graphene-based photodetectors highly promising for many applications. Below we concentrate on long-wavelength and ultrafast photodetectors whose performances are often on par with, or even better than, the best commercially available devices, while offering specific benefits.

**6.1.1 Long-wavelength photodetectors.** The detection of long-wavelength photons is essential for a wide variety of appli-cations. For instance, microwave photodetectors are employed in sensitive applications ranging from radioastronomy to superconducting quantum computing, while mid- to far-infrared (IR) detectors are important for biomedical diagnostics, quality testing, thermal imaging, environmental monitoring, security and free-space optical communications. Most of these applications require long-wavelength detectors that operate at room temperature and exhibit a broad spectral response, fast photoresponse (*i.e.* large bandwidth) and high sensitivity. The latter is characterized by the responsivity $R$, defined as the electrical output of the device per optical power input, and noise-equivalent power (NEP) of the detector, which specifies the minimum incident optical power required to achieve a unitary signal-to-noise ratio over a bandwidth of 1 Hz. Several long-wavelength photodetectors that utilize hot carriers in graphene have been demonstrated during the past decade. While these detectors directly benefit from the broadband and ultrafast photoresponse of graphene, their intrinsic sensitivity is limited by several factors, in particular the relatively low absorption of graphene.

One general approach to increase the sensitivity of graphene photodetectors is to optimize the photodetection mechanism which converts the excess carrier temperature $\Delta T_e$ into an electrical signal. In PTE detectors, this conversion is governed by the Seebeck coefficient $S$, which can be maximized by tuning the chemical potential close to Dirac point but outside the electron–hole puddle regime (see Fig. 15. For this reason, clean and electrically-tunable graphene p–n junctions are often used to increase the responsivity of PTE detectors.[55,82,266,267] In the case of PB devices, the relatively weak $dR/dT$ of graphene (<1%/K)[268] limits the sensitivity of photodetectors based on the readout of the electrical resistance. One way to solve this issue is to artificially increase $dR/dT$, for instance, by nanostructuring graphene[269] or creating a bandgap in bilayer graphene.[270] Another strategy is to employ a different scheme to read out $\Delta T_e$, for example by measuring the Johnson noise of graphene[104,271] or the switching current of graphene-based Josephson junctions.[265]

A second common strategy to increase the sensitivity of both PTE and PB photodetectors is to increase $\Delta T_e$ by enhancing the interaction of graphene with the incident light. This can be accomplished by integrating photonic structures, such as optical cavities, waveguides and photonic crystals, with the photodetector.[272] For THz and mid-IR photodetectors, antennas with various geometries (see Fig. 19a) have been used to compensate for the mismatch between the large area of the incoming radiation and the small photoactive area of the detector.[55,264,273–277] At GHz frequencies, microwave resonators (see Fig. 19b) have been used to couple light with 99% efficiency to graphene bolometers.[32,265] Light–matter interactions can also be enhanced by taking advantage of the polaritonic resonances of graphene or neighbouring materials,[272] such as plasmon-polariton in graphene an nearby metal nanostructures, or hyperbolic phonon-polaritons in hBN. The quasiparticles, which are at the center of a rich field of research,[278] can tightly confine light in the photo-

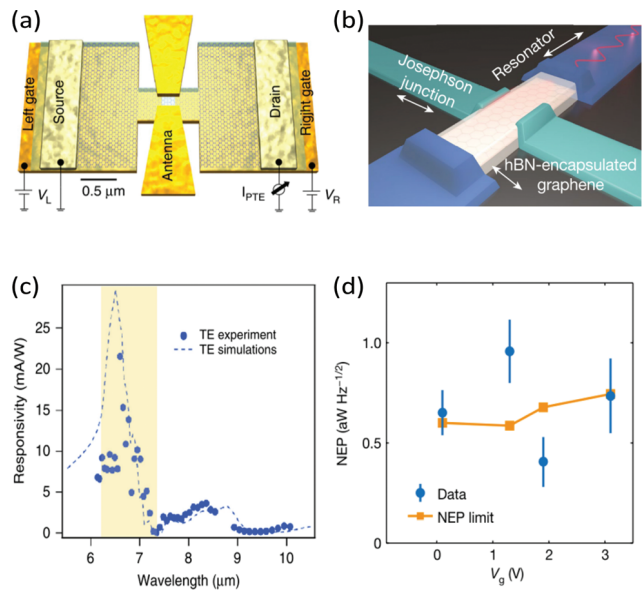

Fig. 19 (a) Schematic of a mid-IR photodetector consisting of a hBN-encapsulated, H-shaped graphene p–n junction with a bow-tie antenna. (b) Schematic of a graphene-based Josephson junction microwave bolometer. The bolometer is embedded in a half-wave resonator to allow DC readout (green) of the Josephson junction and microwave excitation (blue). (c) Measured (dots) and calculated (dashed line) responsivity spectrum of the photodetector shown in (a) for TE-polarization. The yellow shaded region corresponds to the Reststrahlen band of hBN where light absorption in graphene is enhanced due to the propagation hyperbolic phonon-polaritons. (d) NEP measured (blue dots) using the device shown in (b) and the thermal fluctuation limit of the NEP (orange). Panels (a) and (c) are reproduced from ref. 264 and panels (b) and (d) are reproduced with permission from ref. 265 (Copyright 2020 Springer Nature).

detector, resulting in an increased $\Delta T_e$. This approach has been shown to enhance the response of many hot-carrier photodetectors, from the visible to THz spectral range.[279–282] Finally, another avenue to increase $\Delta T_e$ is to reduce the heat transfer coefficient of hot carriers in graphene. For example, this can be accomplished by operating the photodetectors at low temperature[181] or by suspending the graphene layer.[206] However, while this increases the responsivity of the detector, the longer cooling time leads to a longer photoresponse time.

Several of these sensitivity-enhancement strategies have been combined, in order to create high-performance long-wavelength photodetectors that operate at room-temperature. In the mid-IR region, where light absorption in graphene is particularly low due to Pauli blocking, Castilla et al.[264] reported a PTE photodetectors based on clean a hBN-encapsulated graphene p–n junction with a responsivity of ∼27 mA W$^{-1}$ (see Fig. 19c), an NEP of ∼80 pW Hz$^{-0.5}$ and a fast rise time 17 ns. They achieved this by designing a plasmonic antenna efficiently coupled to hyperbolic phonon-polaritons in hBN, which in turn concentrate mid-IR light onto the p–n junction. Safaei et al.[280] demonstrated a plasmon-assisted PTE photodetector using partially nanopatterned graphene, which exhibits a NEP of only ∼7 pW Hz$^{-0.5}$ with a response time of

~100 ns. In the THz range, PTE photodetectors made of hBN-encapsulated graphene coupled to metallic antennas have displayed a very low NEP (~80 pW $Hz^{-0.5}$) and very fast response (~3 ns), in particular compared with other THz detectors at room temperature.[55,283] Finally, two recent studies[32,265] on microwave bolometers based on superconductor–graphene–superconductor junctions reported an NEP below 1 aW $Hz^{-0.5}$ (see Fig. 19d), down to 30 zW $Hz^{-0.5}$, and a response time of 200 ns, when operating in the mK temperature range. All these performance parameters are comparable or superior to other existing long-wavelength detectors, demonstrating the potential of hot-carrier photodetectors for a wide range of applications.

*High-speed telecom photodetectors.* Graphene photodetectors also hold great potential for high-speed photodetection at telecom wavelengths, which is the key process in optical receivers. The latter are a critical part of optical communication systems as they often determine the overall system performance.[284] While most dominant technologies rely on the PV effect in semiconductors like Ge, high-speed graphene photodetectors exploit the ultrafast dynamics of photoexcited carriers in graphene to rapidly convert an optical signal into an electrical one. In the case of PTE and PB photodetectors, an electrical signal can be generated in less than 50 fs (ref. 53) as photoexcited carriers transfer their energy to the electronic bath, leading to the formation of a hot carrier distribution. When the light is turned off, hot carriers cool back to the lattice temperature with a characteristic time $\tau_{cool}$ ~ 1–2 ps. This process determines the intrinsic bandwidth of the PTE and PB detectors. Early studies revealed that the intrinsic bandwidth of graphene photodetectors can exceed 260 GHz,[112,113,285] indicating their potential for data communication applications. We note, however, that the bandwidth of actual devices can be limited by extrinsic factors, including the RC time constant of the device or the bandwidth of the amplifier.

Several chip-integrated graphene-based photodetectors similar to the one shown in Fig. 20a have been reported during the past few years.[242,286–297] In these devices, light in the waveguide is evanescently coupled to the graphene layer. Compared to free-space illumination, the long interaction length between the guided mode and graphene increases the optical absorption well above the usual value of a few percent. Almost 100% optical absorption can be achieved with an interaction length >40 μm. PB photodetectors with a bandwidth larger than 110 GHz and a responsivity of up to ~0.5 A $W^{-1}$ have been demonstrated.[295,296] However, since these photodetectors operate with a bias voltage, they suffer from a large (>100 μA) dark current, resulting in high shot noise and increased power consumption.

In contrast, PTE detectors operate at zero-bias in either current or voltage mode, while maintaining a large photodetection bandwidth and responsivity. Like most conventional photodetectors, the current mode requires a transimpedance amplifier, whereas the voltage configuration can operate using a simpler and less costly voltage amplifier. Responsivities of

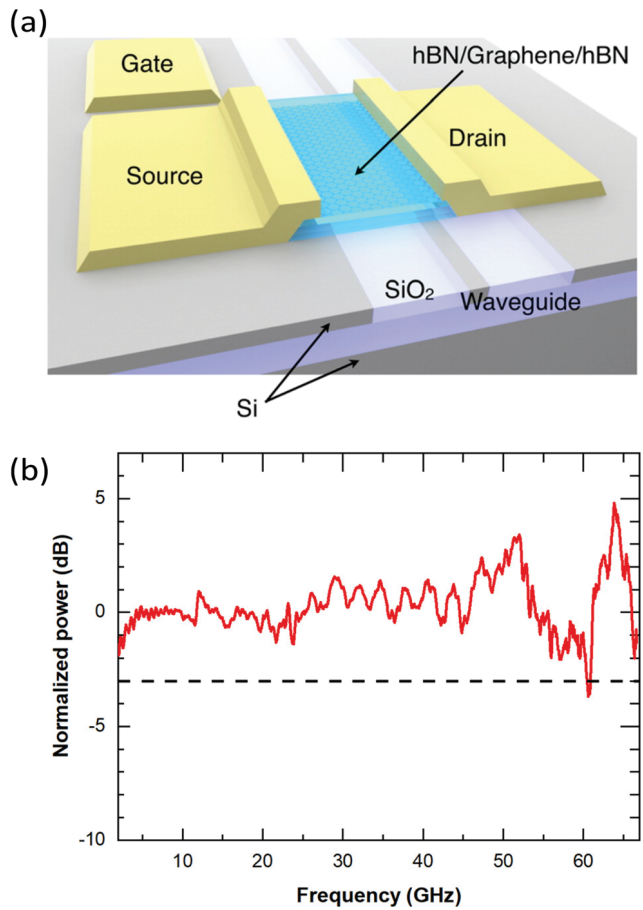

Fig. 20 (a) Schematic of a high-speed photodetector consisting of a hBN-encapsulated graphene layer evanescently coupled to light in the buried silicon waveguide. (b) Spectral response of a PTE photodetector measured up to 67 GHz. The dashed line indicates the 3 dB drop of the photoresponse. No roll off is observed in the in the frequency range of the measurement. Panel (a) is reproduced with permission from ref. 286 (Copyright 2015 American Chemical Society) and panel (b) is reproduced with permission from ref. 287 (Copyright 2020 American Chemical Society).

up to ~0.4 A $W^{-1}$ and 12 V $W^{-1}$ have been reported for PTE photodetectors. These devices typically display a bandwidth larger than 40 GHz (ref. 286, 289, 290 and 292) and the highest to date surpasses 67 GHz (setup-limited, see Fig. 20b).[287] Finally, we note that graphene heterostructures, which are based on interlayer transport mechanisms (see section 5.3), are also promising for high-speed photodetection as they can potentially improve the responsivity while maintaining low dark current and high bandwidth.[242,294] Ultimately, improving the large-scale fabrication[298] and design[13] of graphene-based photodetectors could open the door to faster and more power-efficient receivers in tele- and datacom modules.

### 6.2 Nonlinear optical devices

Owing to its strong and gate-tuneable nonlinear optical response[148,149] and its ease of integration in photonic devices such as fibers,[153] waveguides[152] and micro-resonators,[300,301] graphene has been widely explored for applications in nonlinear optics. Notable examples are saturable absorbers and other optical modulators, frequency converters and sensors.[15,155,259,302,303]

**6.2.1 All-optical modulators.** Optical modulators are fundamental building blocks in a large variety of modern technological applications: phase, frequency, amplitude and polarization modulators are widely used in fibre optic communication systems, ultrafast spectroscopy, metrology, active Q-switching or mode-locking of lasers and quantum information. The scope of integrated optical modulators is to encode information into light with faster response time and lower power consumption compared to electrical interconnects. Graphene can fulfill all these requirements and, in addition, can be easily integrated with standard photonic platforms such as waveguides, microresonators and fibres.[303] Both amplitude[135,300] and phase[304] modulation at tens of GHz have been achieved with graphene-integrated devices based on electro-optic modulation. However, the speed of electro-optic modulators is limited by the capacitance of the device.[13,300] All-optical modulation overcomes this limitation and thus fully exploits the potential of graphene: the ultimate limit for the operation speed in all-optical modulators is given by the cooling dynamics of hot electrons since the absorption and/or phase modulation are defined by Pauli blocking or refractive index changes induced by photoexcited carriers. Thus, speeds well above 100 GHz are possible.

A notable example of all-optical modulation is saturable absorption (see section 4.1.3), a nonlinear and non-parametric third-order process that has been extensively used for passive mode-locking of ultrafast lasers (Fig. 21a).[155,303] Graphene SAs have been used to mode-lock both solid-state[305–307] and fibre lasers[154,308–312] at 800 nm,[307] 1 μm,[311] 1.5 μm,[154,308] 2 μm (ref. 306 and 311) wavelengths and for passive synchronization of separate laser cavities operating at different wavelengths.[313] Saturable absorption based on $\sigma_{intra}$ (see section 4) is possible also at THz frequencies.[173,314] Graphene SAs have enabled the generation of ultrashort pulses with <70 fs duration[307] and lasers with GHz repetition rate.[301,310] Several other approaches have been implemented for the realization of graphene-based all-optical amplitude modulators. With a graphene-clad microfiber (Fig. 21b) Li et al.[299] demonstrated all-optical modulation with 38% modulation depth and ∼2 ps switching time while Liu et al.[315] showed CW all-optical modulation with modulation depth of 5 dB and 13 dB using single- and bi-layer graphene respectively. An all-optical modulator based on a Mach–Zehnder interferometer reached a modulation depth >50% while maintaining an optical transmission through the device of 19%.[316] All-optical modulation can be obtained also with graphene transferred on photonic crystal cavities owing to a combination of light-induced resonance-tuning and SA.[317] Finally, Ono et al.[318] showed ultrafast all-optical modulation with modulation speed of 260 fs and extinction ratio of 3.5 dB using only 3.5 fJ of switching energy in a graphene-loaded deep-subwavelength plasmonic waveguide.

**6.2.2. Frequency converters.** Similar to absorption modulation, in graphene also frequency conversion can be modulated by electric fields, as demonstrated in the case of THG[148] and FWM[149,152] (see section 4.1.3). Recently, ultrafast all-optical modulation with a response time of ∼2.5 ps, thus limited by the hot-electron cooling time, has also been observed.[319] Frequency conversion in the THz range is another application which, through the thermodynamic nonlinearity mechanism, is fully enabled by hot electrons. The attractiveness of graphene in this field comes from its small footprint and high efficiency, as shown by observation of the record value of 1% conversion efficiency obtained with a grating-graphene metamaterial using a moderate field strength of ∼30 kV cm$^{-1}$ (ref. 174) (Fig. 21c).

## 6.3 Light emitting devices

In addition to detecting and modulating light, graphene can also serve as a broadband and ultrafast thermal light emitter. The superior mechanical, thermal and electronic properties of graphene allow it to sustain very high current densities (∼4 × 10$^8$ A cm$^{-2}$),[320] enabling its incandescence via Joule heating. Early studies[25,78] showed that the emission spectra of electri-cally biased graphene devices agree well with Planck's law for the spectral radiance $\mathcal{U}$ of a grey body,

$$\mathcal{U}(\hbar\omega, T_e) = \Xi \frac{2(\hbar\omega)^3}{h^2 c^2} \frac{1}{\exp\left(\frac{\hbar\omega}{k_B T_e}\right) - 1}, \quad (31)$$

where $\hbar\omega$ is the photon energy, $c$ is the speed of light and $\Xi$ is the emissivity of the grey body. An emissivity value of $\sim$2.3% was measured,[25,78] in agreement with Kirchhoff's law stating that the emissivity and absorption are equal. Electronic temperatures $T_e$ on the order of 1500 K were measured using this technique, corresponding to a peak emission in the infrared (see Fig. 22a). Raman spectroscopy measurements further revealed that optical phonons are in thermal equilibrium with the electronic bath, while low-energy acoustic phonons are at a significantly lower temperature. The low thermal radiation efficiency of these early devices ($\sim$10$^{-6}$) was attributed to the fact that Joule heat is mainly dissipated into the metallic contacts and through the SiO$_2$ substrate via surface polar phonon scattering.

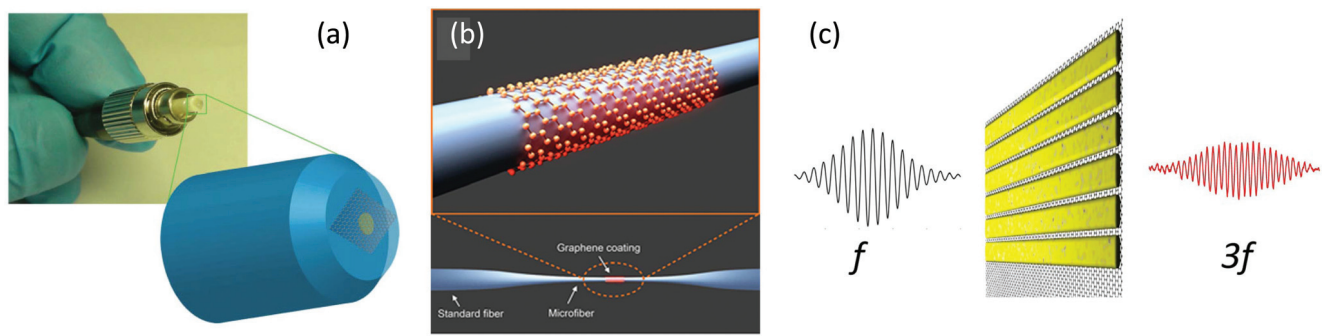

**Fig. 21** Examples of graphene-based nonlinear devices enabled by hot electrons: (a) graphene based SA for mode-locking of fiber lasers. (b) Graphene-clad microfiber all-optical modulator. (c) Grating-graphene metamaterial for THz frequency conversion. Panels (a), (b) and (c) are adapted with permission from ref. 154 (Copyright 2009 Wiley), ref. 299 (Copyright 2014 American Chemical Society) and ref. 174 (Copyright 2020 American Chemical Society) respectively.

Several approaches have been investigated to improve the emission efficiency of graphene emitters. To reduce vertical heat dissipation, Kim et al.[26] fabricated suspended graphene emitters (see Fig. 22b) and demonstrated high thermal emission in the visible range. In these devices, hot carriers with temperatures up to nearly 3000 K are localized at the center the graphene layer electronic, resulting in more efficient ($\sim$4 × 10$^{-3}$) and brighter emission. However, suspended devices must operate in vacuum since graphene quickly oxidizes at high temperature. hBN encapsulation was found to provide excellent protection for graphene even at temperatures above 2000 K.[320–324] These emitters have a measured efficiency of $\sim$1.6 × 10$^{-5}$ and an estimated lifetime exceeding 4 years.[320,321] These results were attributed in part to the efficient cooling of the hot carriers into the hBN mediated by coupling to hybrid polaritonic modes (see section 3.2.3). Finally, another avenue to enhance Joule heating and thermal radiation in graphene is to confine the current flow through a narrow constriction.[324]

Furthermore, the broad spectral emission of graphene emitters can be easily tailored to meet the needs of specific applications. This can be achieved by integrating the graphene emitter in a sub-wavelength optical structure which modifies the electromagnetic local density of state. Different structures have been demonstrated, including photonic microcavities[26,320,321,324,325] and photonic crystals.[323] The latter enabled the strongest modulation of the black-body radiation spectrum in the near-infrared region, which is relevant for optical communications.

Another important advantage of graphene emitters is their high-speed modulation rate, which is facilitated by the ultrafast dynamics of hot carriers. Indeed, recent studies[28,320,323] reported a temporal response of $\sim$100 ps (see Fig. 22c) for various device architectures, corresponding to a modulation bandwidth of $\sim$10 GHz. This exceptional speed, several orders of magnitude faster than conventional thermal emitters, was attributed to the rapid cooling of hot carriers, mainly due to remote heat to the hBN or SiO$_2$ substrates. Real-time optical communications at 50 Mbps was demonstrated by coupling directly the emitter with an optical fiber.[28] These results demonstrate the potential of graphene emitters as chip-integrated light sources for telecom and datacom applications.

## 7 Discussion and outlook

In this section we first discuss what sets apart the properties of hot carriers in graphene, compared to the properties of hot carriers in common metals. We then provide a brief compari-

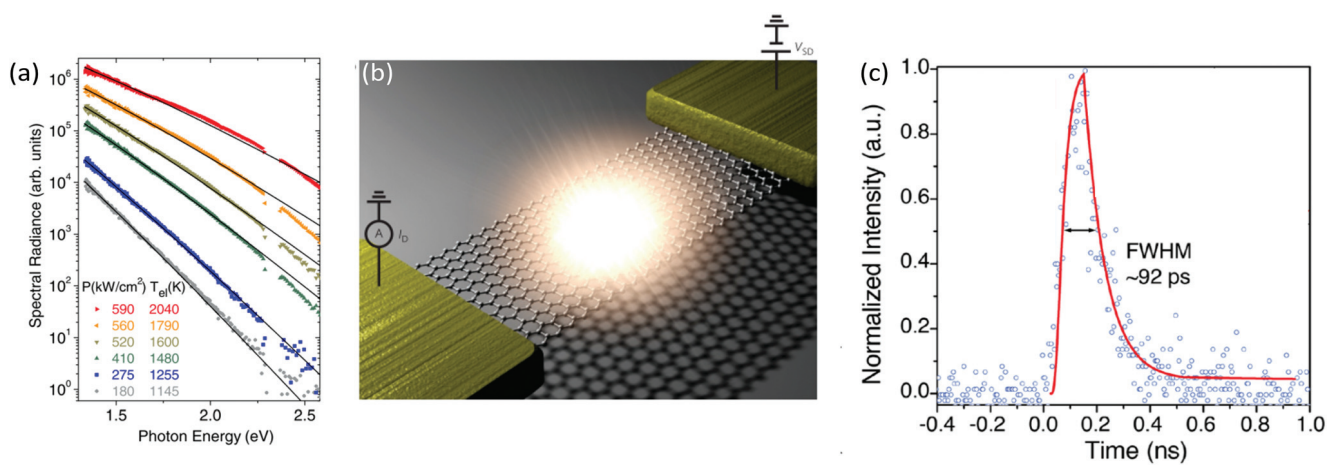

**Fig. 22** (a) Spectral radiance of graphene under electrical bias measured as a function of the electrical power density $P$ dissipated in the channel (dots). The measurements are fitted to Planck's law (eqn (31), solid lines) with the indicated $T_e$. (b) Schematic of a suspended graphene. Under electrical bias, light is emitted mainly from the center of the graphene sheet. (c) Generation of short ($\sim$92 ps) light pulses from a graphene light emitter excited by a 80 ps electrical pulse, which translates into to a bandwidth of 10 GHz. Panel (a) is reproduced with permission from ref. 78 (Copyright 2010 American Physical Society). Panel (b) is reproduced with permission from ref. 26 (Copyright2015 Springer Nature). Panel (c) is reproduced with permission from ref. 320 (Copyright 2018 American Chemical Society).

son with hot-carrier phenomena in other material systems with Dirac electrons. Finally, we will mention some emerging potential applications of hot Dirac carriers.

### 7.1 Graphene vs. common metals

As we have shown throughout this review, hot carriers play an important role in the electrical and optical response of graphene. We note that semiconducting two-dimensional material systems can also exhibit important physical effects related to hot carriers. Nevertheless, the vanishing band gap, linear energy–momentum dispersion relation, and strong carrier–carrier interactions of the Dirac fermions in graphene set graphene apart. One important consequence of these properties is graphene's small electronic heat capacity $C_e$ and heat transfer coefficient $\Gamma_{cool}$, which mainly stem from graphene's low charge carrier density. To emphasize this point, it is instructive to examine and compare the behaviour of hot carriers in other systems. Hot carriers have been studied in metals, in particular gold, for over 50 years.[11,326–335] At room temperature, the electron–phonon coupling in gold leads to a $\Gamma_{cool}$ of $\sim 2.2 \times 10^{16}$ W m$^{-3}$ K$^{-1}$.[336] Even when considering an hypothetical layer of gold as thin as graphene, $\Gamma_{cool,2D} \sim 7$ MW m$^{-2}$ K$^{-1}$ is two orders of magnitude larger than that of graphene in most conditions. For instance, for hBN-encapsulated graphene at room temperature, the estimated $\Gamma_{cool}$ mediated by hyperbolic phonons is <0.1 MW m$^{-2}$ K$^{-1}$ for $|\mu| < 0.2$ eV.[92] Hence, under steady state conditions (see eqn (3)) and for a given input power, the excess electronic temperature $\Delta T_e$ in gold is $\sim$100 times smaller than in graphene. For that reason, hot carrier effects in gold are typically observed in transport measurements only at low temperature (<100 mK), where the electron–phonon coupling is drastically reduced.[328,332] Alternatively, hot carriers can also be generated by short intense laser pulse. In this case, $T_e$ is determined by the heat capacity coefficient $\gamma$ (see eqn (6)), which in the case of gold is 68 J m$^{-3}$ K$^{-2}$. Here again, the equivalent 2D value ($\gamma_{2D} \sim 2 \times 10^{-8}$ J m$^{-2}$ K$^{-2}$) is roughly two orders of magnitude larger than that of graphene (<$8 \times 10^{-10}$ J m$^{-2}$ K$^{-2}$ for $|\mu| < 0.2$ eV, see eqn (4)), a difference which is mainly due to the large charge carrier density (or Fermi energy) in gold, compared to graphene. This implies that for the same input fluence, the initial $T_{e,peak}$ in graphene can be significantly larger than in gold. Interestingly, these hot carriers will cool down with a similar timescale $\tau_{cool} = C_e/\Gamma_{cool} \sim$ 1–2 ps.[331] We also note that the carrier heating time in metals ($\tau_{heat} \sim$ 350–500 fs)[331,334] is significantly longer than in graphene (<100 fs),[37,45] which leads to higher heating efficiency in the latter. All these differences help explain the preponderance of hot carrier effects in graphene as compared to conventional metals.

### 7.2 Other Dirac-electron systems

Besides graphene, there are several other material systems with Dirac electrons, characterized by a linear energy–momentum dispersion relation. It can be expected that similar hot-carrier effects as in graphene play a role in these materials. We will discuss several quantum materials, starting with topological insulators. These are material systems that are semiconducting in the bulk, whereas on the surface (for 3D systems) or edge (for 2D systems), protected conducting states emerge with nearly linear dispersion (see the review of ref. 337). OPTP experiments on $Bi_2Se_3$[338] showed the occurrence of negative photoconductivity (a decrease in THz conductivity after photoexcitation). This is similar to OPTP experiments on graphene, where negative photoconductivity was interpreted as a signature of hot carriers.[34] Time-resolved ARPES measurements with visible pump light performed on $Bi_2Se_3$[339,340] and $Bi_2Te_3$[341] discussed the photo-induced carrier thermalization and cooling dynamics. In contrast with graphene, in these topological-insulator systems, the presence of bulk bands leads to complex dynamics involving an interplay between bulk and surface charges. There have been several attempts to eliminate the effects of bulk charges, for example by studying topological insulators excited by THz and mid-infrared light, such that no interband absorption takes place. For example, Luo et al. recently used this approach to demonstrate faster dynamics for hot surface-state carriers compared to hot bulk carriers, for $Bi_2Se_3$ at 5 K.[324] At room temperature, the experimentally observed dynamics became similar, which is due to efficient surface–bulk coupling, as explained in ref. 339. As a result of the surface–bulk interactions, the precise dynamics of hot Dirac carriers in topological insulators are currently still under debate.

Another quantum material system that hosts Dirac electrons is the 3D Dirac semimetal $Cd_3As_2$.[342] Optical pump–probe measurements showed clear signatures that are consistent with the formation of a hot-carrier distribution on a timescale of $\sim$400 fs.[343] Cooling was observed to be slower than in graphene. An OPTP study showed fluence- and temperature-dependent cooling dynamics with two time constants – one on the order of a few ps, and one that is up to $\sim$8 ps.[344] These dynamics were interpreted within a cooling model that has also been used for graphene: initial fast decay by optical phonon emission, followed by "anharmonic" phonon decay. Interestingly, in ref. 344 a positive THz photoconductivity was observed, which was also observed by Zhang et al.[160] and Cheng et al.[163] In graphene, a positive photoconductivity was only found close to the Dirac point, for $E_F$ below $\sim$100 meV,[57,137,139] as a result of interband thermalization.[35] The origin of the positive photoconductivity observed for $Cd_3As_2$ is still under debate.

Given these similarities in heating–cooling dynamics of different Dirac electron systems, we could expect that similar technological applications as those discussed in section 6 are possible. Topological insulator materials such as $Bi_2Te_3$ have been known for their large Seebeck coefficients (see for example this review by Xu et al.[345]). Not surprisingly, it has been shown that significant photo-thermoelectric currents can be generated,[346,347] similar to graphene. Furthermore, using the 3D semimetal $Cd_3As_2$, ultrafast, broadband photodetection based on the photo-thermoelectric effect has been demonstrated.[348] Also in the field of nonlinear light conversion, different Dirac materials have generated interesting results.

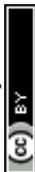

Terahertz nonlinearities, for example, have been shown to originate from Dirac surface states in $Bi_2Se_3$,[349] and, very recently, THz harmonics have been generated from $Cd_3As_2$.[163,350] We should note that in the latter case, the role of the hot carriers is debated, as these results were described using coherent carrier acceleration, rather than the thermodynamic nonlinearity that is responsible for the exceptionally large THz nonlinearity in the case of graphene (see Fig. 14).

# 8 Conclusion

In this review, we have described fundamental properties of hot carriers in graphene, and systems that exploit hot carriers towards useful applications with a large potential to have an important impact on society. Whereas on a fundamental level many physical phenomena are well understood, we conclude that several areas are still subjected to debate, and therefore anticipate the continuous development of novel insights. As an example, we refer to the many competing cooling mechanisms that have been identified in recent years (see section 3). Besides continuing the pursuit of a detailed understanding of the fundamental physics of hot carriers in graphene itself, a highly promising research direction is that of graphene interfaced with other (2D) materials. We have discussed a few examples of such systems, where the substrate of graphene, or adjacent 2D materials, play an important role in determining the overall system properties, in particular related to transport (see section 5). Another highly interesting avenue is that of graphene interfaced with graphene itself. For example, we note that exciting hot-carrier dynamics have been observed for 30°-twisted bilayer graphene,[351] and have been predicted for graphene Moiré superlattices.[352]. Finally, we note that besides the optical and optoelectronic applications that we have focused on in section 6, there are several predictions and preliminary results related to applications, for example hot-carrier transistors[353,354] and hot-carrier enabled energy-harvesting devices.[257,355] Thus, we expect hot Dirac carriers to continue to give rise to interesting physics and promising applications.

# Conflicts of interest

There are no conflicts to declare.

# Acknowledgements

ICN2 was supported by the Severo Ochoa program from Spanish MINECO (grant no. SEV-2017-0706). K. J. T. acknowledges funding from the European Union's Horizon 2020 research and innovation program under grant agreement no. 804349 (ERC StG CUHL), RyC fellowship no. RYC-2017-22330, IAE project PID2019-111673GB-I00, and financial support through the MAINZ Visiting Professorship. G. S. acknowledges funding from the European Union's Horizon 2020 research and innovation program under grant agreement GrapheneCore3 881603, the German Research Foundation DFG (CRC 1375 NOA, project B5) and the Daimler und Benz foundation. M. M. acknowledges support from the Natural Sciences and Engineering Research Council of Canada (PDF-516936-2018) and from the Canada First Research Excellence Fund. A. P. is supported by the European Commission under the EU Horizon 2020 MSCA-RISE-2019 programme (project 873028 HYDROTRONICS). A. P. also acknowledges support of the Leverhulme Trust under the grant RPG-2019-363.

# References

1 P. A. George, J. Strait, J. Dawlaty, S. Shivaraman, M. Chandrashekhar, F. Rana and M. G. Spencer, *Nano Lett.*, 2008, **8**, 4248–4251.
2 D. Sun, Z.-K. Wu, C. Divin, X. Li, C. Berger, W. A. de Heer, P. N. First and T. B. Norris, *Phys. Rev. Lett.*, 2008, **101**, 157402.
3 J. M. Dawlaty, S. Shivaraman, M. Chandrashekhar, F. Rana and M. G. Spencer, *Appl. Phys. Lett.*, 2008, **92**, 042116.
4 I. Meric, M. Y. Han, A. F. Young, B. Ozyilmaz, P. Kim and K. L. Shepard, *Nat. Nanotechnol.*, 2008, **3**, 654–659.
5 S. Butscher, F. Milde, M. Hirtschulz, E. Malić and A. Knorr, *Appl. Phys. Lett.*, 2007, **91**, 203103.
6 E. H. Hwang, B. Y. K. Hu and S. Das Sarma, *Phys. Rev. B: Condens. Matter Mater. Phys.*, 2007, **76**, 1–6.
7 T. Hertel and G. Moos, *Phys. Rev. Lett.*, 2000, **84**, 5002–5005.
8 O. J. Korovyanko, C. X. Sheng, Z. V. Vardeny, A. B. Dalton and R. H. Baughman, *Phys. Rev. Lett.*, 2004, **92**, 4.
9 J. C. Charlier, P. C. Eklund, J. Zhu and A. C. Ferrari, *Top. Appl. Phys.*, 2008, **111**, 673–709.
10 P. Avouris, M. Freitag and V. Perebeinos, *Nat. Photonics*, 2008, **2**, 341–350.
11 M. L. Brongersma, N. J. Halas and P. Nordlander, *Nat. Nanotechnol.*, 2015, **10**, 25–34.
12 F. H. L. Koppens, T. Mueller, P. Avouris, A. C. Ferrari, M. S. Vitiello and M. Polini, *Nat. Nanotechnol.*, 2014, **9**, 780–793.
13 M. Romagnoli, V. Sorianello, M. Midrio, F. H. L. Koppens, C. Huyghebaert, D. Neumaier, P. Galli, W. Templ, A. D'Errico and A. C. Ferrari, *Nat. Rev. Mater.*, 2018, **3**, 392–414.
14 J. You, S. Bongu, Q. Bao and N. Panoiu, *Nanophotonics*, 2018, **8**, 63–97.
15 A. Autere, H. Jussila, Y. Dai, Y. Wang, H. Lipsanen and Z. Sun, *Adv. Mater.*, 2018, **30**, 1705963.
16 J. C. W. Song and L. S. Levitov, *J. Phys.: Condens. Matter*, 2015, **27**, 164201.
17 Z.-Y. Ong and M.-H. Bae, *2D Mater.*, 2019, **6**, 032005.
18 Y. M. Zuev, W. Chang and P. Kim, *Phys. Rev. Lett.*, 2009, **102**, 096807.

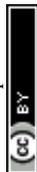

19 P. Wei, W. Bao, Y. Pu, C. N. Lau and J. Shi, *Phys. Rev. Lett.*, 2009, **102**, 166808.
20 K. L. Grosse, M.-H. Bae, F. Lian, E. Pop and W. P. King, *Nat. Nanotechnol.*, 2011, **6**, 287–290.
21 A. C. Betz, S. H. Jhang, E. Pallecchi, R. Ferreira, G. Fève, J.-M. M. Berroir and B. Plaçais, *Nat. Phys.*, 2013, **9**, 109–112.
22 W. Yang, S. Berthou, X. Lu, Q. Wilmart, A. Denis, M. Rosticher, T. Taniguchi, K. Watanabe, G. Fève, J.-M. Berroir, G. Zhang, C. Voisin, E. Baudin and B. Plaçais, *Nat. Nanotechnol.*, 2018, **13**, 47–52.
23 M. Freitag, M. Steiner, Y. Martin, V. Perebeinos, Z. Chen, J. C. Tsang and P. Avouris, *Nano Lett.*, 2009, **9**, 1883–1888.
24 D.-H. Chae, B. Krauss, K. von Klitzing and J. H. Smet, *Nano Lett.*, 2010, **10**, 466–471.
25 M. Freitag, H.-Y. Chiu, M. Steiner, V. Perebeinos and P. Avouris, *Nat. Nanotechnol.*, 2010, **5**, 497–501.
26 Y. D. Kim, H. Kim, Y. Cho, J. H. Ryoo, C.-H. Park, P. Kim, Y. S. Kim, S. Lee, Y. Li, S.-N. Park, Y. Shim Yoo, D. Yoon, V. E. Dorgan, E. Pop, T. F. Heinz, J. Hone, S.-H. Chun, H. Cheong, S. W. Lee, M.-H. Bae and Y. D. Park, *Nat. Nanotechnol.*, 2015, **10**, 676–681.
27 N. H. Mahlmeister, L. M. Lawton, I. J. Luxmoore and G. R. Nash, *Appl. Phys. Express*, 2016, **9**, 012105.
28 Y. Miyoshi, Y. Fukazawa, Y. Amasaka, R. Reckmann, T. Yokoi, K. Ishida, K. Kawahara, H. Ago and H. Maki, *Nat. Commun.*, 2018, **9**, 1279.
29 J. K. Viljas and T. T. Heikkilä, *Phys. Rev. B: Condens. Matter Mater. Phys.*, 2010, **81**, 245404.
30 C. H. Lui, K. F. Mak, J. Shan and T. F. Heinz, *Phys. Rev. Lett.*, 2010, **105**, 127404.
31 K. C. Fong, E. E. Wollman, H. Ravi, W. Chen, A. A. Clerk, M. D. Shaw, H. G. Leduc and K. C. Schwab, *Phys. Rev. X*, 2013, **3**, 041008.
32 R. Kokkoniemi, J.-P. Girard, D. Hazra, A. Laitinen, J. Govenius, R. E. Lake, I. Sallinen, V. Vesterinen, M. Partanen, J. Y. Tan, K. W. Chan, K. Y. Tan, P. Hakonen and M. Möttönen, *Nature*, 2020, **586**, 47–51.
33 M. W. Graham, S.-F. F. Shi, D. C. Ralph, J. Park and P. L. McEuen, *Nat. Phys.*, 2013, **9**, 103–108.
34 K. J. Tielrooij, J. C. W. Song, S. A. Jensen, A. Centeno, A. Pesquera, A. Zurutuza Elorza, M. Bonn, L. S. Levitov and F. H. L. Koppens, *Nat. Phys.*, 2013, **9**, 248–252.
35 A. Tomadin, S. M. Hornett, H. I. Wang, E. M. Alexeev, A. Candini, C. Coletti, D. Turchinovich, M. Kläui, M. Bonn, F. H. L. Koppens, E. Hendry, M. Polini and K.-J. Tielrooij, *Sci. Adv.*, 2018, **4**, eaar5313.
36 A. H. Castro Neto, F. Guinea, N. M. R. Peres, K. S. Novoselov and A. K. Geim, *Rev. Mod. Phys.*, 2009, **81**, 109–162.
37 I. Gierz, J. C. Petersen, M. Mitrano, C. Cacho, I. C. E. Turcu, E. Springate, A. Stöhr, A. Köhler, U. Starke and A. Cavalleri, *Nat. Mater.*, 2013, **12**, 1119–1124.
38 J. C. Johannsen, S. Ulstrup, F. Cilento, A. Crepaldi, M. Zacchigna, C. Cacho, I. C. E. Turcu, E. Springate, F. Fromm, C. Raidel, T. Seyller, F. Parmigiani, M. Grioni and P. Hofmann, *Phys. Rev. Lett.*, 2013, **111**, 027403.
39 G. Rohde, A. Stange, A. Müller, M. Behrendt, L.-P. Oloff, K. Hanff, T. J. Albert, P. Hein, K. Rossnagel and M. Bauer, *Phys. Rev. Lett.*, 2018, **121**, 256401.
40 T. Fang, A. Konar, H. Xing and D. Jena, *Appl. Phys. Lett.*, 2007, **91**, 092109.
41 V. E. Dorgan, A. Behnam, H. J. Conley, K. I. Bolotin and E. Pop, *Nano Lett.*, 2013, **13**, 4581–4586.
42 J. C. W. Song, K. J. Tielrooij, F. H. L. Koppens and L. S. Levitov, *Phys. Rev. B: Condens. Matter Mater. Phys.*, 2013, **87**, 155429.
43 J. C. Johannsen, S. Ulstrup, A. Crepaldi, F. Cilento, M. Zacchigna, J. A. Miwa, C. Cacho, R. T. Chapman, E. Springate, F. Fromm, C. Raidel, T. Seyller, P. D. C. King, F. Parmigiani, M. Grioni and P. Hofmann, *Nano Lett.*, 2015, **15**, 326–331.
44 S. Wu, L. Wang, Y. Lai, W.-Y. Shan, G. Aivazian, X. Zhang, T. Taniguchi, K. Watanabe, D. Xiao, C. Dean, J. Hone, Z. Li and X. Xu, *Sci. Adv.*, 2016, **2**, e1600002.
45 D. Brida, A. Tomadin, C. Manzoni, Y. J. Kim, A. Lombardo, S. Milana, R. R. Nair, K. S. Novoselov, A. C. Ferrari, G. Cerullo and M. Polini, *Nat. Commun.*, 2013, **4**, 1987.
46 F. Kadi, T. Winzer, A. Knorr and E. Malic, *Sci. Rep.*, 2015, **5**, 16841.
47 J. C. König-Otto, M. Mittendorff, T. Winzer, F. Kadi, E. Malic, A. Knorr, C. Berger, W. A. de Heer, A. Pashkin, H. Schneider, M. Helm and S. Winnerl, *Phys. Rev. Lett.*, 2016, **117**, 087401.
48 E. H. Hwang and S. Das Sarma, *Phys. Rev. B: Condens. Matter Mater. Phys.*, 2007, **75**, 205418.
49 A. Tomadin, D. Brida, G. Cerullo, A. C. Ferrari and M. Polini, *Phys. Rev. B: Condens. Matter Mater. Phys.*, 2013, **88**, 035430.
50 T. Winzer, A. Knorr and E. Malic, *Nano Lett.*, 2010, **10**, 4839–4843.
51 T. Plötzing, T. Winzer, E. Malic, D. Neumaier, A. Knorr and H. Kurz, *Nano Lett.*, 2014, **14**, 5371–5375.
52 W. Shockley and H. J. Queisser, *J. Appl. Phys.*, 1961, **32**, 510–519.
53 K. J. Tielrooij, L. Piatkowski, M. Massicotte, A. Woessner, Q. Ma, Y. Lee, K. S. Myhro, C. N. Lau, P. Jarillo-Herrero, N. F. van Hulst and F. H. L. Koppens, *Nat. Nanotechnol.*, 2015, **10**, 437–443.
54 Z. Mics, K.-J. Tielrooij, K. Parvez, S. a. Jensen, I. Ivanov, X. Feng, K. Müllen, M. Bonn and D. Turchinovich, *Nat. Commun.*, 2015, **6**, 7655.
55 S. Castilla, B. Terrés, M. Autore, L. Viti, J. Li, A. Y. Nikitin, I. Vangelidis, K. Watanabe, T. Taniguchi, E. Lidorikis, M. S. Vitiello, R. Hillenbrand, K.-J. Tielrooij and F. H. Koppens, *Nano Lett.*, 2019, **19**, 2765–2773.
56 H. A. Hafez, S. Kovalev, J.-C. Deinert, Z. Mics, B. Green, N. Awari, M. Chen, S. Germanskiy, U. Lehnert, J. Teichert, Z. Wang, K.-J. Tielrooij, Z. Liu, Z. Chen, A. Narita,

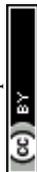

K. Müllen, M. Bonn, M. Gensch and D. Turchinovich, *Nature*, 2018, **561**, 507–511.

57 S. A. Jensen, Z. Mics, I. Ivanov, H. S. Varol, D. Turchinovich, F. H. L. Koppens, M. Bonn and K. J. Tielrooij, *Nano Lett.*, 2014, **14**, 5839–5845.

58 M. Mittendorff, T. Winzer, E. Malic, A. Knorr, C. Berger, W. A. de Heer, H. Schneider, M. Helm and S. Winnerl, *Nano Lett.*, 2014, **14**, 1504–1507.

59 L. Fritz, J. Schmalian, M. Müller and S. Sachdev, *Phys. Rev. B: Condens. Matter Mater. Phys.*, 2008, **78**, 085416.

60 Q. Ma, C. H. Lui, J. C. W. Song, Y. Lin, J. F. Kong, Y. Cao, T. H. Dinh, N. L. Nair, W. Fang, K. Watanabe, T. Taniguchi, S.-Y. Xu, J. Kong, T. Palacios, N. Gedik, N. M. Gabor and P. Jarillo-Herrero, *Nat. Nanotechnol.*, 2019, **14**, 145–150.

61 P. Gallagher, C.-s. Yang, T. Lyu, F. Tian, R. Kou, H. Zhang, K. Watanabe, T. Taniguchi and F. Wang, *Science*, 2019, **162**, eaat8687.

62 M. Kim, S. G. Xu, A. I. Berdyugin, A. Principi, S. Slizovskiy, N. Xin, P. Kumaravadivel, W. Kuang, M. Hamer, R. Krishna Kumar, R. V. Gorbachev, K. Watanabe, T. Taniguchi, I. V. Grigorieva, V. I. Fal'ko, M. Polini and A. K. Geim, *Nat. Commun.*, 2020, **11**, 2339.

63 P. Plochocka, P. Kossacki, A. Golnik, T. Kazimierczuk, C. Berger, W. A. De Heer and M. Potemski, *Phys. Rev. B: Condens. Matter Mater. Phys.*, 2009, **80**, 1–5.

64 L. D. Landau, E. M. Lifshitz and L. P. Pitaevskij, *Course of theoretical physics. vol. 10: Physical kinetics*, 1981.

65 C. Di Castro and R. Raimondi, *Statistical Mechanics and Applications in Condensed Matter*, 2015.

66 N. W. Ashcroft and N. D. Mermin, *Solid State Physics*, Cornell University, 1976.

67 D. Xiao, M.-C. Chang and Q. Niu, *Rev. Mod. Phys.*, 2010, **82**, 1959–2007.

68 S. Chapman, T. G. Cowling and D. Burnett, *The mathematical theory of non-uniform gases: an account of the kinetic theory of viscosity, thermal conduction and diffusion in gases*, 1990.

69 N. Mounet and N. Marzari, *Phys. Rev. B: Condens. Matter Mater. Phys.*, 2005, **71**, 205214.

70 S. Piscanec, M. Lazzeri, F. Mauri, A. C. Ferrari and J. Robertson, *Phys. Rev. Lett.*, 2004, **93**, 185503.

71 S. Piscanec, M. Lazzeri, J. Robertson, A. C. Ferrari and F. Mauri, *Phys. Rev. B: Condens. Matter Mater. Phys.*, 2007, **75**, 035427.

72 T. Sohier, M. Calandra, C.-H. Park, N. Bonini, N. Marzari and F. Mauri, *Phys. Rev. B: Condens. Matter Mater. Phys.*, 2014, **90**, 125414.

73 F. Rana, P. A. George, J. H. Strait, J. Dawlaty, S. Shivaraman, M. Chandrashekhar and M. G. Spencer, *Phys. Rev. B: Condens. Matter Mater. Phys.*, 2009, **79**, 115447.

74 H. Wang, J. H. Strait, P. A. George, S. Shivaraman, V. B. Shields, M. Chandrashekhar, J. Hwang, F. Rana, M. G. Spencer, C. S. Ruiz-Vargas and J. Park, *Appl. Phys. Lett.*, 2010, **96**, 081917.

75 T. Kampfrath, L. Perfetti, F. Schapper, C. Frischkorn and M. Wolf, *Phys. Rev. Lett.*, 2005, **95**, 187403.

76 M. Breusing, S. Kuehn, T. Winzer, E. Malić, F. Milde, N. Severin, J. P. Rabe, C. Ropers, A. Knorr and T. Elsaesser, *Phys. Rev. B: Condens. Matter Mater. Phys.*, 2011, **83**, 153410.

77 L. M. Malard, K. Fai Mak, A. H. Castro Neto, N. M. R. Peres and T. F. Heinz, *New J. Phys.*, 2013, **15**, 015009.

78 S. Berciaud, M. Y. Han, K. F. Mak, L. E. Brus, P. Kim and T. F. Heinz, *Phys. Rev. Lett.*, 2010, **104**, 227401.

79 K. Kang, D. Abdula, D. G. Cahill and M. Shim, *Phys. Rev. B: Condens. Matter Mater. Phys.*, 2010, **81**, 165405.

80 S. Wu, W.-T. Liu, X. Liang, P. J. Schuck, F. Wang, Y. R. Shen and M. Salmeron, *Nano Lett.*, 2012, **12**, 5495–5499.

81 N. Bonini, M. Lazzeri, N. Marzari and F. Mauri, *Phys. Rev. Lett.*, 2007, **99**, 176802.

82 N. M. Gabor, J. C. W. Song, Q. Ma, N. L. Nair, T. Taychatanapat, K. Watanabe, T. Taniguchi, L. S. Levitov and P. Jarillo-Herrero, *Science*, 2011, **334**, 648–652.

83 M. T. Mihnev, F. Kadi, C. J. Divin, T. Winzer, S. Lee, C.-h. Liu, Z. Zhong, C. Berger, W. A. de Heer, E. Malic, A. Knorr and T. B. Norris, *Nat. Commun.*, 2016, **7**, 11617.

84 E. A. A. Pogna, X. Jia, A. Principi, A. Block, L. Banszerus, J. Zhang, X. Liu, T. Sohier, S. Forti, K. Soundarapandian, B. Terrés, J. D. Mehew, C. Trovatello, C. Coletti, F. H. L. Koppens, M. Bonn, N. van Hulst, M. J. Verstraete, H. Peng, Z. Liu, C. Stampfer, G. Cerullo and K.-J. Tielrooij, 2021, arXiv:2103.03527.

85 J. C. W. Song, M. Y. Reizer and L. S. Levitov, *Phys. Rev. Lett.*, 2012, **109**, 106602.

86 K. Kaasbjerg, K. S. Thygesen and K. W. Jacobsen, *Phys. Rev. B: Condens. Matter Mater. Phys.*, 2012, **85**, 165440.

87 R. Bistritzer and A. H. MacDonald, *Phys. Rev. Lett.*, 2009, **102**, 206410.

88 A. C. Betz, F. Vialla, D. Brunel, C. Voisin, M. Picher, A. Cavanna, A. Madouri, G. Fève, J.-M. Berroir, B. Plaçais and E. Pallecchi, *Phys. Rev. Lett.*, 2012, **109**, 056805.

89 M. W. Graham, S.-F. Shi, Z. Wang, D. C. Ralph, J. Park and P. L. McEuen, *Nano Lett.*, 2013, **13**, 5497–5502.

90 T. V. Alencar, M. G. Silva, L. M. Malard and A. M. de Paula, *Nano Lett.*, 2014, **14**, 5621–5624.

91 A. Principi, M. B. Lundeberg, N. C. H. Hesp, K.-J. Tielrooij, F. H. L. Koppens and M. Polini, *Phys. Rev. Lett.*, 2017, **118**, 126804.

92 K.-J. J. Tielrooij, N. C. H. Hesp, A. Principi, M. B. Lundeberg, E. A. A. Pogna, L. Banszerus, Z. Mics, M. Massicotte, P. Schmidt, D. Davydovskaya, D. G. Purdie, I. Goykhman, G. Soavi, A. Lombardo, K. Watanabe, T. Taniguchi, M. Bonn, D. Turchinovich, C. Stampfer, A. C. Ferrari, G. Cerullo, M. Polini and F. H. L. Koppens, *Nat. Nanotechnol.*, 2018, **13**, 41–46.

93 T. Low, V. Perebeinos, R. Kim, M. Freitag and P. Avouris, *Phys. Rev. B: Condens. Matter Mater. Phys.*, 2012, **86**, 045413.

94 Z. Jacob, J. Y. Kim, G. V. Naik, A. Boltasseva, E. E. Narimanov and V. M. Shalaev, *Appl. Phys. B: Lasers Opt.*, 2010, **100**, 215–218.

95 M. N. Gjerding, R. Petersen, T. G. Pedersen, N. A. Mortensen and K. S. Thygesen, *Nat. Commun.*, 2017, **8**, 320.
96 C. R. Dean, a. F. Young, I. Meric, C. Lee, L. Wang, S. Sorgenfrei, K. Watanabe, T. Taniguchi, P. Kim, K. L. Shepard and J. Hone, *Nat. Nanotechnol.*, 2010, **5**, 722–726.
97 L. Wang, I. Meric, P. Y. Huang, Q. Gao, Y. Gao, H. Tran, T. Taniguchi, K. Watanabe, L. M. Campos, D. a. Muller, J. Guo, P. Kim, J. Hone, K. L. Shepard and C. R. Dean, *Science*, 2013, **342**, 614–617.
98 J. D. Caldwell, A. V. Kretinin, Y. Chen, V. Giannini, M. M. Fogler, Y. Francescato, C. T. Ellis, J. G. Tischler, C. R. Woods, A. J. Giles, M. Hong, K. Watanabe, T. Taniguchi, S. A. Maier and K. S. Novoselov, *Nat. Commun.*, 2014, **5**, 1–9.
99 J. Crossno, J. K. Shi, K. Wang, X. Liu, A. Harzheim, A. Lucas, S. Sachdev, P. Kim, T. Taniguchi, K. Watanabe, T. A. Ohki and K. C. Fong, *Science*, 2016, **351**, 1058–1061.
100 J. M. Hamm, A. F. Page, J. Bravo-Abad, F. J. Garcia-Vidal and O. Hess, *Phys. Rev. B*, 2016, **93**, 041408.
101 L. Kim, S. Kim, P. K. Jha, V. W. Brar and H. A. Atwater, 2020, arXiv:2001.11052.
102 R. Yu, A. Manjavacas and F. J. García de Abajo, *Nat. Commun.*, 2017, **8**, 2.
103 D. Halbertal, M. Ben Shalom, A. Uri, K. Bagani, A. Y. Meltzer, I. Marcus, Y. Myasoedov, J. Birkbeck, L. S. Levitov, A. K. Geim and E. Zeldov, *Science*, 2017, **358**, 1303–1306.
104 K. C. Fong and K. C. Schwab, *Phys. Rev. X*, 2012, **2**, 031006.
105 M. R. Arai, *Appl. Phys. Lett.*, 1983, **42**, 906–908.
106 M. M. Jadidi, R. J. Suess, C. Tan, X. Cai, K. Watanabe, T. Taniguchi, A. B. Sushkov, M. Mittendorff, J. Hone, H. D. Drew, M. S. Fuhrer and T. E. Murphy, *Phys. Rev. Lett.*, 2016, **117**, 257401.
107 A. M. Weiner, *Ultrafast Time-Resolved Spectroscopy*, John Wiley & Sons Ltd, 2009, ch. 9, pp. 422–506.
108 R. Ulbricht, E. Hendry, J. Shan, T. F. Heinz and M. Bonn, *Rev. Mod. Phys.*, 2011, **83**, 543–586.
109 G. Soavi, F. Scotognella, G. Lanzani and G. Cerullo, *Adv. Opt. Mater.*, 2016, **4**, 1670–1688.
110 A. Grupp, A. Budweg, M. P. Fischer, J. Allerbeck, G. Soavi, A. Leitenstorfer and D. Brida, *J. Opt.*, 2018, **20**, 014005.
111 C. L. Smallwood, R. A. Kaindl and A. Lanzara, *EPL*, 2016, **115**, 27001.
112 A. Urich, K. Unterrainer and T. Mueller, *Nano Lett.*, 2011, **11**, 2804–2808.
113 D. Sun, G. Aivazian, A. M. Jones, J. S. Ross, W. Yao, D. Cobden and X. Xu, *Nat. Nanotechnol.*, 2012, **7**, 114–118.
114 K. J. Tielrooij, M. Massicotte, L. Piatkowski, A. Woessner, Q. Ma, P. Jarillo-Herrero, N. F. V. Hulst and F. H. L. Koppens, *J. Phys.: Condens. Matter*, 2015, **27**, 164207.
115 N. M. R. Peres, *Rev. Mod. Phys.*, 2010, **82**, 2673–2700.
116 T. Ando, Y. Zheng and H. Suzuura, *J. Phys. Soc. Jpn.*, 2002, **71**, 1318–1324.
117 S. Ryu, C. Mudry, A. Furusaki and A. W. W. Ludwig, *Phys. Rev. B: Condens. Matter Mater. Phys.*, 2007, **75**, 205344.
118 V. P. Gusynin and S. G. Sharapov, *Phys. Rev. B: Condens. Matter Mater. Phys.*, 2006, **73**, 245411.
119 L. A. Falkovsky and A. A. Varlamov, *Eur. Phys. J. B*, 2007, **56**, 281–284.
120 S. A. Mikhailov and K. Ziegler, *Phys. Rev. Lett.*, 2007, **99**, 016803.
121 L. Yang, J. Deslippe, C.-H. Park, M. L. Cohen and S. G. Louie, *Phys. Rev. Lett.*, 2009, **103**, 186802.
122 S. Mikhailov, *Microelectron. J.*, 2009, **40**, 712–715.
123 S. Das Sarma, S. Adam, E. H. Hwang and E. Rossi, *Rev. Mod. Phys.*, 2011, **83**, 407–470.
124 R. R. Nair, P. Blake, A. N. Grigorenko, K. S. Novoselov, T. J. Booth, T. Stauber, N. M. R. Peres and A. K. Geim, *Science*, 2008, **320**, 1308–1308.
125 K. F. Mak, M. Y. Sfeir, Y. Wu, C. H. Lui, J. A. Misewich and T. F. Heinz, *Phys. Rev. Lett.*, 2008, **101**, 196405.
126 H. Choi, F. Borondics, D. A. Siegel, S. Y. Zhou, M. C. Martin, A. Lanzara and R. A. Kaindl, *Appl. Phys. Lett.*, 2009, **94**, 172102.
127 K. F. Mak, J. Shan and T. F. Heinz, *Phys. Rev. Lett.*, 2011, **106**, 046401.
128 J. Horng, C.-F. Chen, B. Geng, C. Girit, Y. Zhang, Z. Hao, H. A. Bechtel, M. Martin, A. Zettl, M. F. Crommie, Y. R. Shen and F. Wang, *Phys. Rev. B: Condens. Matter Mater. Phys.*, 2011, **83**, 165113.
129 I. Ivanov, M. Bonn, Z. Mics and D. Turchinovich, *EPL*, 2015, **111**, 67001.
130 H. Yan, F. Xia, W. Zhu, M. Freitag, C. Dimitrakopoulos, A. A. Bol, G. Tulevski and P. Avouris, *ACS Nano*, 2011, **5**, 9854–9860.
131 K. F. Mak, L. Ju, F. Wang and T. F. Heinz, *Solid State Commun.*, 2012, **152**, 1341–1349.
132 S. Winnerl, M. Orlita, P. Plochocka, P. Kossacki, M. Potemski, T. Winzer, E. Malic, A. Knorr, M. Sprinkle, C. Berger, W. A. de Heer, H. Schneider and M. Helm, *Phys. Rev. Lett.*, 2011, **107**, 237401.
133 A. B. Kuzmenko, E. van Heumen, F. Carbone and D. van der Marel, *Phys. Rev. Lett.*, 2008, **100**, 117401.
134 C.-F. Chen, C.-H. Park, B. W. Boudouris, J. Horng, B. Geng, C. Girit, A. Zettl, M. F. Crommie, R. A. Segalman, S. G. Louie and F. Wang, *Nature*, 2011, **471**, 617–620.
135 M. Liu, X. Yin, E. Ulin-Avila, B. Geng, T. Zentgraf, L. Ju, F. Wang and X. Zhang, *Nature*, 2011, **474**, 64–67.
136 A. J. Frenzel, C. H. Lui, W. Fang, N. L. Nair, P. K. Herring, P. Jarillo-Herrero, J. Kong and N. Gedik, *Appl. Phys. Lett.*, 2013, **102**, 113111.
137 A. J. Frenzel, C. H. Lui, Y. C. Shin, J. Kong and N. Gedik, *Phys. Rev. Lett.*, 2014, **113**, 056602.
138 G. Jnawali, Y. Rao, H. Yan and T. F. Heinz, *Nano Lett.*, 2013, **13**, 524–530.
139 S.-F. Shi, T.-T. Tang, B. Zeng, L. Ju, Q. Zhou, A. Zettl and F. Wang, *Nano Lett.*, 2014, **14**, 1578–1582.
140 S. Savasta, O. Di Stefano and F. Nori, *Nanophotonics*, 2020, **10**, 465–476.

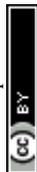

141 H. Rostami and M. Polini, *Phys. Rev. B*, 2016, **93**, 161411.
142 J. L. Cheng, N. Vermeulen and J. E. Sipe, *Sci. Rep.*, 2017, **7**, 43843.
143 A. Marini, J. D. Cox and F. J. García de Abajo, *Phys. Rev. B*, 2017, **95**, 125408.
144 S. A. Mikhailov, *Phys. Rev. B*, 2019, **100**, 115416.
145 N. Yoshikawa, T. Tamaya and K. Tanaka, *Science*, 2017, **356**, 736–738.
146 N. Kumar, J. Kumar, C. Gerstenkorn, R. Wang, H.-Y. Chiu, A. L. Smirl and H. Zhao, *Phys. Rev. B: Condens. Matter Mater. Phys.*, 2013, **87**, 121406.
147 S.-Y. Hong, J. I. Dadap, N. Petrone, P.-C. Yeh, J. Hone and R. M. Osgood, *Phys. Rev. X*, 2013, **3**, 021014.
148 G. Soavi, G. Wang, H. Rostami, D. G. Purdie, D. De Fazio, T. Ma, B. Luo, J. Wang, A. K. Ott, D. Yoon, S. A. Bourelle, J. E. Muench, I. Goykhman, S. Dal Conte, M. Celebrano, A. Tomadin, M. Polini, G. Cerullo and A. C. Ferrari, *Nat. Nanotechnol.*, 2018, **13**, 583–588.
149 T. Jiang, D. Huang, J. Cheng, X. Fan, Z. Zhang, Y. Shan, Y. Yi, Y. Dai, L. Shi, K. Liu, C. Zeng, J. Zi, J. E. Sipe, Y.-R. Shen, W.-T. Liu and S. Wu, *Nat. Photonics*, 2018, **12**, 430–436.
150 G. Soavi, G. Wang, H. Rostami, A. Tomadin, O. Balci, I. Paradisanos, E. A. A. Pogna, G. Cerullo, E. Lidorikis, M. Polini and A. C. Ferrari, *ACS Photonics*, 2019, **6**, 2841–2849.
151 E. Hendry, P. J. Hale, J. Moger, A. K. Savchenko and S. A. Mikhailov, *Phys. Rev. Lett.*, 2010, **105**, 097401.
152 K. Alexander, N. A. Savostianova, S. A. Mikhailov, B. Kuyken and D. Van Thourhout, *ACS Photonics*, 2017, **4**, 3039–3044.
153 N. An, T. Tan, Z. Peng, C. Qin, Z. Yuan, L. Bi, C. Liao, Y. Wang, Y. Rao, G. Soavi and B. Yao, *Nano Lett.*, 2020, **20**, 6473–6480.
154 Q. Bao, H. Zhang, Y. Wang, Z. Ni, Y. Yan, Z. X. Shen, K. P. Loh and D. Y. Tang, *Adv. Funct. Mater.*, 2009, **19**, 3077–3083.
155 A. Martinez and Z. Sun, *Nat. Photonics*, 2013, **7**, 842–845.
156 E. Dremetsika, B. Dlubak, S.-P. Gorza, C. Ciret, M.-B. Martin, S. Hofmann, P. Seneor, D. Dolfi, S. Massar, P. Emplit and P. Kockaert, *Opt. Lett.*, 2016, **41**, 3281.
157 J. J. Dean and H. M. van Driel, *Appl. Phys. Lett.*, 2009, **95**, 261910.
158 A. Y. Bykov, T. V. Murzina, M. G. Rybin and E. D. Obraztsova, *Phys. Rev. B: Condens. Matter Mater. Phys.*, 2012, **85**, 121413.
159 Y. Q. An, F. Nelson, J. U. Lee and A. C. Diebold, *Nano Lett.*, 2013, **13**, 2104–2109.
160 Y. Zhang, D. Huang, Y. Shan, T. Jiang, Z. Zhang, K. Liu, L. Shi, J. Cheng, J. E. Sipe, W.-T. Liu and S. Wu, *Phys. Rev. Lett.*, 2019, **122**, 047401.
161 S. A. Mikhailov, *EPL*, 2007, **79**, 27002.
162 S. A. Mikhailov and K. Ziegler, *J. Phys.: Condens. Matter*, 2008, **20**, 384204.
163 B. Cheng, N. Kanda, T. N. Ikeda, T. Matsuda, P. Xia, T. Schumann, S. Stemmer, J. Itatani, N. P. Armitage and R. Matsunaga, *Phys. Rev. Lett.*, 2020, **124**, 117402.
164 S. A. Mikhailov, *Phys. Rev. B*, 2016, **93**, 085403.
165 J. L. Cheng, N. Vermeulen and J. E. Sipe, *Phys. Rev. B: Condens. Matter Mater. Phys.*, 2015, **91**, 235320.
166 K. Alexander, N. A. Savostianova, S. A. Mikhailov, D. Van Thourhout and B. Kuyken, *ACS Photonics*, 2018, **5**, 4944–4950.
167 A. R. Wright, X. G. Xu, J. C. Cao and C. Zhang, *Appl. Phys. Lett.*, 2009, **95**, 072101.
168 S. Shareef, Y. S. Ang and C. Zhang, *J. Opt. Soc. Am. B*, 2012, **29**, 274.
169 I. Al-Naib, J. E. Sipe and M. M. Dignam, *Phys. Rev. B: Condens. Matter Mater. Phys.*, 2014, **90**, 245423.
170 M. J. Paul, Y. C. Chang, Z. J. Thompson, A. Stickel, J. Wardini, H. Choi, E. D. Minot, B. Hou, J. A. Nees, T. B. Norris and Y.-S. Lee, *New J. Phys.*, 2013, **15**, 085019.
171 P. Bowlan, E. Martinez-Moreno, K. Reimann, T. Elsaesser and M. Woerner, *Phys. Rev. B: Condens. Matter Mater. Phys.*, 2014, **89**, 041408.
172 X. Li, E. A. Barry, J. M. Zavada, M. B. Nardelli and K. W. Kim, *Appl. Phys. Lett.*, 2010, **97**, 082101.
173 H. A. Hafez, S. Kovalev, K. Tielrooij, M. Bonn, M. Gensch and D. Turchinovich, *Adv. Opt. Mater.*, 2020, **8**, 1900771.
174 J.-C. Deinert, D. Alcaraz Iranzo, R. Pérez, X. Jia, H. A. Hafez, I. Ilyakov, N. Awari, M. Chen, M. Bawatna, A. N. Ponomaryov, S. Germanskiy, M. Bonn, F. H. Koppens, D. Turchinovich, M. Gensch, S. Kovalev and K.-J. Tielrooij, *ACS Nano*, 2021, **15**, 1145.
175 Z. Sun, D. N. Basov and M. M. Fogler, *Phys. Rev. B*, 2018, **97**, 075432.
176 A. Block, A. Principi, N. C. H. Hesp, A. W. Cummings, M. Liebel, K. Watanabe, T. Taniguchi, S. Roche, F. H. L. Koppens, N. F. van Hulst and K.-J. Tielrooij, 2020, arXiv:2008.04189 [cond-mat.mes-hall].
177 R. Jago, R. Perea-Causin, S. Brem and E. Malic, *Nanoscale*, 2019, **11**, 10017–10022.
178 B. A. Ruzicka, S. Wang, L. K. Werake, B. Weintrub, K. P. Loh and H. Zhao, *Phys. Rev. B: Condens. Matter Mater. Phys.*, 2010, **82**, 195414.
179 J. C. W. Song, M. S. Rudner, C. M. Marcus and L. S. Levitov, *Nano Lett.*, 2011, **11**, 4688–4692.
180 M. B. Lundeberg, Y. Gao, A. Woessner, C. Tan, P. Alonso-González, K. Watanabe, T. Taniguchi, J. Hone, R. Hillenbrand and F. H. L. Koppens, *Nat. Mater.*, 2017, **16**, 204–207.
181 Q. Ma, N. M. Gabor, T. I. Andersen, N. L. Nair, K. Watanabe, T. Taniguchi and P. Jarillo-Herrero, *Phys. Rev. Lett.*, 2014, **112**, 247401.
182 D. A. Bandurin, I. Torre, R. Krishna Kumar, M. Ben Shalom, A. Tomadin, A. Principi, G. H. Auton, E. Khestanova, K. S. Novoselov, I. V. Grigorieva, L. A. Ponomarenko, A. K. Geim and M. Polini, *Science*, 2016, **351**, 1055–1058.
183 S. Huberman, R. A. Duncan, K. Chen, B. Song, V. Chiloyan, Z. Ding, A. A. Maznev, G. Chen and K. A. Nelson, *Science*, 2019, **364**, 375–379.
184 D. Li, Y. Gong, Y. Chen, J. Lin, Q. Khan, Y. Zhang, Y. Li, H. Zhang and H. Xie, *Nano-Micro Lett.*, 2020, **12**, 36.

185 Y. Li, Y. Tian, M. Sun, T. Tu, Z. Ju, G. Gou, Y. Zhao, Z. Yan, F. Wu, D. Xie, H. Tian, Y. Yang and T. Ren, *Adv. Funct. Mater.*, 2020, **30**, 1903888.
186 J. Wu, Y. Chen, J. Wu and K. Hippalgaonkar, *Adv. Electron. Mater.*, 2018, **4**, 1800248.
187 J. Wang, X. Mu and M. Sun, *Nanomaterials*, 2019, **9**, 218.
188 P. Dollfus, V. Hung Nguyen and J. Saint-Martin, *J. Phys.: Condens. Matter*, 2015, **27**, 133204.
189 Q. Bao, H. Zhang, B. Wang, Z. Ni, C. H. Y. X. Lim, Y. Wang, D. Y. Tang and K. P. Loh, *Nat. Photonics*, 2011, **5**, 411–415.
190 F. Ghahari, H.-Y. Y. Xie, T. Taniguchi, K. Watanabe, M. S. Foster and P. Kim, *Phys. Rev. Lett.*, 2016, **116**, 1–5.
191 J. G. Checkelsky and N. P. Ong, *Phys. Rev. B: Condens. Matter Mater. Phys.*, 2009, **80**, 081413.
192 J. F. Sierra, I. Neumann, M. V. Costache and S. O. Valenzuela, *Nano Lett.*, 2015, **15**, 4000–4005.
193 D. Wang and J. Shi, *Phys. Rev. B: Condens. Matter Mater. Phys.*, 2011, **83**, 113403.
194 E. H. Hwang, E. Rossi and S. Das Sarma, *Phys. Rev. B: Condens. Matter Mater. Phys.*, 2009, **80**, 235415.
195 X. Wu, Y. Hu, M. Ruan, N. K. Madiomanana, C. Berger and W. A. de Heer, *Appl. Phys. Lett.*, 2011, **99**, 133102.
196 J. Duan, X. Wang, X. Lai, G. Li, K. Watanabe, T. Taniguchi, M. Zebarjadi and E. Y. Andrei, *Proc. Natl. Acad. Sci. U. S. A.*, 2016, **113**, 14272–14276.
197 M. S. Foster and I. L. Aleiner, *Phys. Rev. B: Condens. Matter Mater. Phys.*, 2009, **79**, 085415.
198 M. Müller, L. Fritz and S. Sachdev, *Phys. Rev. B: Condens. Matter Mater. Phys.*, 2008, **78**, 115406.
199 T. I. Andersen, T. B. Smith and A. Principi, *Phys. Rev. Lett.*, 2019, **122**, 166802.
200 A. Block, Ph.D. thesis, UPC, Institut de Ciències Fotòniques, 2019.
201 J. C. W. Song and L. S. Levitov, *Phys. Rev. B: Condens. Matter Mater. Phys.*, 2014, **90**, 075415.
202 M. B. Lundeberg and F. H. L. Koppens, 2020, arXiv:2011.04311.
203 X. Xu, N. M. Gabor, J. S. Alden, A. M. van der Zande and P. L. McEuen, *Nano Lett.*, 2010, **10**, 562–566.
204 T. J. Echtermeyer, P. S. Nene, M. Trushin, R. V. Gorbachev, A. L. Eiden, S. Milana, Z. Sun, J. Schliemann, E. Lidorikis, K. S. Novoselov and A. C. Ferrari, *Nano Lett.*, 2014, **14**, 3733–3742.
205 V. Shautsova, T. Sidiropoulos, X. Xiao, N. A. Güsken, N. C. G. Black, A. M. Gilbertson, V. Giannini, S. A. Maier, L. F. Cohen and R. F. Oulton, *Nat. Commun.*, 2018, **9**, 5190.
206 M. Freitag, T. Low, F. Xia and P. Avouris, *Nat. Photonics*, 2013, **7**, 53–59.
207 I. J. Vera-Marun, J. J. van den Berg, F. K. Dejene and B. J. van Wees, *Nat. Commun.*, 2016, **7**, 11525.
208 X. Hu, X. Gong, M. Zhang, H. Lu, Z. Xue, Y. Mei, P. K. Chu, Z. An and Z. Di, *Small*, 2020, **16**, 1907170.
209 A. Harzheim, J. Spiece, C. Evangeli, E. McCann, V. Falko, Y. Sheng, J. H. Warner, G. A. D. Briggs, J. A. Mol, P. Gehring and O. V. Kolosov, *Nano Lett.*, 2018, **18**, 7719–7725.
210 K. Zarembo, *JETP Lett.*, 2020, **111**, 157–161.
211 H. Cao, G. Aivazian, Z. Fei, J. Ross, D. H. Cobden and X. Xu, *Nat. Phys.*, 2016, **12**, 236–239.
212 K. S. Novoselov, A. Mishchenko, A. Carvalho and A. H. Castro Neto, *Science*, 2016, **353**, aac9439.
213 A. K. Geim and I. V. Grigorieva, *Nature*, 2013, **499**, 419–425.
214 R. Frisenda, E. Navarro-Moratalla, P. Gant, D. Pérez De Lara, P. Jarillo-Herrero, R. V. Gorbachev and A. Castellanos-Gomez, *Chem. Soc. Rev.*, 2018, **47**, 53–68.
215 Y. Cao, V. Fatemi, S. Fang, K. Watanabe, T. Taniguchi, E. Kaxiras and P. Jarillo-Herrero, *Nature*, 2018, **556**, 43–50.
216 Q. Ma, T. I. Andersen, N. L. Nair, N. M. Gabor, M. Massicotte, C. H. Lui, A. F. Young, W. Fang, K. Watanabe, T. Taniguchi, J. Kong, N. Gedik, F. H. L. Koppens and P. Jarillo-Herrero, *Nat. Phys.*, 2016, **12**, 455–459.
217 S. Dushman, *Phys. Rev.*, 1923, **21**, 623–636.
218 O. Richardson, *London, Edinburgh Dublin Philos. Mag. J. Sci.*, 1912, **23**, 594–627.
219 R. H. Fowler and L. Nordheim, *Proc. R. Soc. London, Ser. A*, 1928, **119**, 173–181.
220 E. L. Murphy and R. H. Good, *Phys. Rev.*, 1956, **102**, 1464–1473.
221 J. G. Simmons, *J. Appl. Phys.*, 1963, **34**, 1793–1803.
222 F. Capasso, K. Mohammed and A. Cho, *IEEE J. Quantum Electron.*, 1986, **22**, 1853–1869.
223 Y. Ang, S.-J. Liang and L. Ang, *MRS Bull*, 2017, **42**, 505–510.
224 S. Bruzzone, D. Logoteta, G. Fiori and G. Iannaccone, *Sci. Rep.*, 2015, **5**, 14519.
225 R. Ahsan, M. A. Sakib, H. U. Chae and R. Kapadia, *Phys. Rev. Appl.*, 2020, **13**, 024060.
226 Y. S. Ang, Y. Chen, C. Tan and L. K. Ang, *Phys. Rev. Appl.*, 2019, **12**, 014057.
227 Y. S. Ang, H. Y. Yang and L. K. Ang, *Phys. Rev. Lett.*, 2018, **121**, 056802.
228 S.-J. Liang and L. K. Ang, *Phys. Rev. Appl.*, 2015, **3**, 014002.
229 D. Sinha and J. U. Lee, *Nano Lett.*, 2014, **14**, 4660–4664.
230 S. Tongay, M. Lemaitre, X. Miao, B. Gila, B. R. Appleton and a. F. Hebard, *Phys. Rev. X*, 2012, **2**, 011002.
231 Y. Liu, N. O. Weiss, X. Duan, H.-c. Cheng, Y. Huang and X. Duan, *Nat. Rev. Mater.*, 2016, **1**, 16042.
232 L. Britnell, R. V. Gorbachev, R. Jalil, B. D. Belle, F. Schedin, M. I. Katsnelson, L. Eaves, S. V. Morozov, A. S. Mayorov, N. M. R. Peres, A. H. Castro Neto, J. Leist, A. K. Geim, L. A. Ponomarenko and K. S. Novoselov, *Nano Lett.*, 2012, **12**, 1707–1710.
233 L. Britnell, R. V. Gorbachev, R. Jalil, B. D. Belle, F. Schedin, A. Mishchenko, T. Georgiou, M. I. Katsnelson, L. Eaves, S. V. Morozov, N. M. R. Peres, J. Leist, a. K. Geim, K. S. Novoselov and L. a. Ponomarenko, *Science*, 2012, **335**, 947–950.

234 T. Georgiou, R. Jalil, B. D. Belle, L. Britnell, R. V. Gorbachev, S. V. Morozov, Y.-J. Kim, A. Gholinia, S. J. Haigh, O. Makarovsky, L. Eaves, L. A. Ponomarenko, A. K. Geim, K. S. Novoselov and A. Mishchenko, *Nat. Nanotechnol.*, 2013, **8**, 100–103.
235 A. Mishchenko, J. S. Tu, Y. Cao, R. V. Gorbachev, J. R. Wallbank, M. T. Greenaway, V. E. Morozov, S. V. Morozov, M. J. Zhu, S. L. Wong, F. Withers, C. R. Woods, Y.-J. Kim, K. Watanabe, T. Taniguchi, E. E. Vdovin, O. Makarovsky, T. M. Fromhold, V. I. Fal'ko, A. K. Geim, L. Eaves and K. S. Novoselov, *Nat. Nanotechnol.*, 2014, **9**, 808–813.
236 L. Britnell, R. V. Gorbachev, A. K. Geim, L. A. Ponomarenko, A. Mishchenko, M. T. Greenaway, T. M. Fromhold, K. S. Novoselov and L. Eaves, *Nat. Commun.*, 2013, **4**, 1794.
237 B. Amorim, R. M. Ribeiro and N. M. R. Peres, *Phys. Rev. B*, 2016, **93**, 235403.
238 L. Brey, *Phys. Rev. Appl.*, 2014, **2**, 014003.
239 J. F. Rodriguez-Nieva, M. S. Dresselhaus and L. S. Levitov, *Nano Lett.*, 2015, **15**, 1451–1456.
240 J. F. Rodriguez-Nieva, M. S. Dresselhaus and J. C. W. Song, *Nano Lett.*, 2016, **16**, 6036–6041.
241 M. Massicotte, P. Schmidt, F. Vialla, K. Watanabe, T. Taniguchi, K. J. Tielrooij and F. H. L. Koppens, *Nat. Commun.*, 2016, **7**, 12174.
242 I. Goykhman, U. Sassi, B. Desiatov, N. Mazurski, S. Milana, D. de Fazio, A. Eiden, J. Khurgin, J. Shappir, U. Levy and A. C. Ferrari, *Nano Lett.*, 2016, **16**, 3005–3013.
243 A. Hartstein and Z. A. Weinberg, *Phys. Rev. B: Condens. Matter Mater. Phys.*, 1979, **20**, 1335–1338.
244 C. Heide, M. Hauck, T. Higuchi, J. Ristein, L. Ley, H. B. Weber and P. Hommelhoff, *Nat. Photonics*, 2020, **14**, 219–222.
245 S. Fu, I. du Fossé, X. Jia, J. Xu, X. Yu, H. Zhang, W. Zheng, S. Krasel, Z. Chen, Z. M. Wang, K.-J. Tielrooij, M. Bonn, A. J. Houtepen and H. I. Wang, *Sci. Adv.*, 2021, **7**, eabd9061.
246 L. Yuan, T.-F. Chung, A. Kuc, Y. Wan, Y. Xu, Y. P. Chen, T. Heine and L. Huang, *Sci. Adv.*, 2018, **4**, e1700324.
247 Y. Chen, Y. Li, Y. Zhao, H. Zhou and H. Zhu, *Sci. Adv.*, 2019, **5**, eaax9958.
248 N. Poudel, S.-J. Liang, D. Choi, B. Hou, L. Shen, H. Shi, L. K. Ang, L. Shi and S. Cronin, *Sci. Rep.*, 2017, **7**, 14148.
249 C.-C. C. Chen, Z. Li, L. Shi and S. B. Cronin, *Nano Res.*, 2015, **8**, 666–672.
250 M. G. Rosul, D. Lee, D. H. Olson, N. Liu, X. Wang, P. E. Hopkins, K. Lee and M. Zebarjadi, *Sci. Adv.*, 2019, **5**, eaax7827.
251 D. Vashaee and A. Shakouri, *J. Appl. Phys.*, 2004, **95**, 1233–1245.
252 G. D. Mahan and L. M. Woods, *Phys. Rev. Lett.*, 1998, **80**, 4016–4019.
253 S. Misra, M. Upadhyay Kahaly and S. K. Mishra, *J. Appl. Phys.*, 2017, **121**, 065102.
254 X. Wang, M. Zebarjadi and K. Esfarjani, *Sci. Rep.*, 2018, **8**, 9303.
255 X. Wang, M. Zebarjadi and K. Esfarjani, *Nanoscale*, 2016, **8**, 14695–14704.
256 S.-J. Liang, B. Liu, W. Hu, K. Zhou and L. K. Ang, *Sci. Rep.*, 2017, **7**, 46211.
257 H. Yuan, D. C. Riley, Z.-X. Shen, P. A. Pianetta, N. A. Melosh and R. T. Howe, *Nano Energy*, 2017, **32**, 67–72.
258 A. Block, M. Liebel, R. Yu, M. Spector, Y. Sivan, F. J. García de Abajo and N. F. van Hulst, *Sci. Adv.*, 2019, **5**, eaav8965.
259 T. Tan, X. Jiang, C. Wang, B. Yao and H. Zhang, *Sci. Adv.*, 2020, **7**, 2000058.
260 C.-H. Liu, J. Zheng, Y. Chen, T. Fryett and A. Majumdar, *Opt. Mater. Express*, 2019, **9**, 384.
261 Q. Bao and K. P. Loh, *ACS Nano*, 2012, **6**, 3677–3694.
262 X. Lu, L. Sun, P. Jiang and X. Bao, *Adv. Mater.*, 2019, **31**, 1–26.
263 X. Du, D. E. Prober, H. Vora and C. B. Mckitterick, *Graphene 2D Mater.*, 2014, **1**, 1–22.
264 S. Castilla, I. Vangelidis, V.-V. Pusapati, J. Goldstein, M. Autore, T. Slipchenko, K. Rajendran, S. Kim, K. Watanabe, T. Taniguchi, L. Martín-Moreno, D. Englund, K.-J. Tielrooij, R. Hillenbrand, E. Lidorikis and F. H. L. Koppens, *Nat. Commun.*, 2020, **11**, 4872.
265 G.-H. Lee, D. K. Efetov, W. Jung, L. Ranzani, E. D. Walsh, T. A. Ohki, T. Taniguchi, K. Watanabe, P. Kim, D. Englund and K. C. Fong, *Nature*, 2020, **586**, 42–46.
266 P. K. Herring, A. L. Hsu, N. M. Gabor, Y. C. Shin, J. Kong, T. Palacios and P. Jarillo-Herrero, *Nano Lett.*, 2014, **14**, 901–907.
267 M. Jung, P. Rickhaus, S. Zihlmann, P. Makk and C. Schönenberger, *Nano Lett.*, 2016, **16**, 6988–6993.
268 S. Yuan, R. Yu, C. Ma, B. Deng, Q. Guo, X. Chen, C. Li, C. Chen, K. Watanabe, T. Taniguchi, F. J. García de Abajo and F. Xia, *ACS Photonics*, 2020, **7**, 1206–1215.
269 A. El Fatimy, R. L. Myers-Ward, A. K. Boyd, K. M. Daniels, D. K. Gaskill and P. Barbara, *Nat. Nanotechnol.*, 2016, **11**, 335–338.
270 J. Yan, M.-H. Kim, J. A. Elle, A. B. Sushkov, G. S. Jenkins, H. M. Milchberg, M. S. Fuhrer and H. D. Drew, *Nat. Nanotechnol.*, 2012, **7**, 472–478.
271 D. K. Efetov, R.-J. Shiue, Y. Gao, B. Skinner, E. D. Walsh, H. Choi, J. Zheng, C. Tan, G. Grosso, C. Peng, J. Hone, K. C. Fong and D. Englund, *Nat. Nanotechnol.*, 2018, **13**, 797–801.
272 F. Xia, H. Wang, D. Xiao, M. Dubey and A. Ramasubramaniam, *Nat. Photonics*, 2014, **8**, 899–907.
273 Y. Yao, R. Shankar, P. Rauter, Y. Song, J. Kong, M. Loncar and F. Capasso, *Nano Lett.*, 2014, **14**, 3749–3754.
274 G. Skoblin, J. Sun and A. Yurgens, *Appl. Phys. Lett.*, 2018, **112**, 063501.
275 L. Vicarelli, M. S. Vitiello, D. Coquillat, A. Lombardo, A. C. Ferrari, W. Knap, M. Polini, V. Pellegrini and A. Tredicucci, *Nat. Mater.*, 2012, **11**, 865–871.
276 G. Auton, D. B. But, J. Zhang, E. Hill, D. Coquillat, C. Consejo, P. Nouvel, W. Knap, L. Varani, F. Teppe, J. Torres and A. Song, *Nano Lett.*, 2017, **17**, 7015–7020.

277 M. Mittendorff, S. Winnerl, J. Kamann, J. Eroms, D. Weiss, H. Schneider and M. Helm, *Appl. Phys. Lett.*, 2013, **103**, 021113.
278 T. Low, A. Chaves, J. D. Caldwell, A. Kumar, N. X. Fang, P. Avouris, T. F. Heinz, F. Guinea, L. Martin-Moreno and F. Koppens, *Nat. Mater.*, 2017, **16**, 182–194.
279 Y. Liu, R. Cheng, L. Liao, H. Zhou, J. Bai, G. Liu, L. Liu, Y. Huang and X. Duan, *Nat. Commun.*, 2011, **2**, 579.
280 A. Safaei, S. Chandra, M. W. Shabbir, M. N. Leuenberger and D. Chanda, *Nat. Commun.*, 2019, **10**, 3498.
281 D. A. Bandurin, D. Svintsov, I. Gayduchenko, S. G. Xu, A. Principi, M. Moskotin, I. Tretyakov, D. Yagodkin, S. Zhukov, T. Taniguchi, K. Watanabe, I. V. Grigorieva, M. Polini, G. N. Goltsman, A. K. Geim and G. Fedorov, *Nat. Commun.*, 2018, **9**, 5392.
282 M. Freitag, T. Low, W. Zhu, H. Yan, F. Xia and P. Avouris, *Nat. Commun.*, 2013, **4**, 1951.
283 L. Viti, D. G. Purdie, A. Lombardo, A. C. Ferrari and M. S. Vitiello, *Nano Lett.*, 2020, **20**, 3169–3177.
284 T. Van Muoi, *Optoelectronic Technology and Lightwave Communications Systems*, Springer Netherlands, Dordrecht, 1989, vol. 12, pp. 441–472.
285 F. Xia, T. Mueller, Y.-m. Lin, A. Valdes-Garcia and P. Avouris, *Nat. Nanotechnol.*, 2009, **4**, 839–843.
286 R.-J. Shiue, Y. Gao, Y. Wang, C. Peng, A. D. Robertson, D. K. Efetov, S. Assefa, F. H. L. Koppens, J. Hone and D. Englund, *Nano Lett.*, 2015, **15**, 7288–7293.
287 V. Mišeikis, S. Marconi, M. A. Giambra, A. Montanaro, L. Martini, F. Fabbri, S. Pezzini, G. Piccinini, S. Forti, B. Terrés, I. Goykhman, L. Hamidouche, P. Legagneux, V. Sorianello, A. C. Ferrari, F. H. L. Koppens, M. Romagnoli and C. Coletti, *ACS Nano*, 2020, **14**, 11190–11204.
288 A. Pospischil, M. Humer, M. M. Furchi, D. Bachmann, R. Guider, T. Fromherz and T. Mueller, *Nat. Photonics*, 2013, **7**, 892–896.
289 J. E. Muench, A. Ruocco, M. A. Giambra, V. Miseikis, D. Zhang, J. Wang, H. F. Y. Watson, G. C. Park, S. Akhavan, V. Sorianello, M. Midrio, A. Tomadin, C. Coletti, M. Romagnoli, A. C. Ferrari and I. Goykhman, *Nano Lett.*, 2019, **19**, 7632–7644.
290 S. Schuler, D. Schall, D. Neumaier, L. Dobusch, O. Bethge, B. Schwarz, M. Krall and T. Mueller, *Nano Lett.*, 2016, **16**, 7107–7112.
291 S. Schuler, D. Schall, D. Neumaier, B. Schwarz, K. Watanabe, T. Taniguchi and T. Mueller, *ACS Photonics*, 2018, **5**, 4758–4763.
292 D. Schall, D. Neumaier, M. Mohsin, B. Chmielak, J. Bolten, C. Porschatis, A. Prinzen, C. Matheisen, W. Kuebart, B. Junginger, W. Templ, A. L. Giesecke and H. Kurz, *ACS Photonics*, 2014, **1**, 781–784.
293 X. Gan, R.-J. Shiue, Y. Gao, I. Meric, T. F. Heinz, K. Shepard, J. Hone, S. Assefa and D. Englund, *Nat. Photonics*, 2013, **7**, 883–887.
294 Y. Gao, G. Zhou, H. K. Tsang and C. Shu, *Optica*, 2019, **6**, 514.
295 D. Schall, E. Pallecchi, G. Ducournau, V. Avramovic, M. Otto and D. Neumaier, Optical Fiber Communication Conference, Washington, D.C., 2018, p. M2I.4.
296 P. Ma, Y. Salamin, B. Baeuerle, A. Josten, W. Heni, A. Emboras and J. Leuthold, *ACS Photonics*, 2019, **6**, 154–161.
297 D. Schall, C. Porschatis, M. Otto and D. Neumaier, *J. Phys. D: Appl. Phys.*, 2017, **50**, 124004.
298 D. Akinwande, C. Huyghebaert, C.-H. Wang, M. I. Serna, S. Goossens, L.-J. Li, H.-S. P. Wong and F. H. L. Koppens, *Nature*, 2019, **573**, 507–518.
299 W. Li, B. Chen, C. Meng, W. Fang, Y. Xiao, X. Li, Z. Hu, Y. Xu, L. Tong, H. Wang, W. Liu, J. Bao and Y. R. Shen, *Nano Lett.*, 2014, **14**, 955–959.
300 C. T. Phare, Y.-H. Daniel Lee, J. Cardenas and M. Lipson, *Nat. Photonics*, 2015, **9**, 511–514.
301 B. Yao, S.-W. Huang, Y. Liu, A. K. Vinod, C. Choi, M. Hoff, Y. Li, M. Yu, Z. Feng, D.-L. Kwong, Y. Huang, Y. Rao, X. Duan and C. W. Wong, *Nature*, 2018, **558**, 410–414.
302 F. Bonaccorso, Z. Sun, T. Hasan and A. C. Ferrari, *Nat. Photonics*, 2010, **4**, 611–622.
303 Z. Sun, A. Martinez and F. Wang, *Nat. Photonics*, 2016, **10**, 227–238.
304 V. Sorianello, M. Midrio, G. Contestabile, I. Asselberghs, J. Van Campenhout, C. Huyghebaert, I. Goykhman, A. K. Ott, A. C. Ferrari and M. Romagnoli, *Nat. Photonics*, 2018, **12**, 40–44.
305 J.-L. Xu, X.-L. Li, Y.-Z. Wu, X.-P. Hao, J.-L. He and K.-J. Yang, *Opt. Lett.*, 2011, **36**, 1948.
306 G. Q. Xie, J. Ma, P. Lv, W. L. Gao, P. Yuan, L. J. Qian, H. H. Yu, H. J. Zhang, J. Y. Wang and D. Y. Tang, *Opt. Mater. Express*, 2012, **2**, 878.
307 I. H. Baek, H. W. Lee, S. Bae, B. H. Hong, Y. H. Ahn, D.-I. Yeom and F. Rotermund, *Appl. Phys. Express*, 2012, **5**, 032701.
308 Z. Sun, T. Hasan, F. Torrisi, D. Popa, G. Privitera, F. Wang, F. Bonaccorso, D. M. Basko and A. C. Ferrari, *ACS Nano*, 2010, **4**, 803–810.
309 P. L. Huang, S.-C. Lin, C.-Y. Yeh, H.-H. Kuo, S-H. Huang, G.-R. Lin, L.-J. Li, C.-Y. Su and W.-H. Cheng, *Opt. Express*, 2012, **20**, 2460.
310 A. Martinez and S. Yamashita, *Appl. Phys. Lett.*, 2012, **101**, 041118.
311 Bo Fu, Yi Hua, X. Xiao, H. Zhu, Z. Sun and C. Yang, *IEEE J. Sel. Top. Quantum Electron.*, 2014, **20**, 411–415.
312 B. Yao, G. Soavi, T. Ma, X. Zhang, B. Fu, D. Yoon, S. A. Hussain, A. Lombardo, D. Popa and A. C. Ferrari, Conference on Lasers and Electro-Optics, Washington, D. C., 2018, p. SF3K.6.
313 D. Popa, D. Viola, G. Soavi, B. Fu, L. Lombardi, S. Hodge, D. Polli, T. Scopigno, G. Cerullo and A. C. Ferrari, 2017 Conference on Lasers and Electro-Optics Europe & European Quantum Electronics Conference (CLEO/Europe-EQEC), 2017, pp. 1–1.
314 V. Bianchi, T. Carey, L. Viti, L. Li, E. H. Linfield, A. G. Davies, A. Tredicucci, D. Yoon, P. G. Karagiannidis,

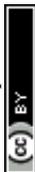

L. Lombardi, F. Tomarchio, A. C. Ferrari, F. Torrisi and M. S. Vitiello, *Nat. Commun.*, 2017, **8**, 15763.

315 Z.-B. Liu, M. Feng, W.-S. Jiang, W. Xin, P. Wang, Q.-W. Sheng, Y.-G. Liu, D. N. Wang, W.-Y. Zhou and J.-G. Tian, *Laser Phys. Lett.*, 2013, **10**, 065901.

316 S. Yu, X. Wu, K. Chen, B. Chen, X. Guo, D. Dai, L. Tong, W. Liu and Y. Ron Shen, *Optica*, 2016, **3**, 541.

317 Z. Shi, L. Gan, T.-H. Xiao, H.-L. Guo and Z.-Y. Li, *ACS Photonics*, 2015, **2**, 1513–1518.

318 M. Ono, M. Hata, M. Tsunekawa, K. Nozaki, H. Sumikura, H. Chiba and M. Notomi, *Nat. Photonics*, 2020, **14**, 37–43.

319 Y. Cheng, H. Hong, H. Zhao, C. Wu, Y. Pan, C. Liu, Y. Zuo, Z. Zhang, J. Xie, J. Wang, D. Yu, Y. Ye, S. Meng and K. Liu, *Nano Lett.*, 2020, **20**, 8053–8058.

320 Y. D. Kim, Y. Gao, R.-J. Shiue, L. Wang, O. B. Aslan, M.-H. Bae, H. Kim, D. Seo, H.-J. Choi, S. H. Kim, A. Nemilentsau, T. Low, C. Tan, D. K. Efetov, T. Taniguchi, K. Watanabe, K. L. Shepard, T. F. Heinz, D. Englund and J. Hone, *Nano Lett.*, 2018, **18**, 934–940.

321 S.-K. Son, M. Šiškins, C. Mullan, J. Yin, V. G. Kravets, A. Kozikov, S. Ozdemir, M. Alhazmi, M. Holwill, K. Watanabe, T. Taniguchi, D. Ghazaryan, K. S. Novoselov, V. I. Fal'ko and A. Mishchenko, *2D Mater.*, 2017, **5**, 011006.

322 H. R. Barnard, E. Zossimova, N. H. Mahlmeister, L. M. Lawton, I. J. Luxmoore and G. R. Nash, *Appl. Phys. Lett.*, 2016, **108**, 131110.

323 R.-J. Shiue, Y. Gao, C. Tan, C. Peng, J. Zheng, D. K. Efetov, Y. D. Kim, J. Hone and D. Englund, *Nat. Commun.*, 2019, **10**, 109.

324 F. Luo, Y. Fan, G. Peng, S. Xu, Y. Yang, K. Yuan, J. Liu, W. Ma, W. Xu, Z. H. Zhu, X.-A. Zhang, A. Mishchenko, Y. Ye, H. Huang, Z. Han, W. Ren, K. S. Novoselov, M. Zhu and S. Qin, *ACS Photonics*, 2019, **6**, 2117–2125.

325 M. Engel, M. Steiner, A. Lombardo, A. C. Ferrari, H. v. Löhneysen, P. Avouris and R. Krupke, *Nat. Commun.*, 2012, **3**, 906.

326 S. Anisimov, B. Kapeliovich and T. Perel'Man, *Sov. J. Exp. Theor. Phys.*, 1974, **39**, 776–781.

327 J. G. Fujimoto, J. M. Liu, E. P. Ippen and N. Bloembergen, *Phys. Rev. Lett.*, 1984, **53**, 1837–1840.

328 M. L. Roukes, M. R. Freeman, R. S. Germain, R. C. Richardson and M. B. Ketchen, *Phys. Rev. Lett.*, 1985, **55**, 422–425.

329 R. W. Schoenlein, W. Z. Lin, J. G. Fujimoto and G. L. Eesley, *Phys. Rev. Lett.*, 1987, **58**, 1680–1683.

330 H. E. Elsayed-Ali, T. B. Norris, M. A. Pessot and G. A. Mourou, *Phys. Rev. Lett.*, 1987, **58**, 1212–1215.

331 C.-K. Sun, F. Vallée, L. H. Acioli, E. P. Ippen and J. G. Fujimoto, *Phys. Rev. B: Condens. Matter Mater. Phys.*, 1994, **50**, 15337–15348.

332 F. C. Wellstood, C. Urbina and J. Clarke, *Phys. Rev. B: Condens. Matter Mater. Phys.*, 1994, **49**, 5942–5955.

333 N. Del Fatti, C. Voisin, M. Achermann, S. Tzortzakis, D. Christofilos and F. Vallée, *Phys. Rev. B: Condens. Matter Mater. Phys.*, 2000, **61**, 16956–16966.

334 G. Della Valle, M. Conforti, S. Longhi, G. Cerullo and D. Brida, *Phys. Rev. B: Condens. Matter Mater. Phys.*, 2012, **86**, 155139.

335 Y. Dubi and Y. Sivan, *Light: Sci. Appl.*, 2019, **8**, 89.

336 J. Hohlfeld, S.-S. Wellershoff, J. Güdde, U. Conrad, V. Jähnke and E. Matthias, *Chem. Phys.*, 2000, **251**, 237–258.

337 M. Z. Hasan and C. L. Kane, *Rev. Mod. Phys.*, 2010, **82**, 3045–3067.

338 S. Sim, M. Brahlek, N. Koirala, S. Cha, S. Oh and H. Choi, *Phys. Rev. B: Condens. Matter Mater. Phys.*, 2014, **89**, 165137.

339 Y. H. Wang, D. Hsieh, E. J. Sie, H. Steinberg, D. R. Gardner, Y. S. Lee, P. Jarillo-Herrero and N. Gedik, *Phys. Rev. Lett.*, 2012, **109**, 127401.

340 J. A. Sobota, S. Yang, J. G. Analytis, Y. L. Chen, I. R. Fisher, P. S. Kirchmann and Z.-X. Shen, *Phys. Rev. Lett.*, 2012, **108**, 117403.

341 M. Hajlaoui, E. Papalazarou, J. Mauchain, G. Lantz, N. Moisan, D. Boschetto, Z. Jiang, I. Miotkowski, Y. P. Chen, A. Taleb-Ibrahimi, L. Perfetti and M. Marsi, *Nano Lett.*, 2012, **12**, 3532–3536.

342 M. Neupane, S.-Y. Xu, R. Sankar, N. Alidoust, G. Bian, C. Liu, I. Belopolski, T.-R. Chang, H.-T. Jeng, H. Lin, A. Bansil, F. Chou and M. Z. Hasan, *Nat. Commun.*, 2014, **5**, 3786.

343 C. Zhu, X. Yuan, F. Xiu, C. Zhang, Y. Xu, R. Zhang, Y. Shi and F. Wang, *Appl. Phys. Lett.*, 2017, **111**, 091101.

344 W. Lu, J. Ling, F. Xiu and D. Sun, *Phys. Rev. B*, 2018, **98**, 104310.

345 N. Xu, Y. Xu and J. Zhu, *npj Quantum Mater.*, 2017, **2**, 51.

346 L. Viti, D. Coquillat, A. Politano, K. A. Kokh, Z. S. Aliev, M. B. Babanly, O. E. Tereshchenko, W. Knap, E. V. Chulkov and M. S. Vitiello, *Nano Lett.*, 2016, **16**, 80–87.

347 S. Mashhadi, D. L. Duong, M. Burghard and K. Kern, *Nano Lett.*, 2017, **17**, 214–219.

348 Q. Wang, C.-Z. Li, S. Ge, J.-G. Li, W. Lu, J. Lai, X. Liu, J. Ma, D.-P. Yu, Z.-M. Liao and D. Sun, *Nano Lett.*, 2017, **17**, 834–841.

349 F. Giorgianni, E. Chiadroni, A. Rovere, M. Cestelli-Guidi, A. Perucchi, M. Bellaveglia, M. Castellano, D. Di Giovenale, G. Di Pirro, M. Ferrario, R. Pompili, C. Vaccarezza, F. Villa, A. Cianchi, A. Mostacci, M. Petrarca, M. Brahlek, N. Koirala, S. Oh and S. Lupi, *Nat. Commun.*, 2016, **7**, 11421.

350 S. Kovalev, R. M. A. Dantas, S. Germanskiy, J.-C. Deinert, B. Green, I. Ilyakov, N. Awari, M. Chen, M. Bawatna, J. Ling, F. Xiu, P. H. M. van Loosdrecht, P. Surówka, T. Oka and Z. Wang, *Nat. Commun.*, 2020, **11**, 2451.

351 T. Suzuki, T. Iimori, S. J. Ahn, Y. Zhao, M. Watanabe, J. Xu, M. Fujisawa, T. Kanai, N. Ishii, J. Itatani, K. Suwa, H. Fukidome, S. Tanaka, J. R. Ahn, K. Okazaki, S. Shin, F. Komori and I. Matsuda, *ACS Nano*, 2019, **13**, 11981–11987.

352 H. Ishizuka, A. Fahimniya, F. Guinea and L. Levitov, 2020, arXiv:2011.01701.
353 S. Vaziri, G. Lupina, C. Henkel, A. D. Smith, M. Östling, J. Dabrowski, G. Lippert, W. Mehr and M. C. Lemme, *Nano Lett.*, 2013, **13**, 1435–1439.
354 C. Zeng, E. B. Song, M. Wang, S. Lee, C. M. Torres, J. Tang, B. H. Weiller and K. L. Wang, *Nano Lett.*, 2013, **13**, 2370–2375.
355 V. B. Svetovoy and G. Palasantzas, *Phys. Rev. Appl.*, 2014, **2**, 034006.